\documentclass[twocolumn]{aastex63}

\usepackage{chngcntr}
\usepackage{longtable}
\usepackage{placeins}

\usepackage{amsmath}

\def \zis{z_{\rm IS}}
\def \zwind{z_{\rm wind}}
\def \ziswind{z_{\rm IS + wind}}
\def \zlya{z_{\rm Ly\alpha}}
\def \zcii{z_{\rm [C{\scriptsize\,II}]}}
\def \hbeta{H$\beta$}
\def \ebmv{E(B-V)}
\def \ebmvs{ E_{s}{\rm (B-V)} }
\def \ebmvn{ E_{n}{\rm (B-V)} }
\def \halpha{H$\alpha$}
\def \Msol{{M}_{\odot}}

\def \Lsol{{\rm L}_{\odot}}
\def \Zsol{{\rm Z}_{\odot}}

\def \logm{\log({\rm M}/{\rm \Msol})}
\def \logsfr{\log({\rm SFR}/{[\rm \Msol\,yr^{-1}]}) }
\def \lya{Ly$\alpha$}

\def \h2{{\rm H_{2}}}

\def \hbeta{H$\beta$}
\def \halpha{H$\alpha$}
\def \siii{Si{\scriptsize\,II}}
\def \siiii{Si{\scriptsize\,III}}
\def \siiv{Si{\scriptsize\,IV}}
\def \cii{C{\scriptsize\,II}}
\def \ciii{C{\scriptsize\,II}}
\def \Cii{[C{\scriptsize\,II}]}

\def \civ{C{\scriptsize\,IV}}

\def \oi{O{\scriptsize\,I}}
\def \oii{[O{\scriptsize\,II}]}
\def \oiii{[O{\scriptsize\,III}]}
\def \heii{He{\scriptsize\,II}}
\def \nv{N{\scriptsize\,V}}
\def \feii{Fe{\scriptsize\,II}}

\def \nii{[N{\scriptsize\,II}]}

\def \logLha{\log(L_{\rm H\alpha}/{\rm L_{\rm \odot} })}

\def \dn4000{D_{{\rm n}}(4000) }

\def \iracA{[$3.6\,{\rm \mu m}$]$-$[$4.5\,{\rm \mu m}$]}

\def \ALPINE{\textit{ALPINE}}

\def \alpinedataurl{\url{http://www.astro.caltech.edu/~afaisst/}}

\graphicspath{{./}{figures/}}

\submitjournal{ApJS}

\shorttitle{The ALPINE-ALMA [CII] Survey: Ancillary Data}
\shortauthors{Faisst et al.}

\begin{document}

\title{\sc \large The ALPINE-ALMA [CII] Survey:\\
Multi-Wavelength Ancillary Data and Basic Physical Measurements}

\correspondingauthor{Andreas L. Faisst}
\email{afaisst@ipac.caltech.edu}

\author[0000-0002-9382-9832]{A. L. Faisst}
\affiliation{IPAC, M/C 314-6, California Institute of Technology, 1200 East California Boulevard, Pasadena, CA 91125, USA}

\author[0000-0001-7144-7182]{D. Schaerer}
\affiliation{Observatoire de Gen\`eve, Universit\'e de Gen\`eve, 51 Ch. des Maillettes, 1290 Versoix, Switzerland}
\affiliation{Institut de Recherche en Astrophysique et Plan\'etologie $-$ IRAP, CNRS, Universit\'e de Toulouse, UPS-OMP, 14, avenue E. Belin, F31400 Toulouse, France}

\author[0000-0002-1428-7036]{B. C. Lemaux}
\affiliation{Department of Physics, University of California, Davis, One Shields Ave., Davis, CA 95616, USA}

\author[0000-0001-5851-6649]{P. A. Oesch}
\affiliation{Observatoire de Gen\`eve, Universit\'e de Gen\`eve, 51 Ch. des Maillettes, 1290 Versoix, Switzerland}

\author[0000-0001-7440-8832]{Y. Fudamoto}
\affiliation{Observatoire de Gen\`eve, Universit\'e de Gen\`eve, 51 Ch. des Maillettes, 1290 Versoix, Switzerland}

\author[0000-0002-6716-4400]{P. Cassata}
\affiliation{Dipartimento di Fisica e Astronomia, Universit\`a di Padova, vicolo dell'Osservatorio, 3 I-35122 Padova, Italy}
\affiliation{INAF, Osservatorio Astronomico di Padova, vicolo dell'Osservatorio 5, I-35122 Padova, Italy}


\author[0000-0002-3915-2015]{M. B\'ethermin}
\affiliation{Aix Marseille Universit\'e, CNRS, LAM (Laboratoire d'Astrophysique de Marseille) UMR 7326, 13388, Marseille, France}

\author[0000-0003-3578-6843]{P. L. Capak}
\affiliation{IPAC, M/C 314-6, California Institute of Technology, 1200 East California Boulevard, Pasadena, CA 91125, USA}
\affiliation{The Cosmic Dawn Center, University of Copenhagen, Vibenshuset, Lyngbyvej 2, DK-2100 Copenhagen, Denmark}
\affiliation{Niels Bohr Institute, University of Copenhagen, Lyngbyvej 2, DK-2100 Copenhagen, Denmark}

\author[0000-0001-5891-2596]{O. Le F\`evre}
\affiliation{Aix Marseille Universit\'e, CNRS, LAM (Laboratoire d'Astrophysique de Marseille) UMR 7326, 13388, Marseille, France}

\author[0000-0002-0000-6977]{J. D. Silverman}
\affiliation{Kavli Institute for the Physics and Mathematics of the Universe, The University of Tokyo, Kashiwa, Japan 277-8583 (Kavli IPMU, WPI)}
\affiliation{Department of Astronomy, School of Science, The University of Tokyo, 7-3-1 Hongo, Bunkyo, Tokyo 113-0033, Japan}

\author[0000-0003-1710-9339]{L. Yan}
\affiliation{The Caltech Optical Observatories, California Institute of Technology, Pasadena, CA 91125, USA}


\author{M. Ginolfi}
\affiliation{Observatoire de Gen\`eve, Universit\'e de Gen\`eve, 51 Ch. des Maillettes, 1290 Versoix, Switzerland}

\author[0000-0002-6610-2048]{A. M. Koekemoer}
\affiliation{Space Telescope Science Institute, 3700 San Martin Drive, Baltimore, MD 21218, USA}

\author{L. Morselli}
\affiliation{Dipartimento di Fisica e Astronomia, Universit\`a di Padova, vicolo dell'Osservatorio, 3 I-35122 Padova, Italy}
\affiliation{INAF, Osservatorio Astronomico di Padova, vicolo dell'Osservatorio 5, I-35122 Padova, Italy}


\author[0000-0001-5758-1000]{R. Amor\'in}
\affiliation{Instituto de Investigaci\'on Multidisciplinar en Ciencia y Tecnolog\'ia, Universidad de La Serena, Ra\'ul Bitr\'an 1305, La Serena, Chile}
\affiliation{Departamento de Astronom\'ia, Universidad de La Serena, Av. Juan Cisternas 1200 Norte, La Serena, Chile}

\author[0000-0002-8900-0298]{S. Bardelli}
\affiliation{INAF - Osservatorio di Astrofisica e Scienza dello Spazio di Bologna, via Gobetti 93/3, I-40129, Bologna, Italy}

\author[0000-0003-0946-6176]{M. Boquien}
\affiliation{Centro de Astronom\'ia (CITEVA), Universidad de Antofagasta, Avenida Angamos 601, Antofagasta, Chile}

\author[0000-0003-2680-005X]{G. Brammer}
\affiliation{The Cosmic Dawn Center, University of Copenhagen, Vibenshuset, Lyngbyvej 2, DK-2100 Copenhagen, Denmark}

\author{A. Cimatti}
\affiliation{Universit\`a di Bologna - Dipartimento di Fisica e Astronomia, Via Gobetti 93/2 - I-40129, Bologna, Italy}
\affiliation{INAF - Osservatorio Astrofisico di Arcetri, Largo E. Fermi 5, I-50125, Firenze, Italy}

\author[0000-0003-0348-2917]{M. Dessauges-Zavadsky}
\affiliation{Observatoire de Gen\`eve, Universit\'e de Gen\`eve, 51 Ch. des Maillettes, 1290 Versoix, Switzerland}

\author[0000-0001-7201-5066]{S. Fujimoto}
\affiliation{The Cosmic Dawn Center, University of Copenhagen, Vibenshuset, Lyngbyvej 2, DK-2100 Copenhagen, Denmark}
\affiliation{Niels Bohr Institute, University of Copenhagen, Lyngbyvej 2, DK-2100 Copenhagen, Denmark}

\author[0000-0002-5836-4056]{C. Gruppioni}
\affiliation{INAF - Osservatorio di Astrofisica e Scienza dello Spazio di Bologna, via Gobetti 93/3, I-40129, Bologna, Italy}

\author[0000-0001-6145-5090]{N. P. Hathi}
\affiliation{Space Telescope Science Institute, 3700 San Martin Drive, Baltimore, MD 21218, USA}

\author[0000-0003-2226-5395]{S. Hemmati}
\affiliation{Jet Propulsion Laboratory, California Institute of Technology, Pasadena, CA 91109, USA}

\author{E. Ibar}
\affiliation{Instituto de F\'isica y Astronom\'ia, Universidad de Valpara\'iso, Avda. Gran Breta\~na 1111, Valpara\'iso, Chile}

\author[0000-0002-0267-9024]{G. C. Jones}
\affiliation{Cavendish Laboratory, University of Cambridge, 19 J. J. Thomson Ave., Cambridge CB3 0HE, UK}
\affiliation{Kavli Institute for Cosmology, University of Cambridge, Madingley Road, Cambridge CB3 0HA, UK}

\author{Y. Khusanova}
\affiliation{Aix Marseille Universit\'e, CNRS, LAM (Laboratoire d'Astrophysique de Marseille) UMR 7326, 13388, Marseille, France}
\affiliation{Max-Planck-Institut f\"ur Astronomie, K\"onigstuhl 17, D-69117 Heidelberg, Germany}

\author{F. Loiacono}
\affiliation{Universit\`a di Bologna - Dipartimento di Fisica e Astronomia, Via Gobetti 93/2 - I-40129, Bologna, Italy}
\affiliation{INAF - Osservatorio di Astrofisica e Scienza dello Spazio di Bologna, via Gobetti 93/3, I-40129, Bologna, Italy}

\author[0000-0002-7412-647X]{F. Pozzi}
\affiliation{Universit\`a di Bologna - Dipartimento di Fisica e Astronomia, Via Gobetti 93/2 - I-40129, Bologna, Italy}

\author[0000-0003-4352-2063]{M. Talia}
\affiliation{Universit\`a di Bologna - Dipartimento di Fisica e Astronomia, Via Gobetti 93/2 - I-40129, Bologna, Italy}
\affiliation{INAF - Osservatorio di Astrofisica e Scienza dello Spazio di Bologna, via Gobetti 93/3, I-40129, Bologna, Italy}

\author{L. A. M. Tasca}
\affiliation{Aix Marseille Universit\'e, CNRS, LAM (Laboratoire d'Astrophysique de Marseille) UMR 7326, 13388, Marseille, France}

\author[0000-0001- 9585-1462]{D. A. Riechers}
\affiliation{Department of Astronomy, Cornell University, Space Sciences Building, Ithaca, NY 14853, USA}
\affiliation{Max-Planck-Institut f\"ur Astronomie, K\"onigstuhl 17, D-69117 Heidelberg, Germany}

\author[0000-0002-9415-2296]{G. Rodighiero}
\affiliation{Dipartimento di Fisica e Astronomia, Universit\`a di Padova, vicolo dell'Osservatorio, 3 I-35122 Padova, Italy}
\affiliation{INAF, Osservatorio Astronomico di Padova, vicolo dell'Osservatorio 5, I-35122 Padova, Italy}

\author{M. Romano}
\affiliation{Dipartimento di Fisica e Astronomia, Universit\`a di Padova, vicolo dell'Osservatorio, 3 I-35122 Padova, Italy}
\affiliation{INAF, Osservatorio Astronomico di Padova, vicolo dell'Osservatorio 5, I-35122 Padova, Italy}

\author[0000-0002-0438-3323]{N. Scoville}
\affiliation{California Institute of Technology, MC 249-17, 1200 East California Boulevard, Pasadena, CA 91125, USA}

\author[0000-0003-3631-7176]{S. Toft}
\affiliation{The Cosmic Dawn Center, University of Copenhagen, Vibenshuset, Lyngbyvej 2, DK-2100 Copenhagen, Denmark}
\affiliation{Niels Bohr Institute, University of Copenhagen, Lyngbyvej 2, DK-2100 Copenhagen, Denmark}

\author[0000-0002-3258-3672]{L. Vallini}
\affiliation{Leiden Observatory, Leiden University, PO Box 9500, 2300 RA Leiden, The Netherlands}

\author{D. Vergani}
\affiliation{INAF - Osservatorio di Astrofisica e Scienza dello Spazio di Bologna, via Gobetti 93/3, I-40129, Bologna, Italy}

\author[0000-0002-2318-301X]{G. Zamorani}
\affiliation{INAF - Osservatorio di Astrofisica e Scienza dello Spazio di Bologna, via Gobetti 93/3, I-40129, Bologna, Italy}

\author[0000-0002-5845-8132]{E. Zucca}
\affiliation{INAF - Osservatorio di Astrofisica e Scienza dello Spazio di Bologna, via Gobetti 93/3, I-40129, Bologna, Italy}




\begin{abstract}
We present the ancillary data and basic physical measurements for the galaxies in the \textit{ALMA Large Program to Investigate C$^+$ at Early Times (ALPINE)} survey $-$ the first large multi-wavelength survey which aims at characterizing the gas and dust properties of $118$ main-sequence galaxies at redshifts $4.4 < z < 5.9$ via the measurement of \Cii~emission at $158\,{\rm \mu m}$ ($64\%$ at $>3.5\sigma$) and the surrounding far-infrared (FIR) continuum in conjunction with a wealth of optical and near-infrared data.
We outline in detail the spectroscopic data and selection of the galaxies as well as the ground- and space-based imaging products. In addition, we provide several basic measurements including stellar masses, star formation rates (SFR), rest-frame ultra-violet (UV) luminosities, UV continuum slopes ($\beta$), and absorption line redshifts, as well as \halpha~emission derived from Spitzer colors.
We find that the \textit{ALPINE} sample is representative of the $4 < z < 6$ galaxy population selected by photometric methods and only slightly biased towards bluer colors ($\Delta\beta \sim 0.2$).
Using \Cii~as tracer of the systemic redshift (confirmed for one galaxy at $z=4.5$ out of 118 for which we obtained optical \oii$\lambda3727{\rm \AA}$~emission), we confirm red shifted \lya~emission and blue shifted absorption lines similar to findings at lower redshifts. By stacking the rest-frame UV spectra in the \Cii~rest-frame we find that the absorption lines in galaxies with high specific SFR are more blue shifted, which could be indicative of stronger winds and outflows.
\end{abstract}



\section{Introduction} \label{sec:intro}

\begin{figure*}[t!]
\includegraphics[width=2.0\columnwidth, angle=0]{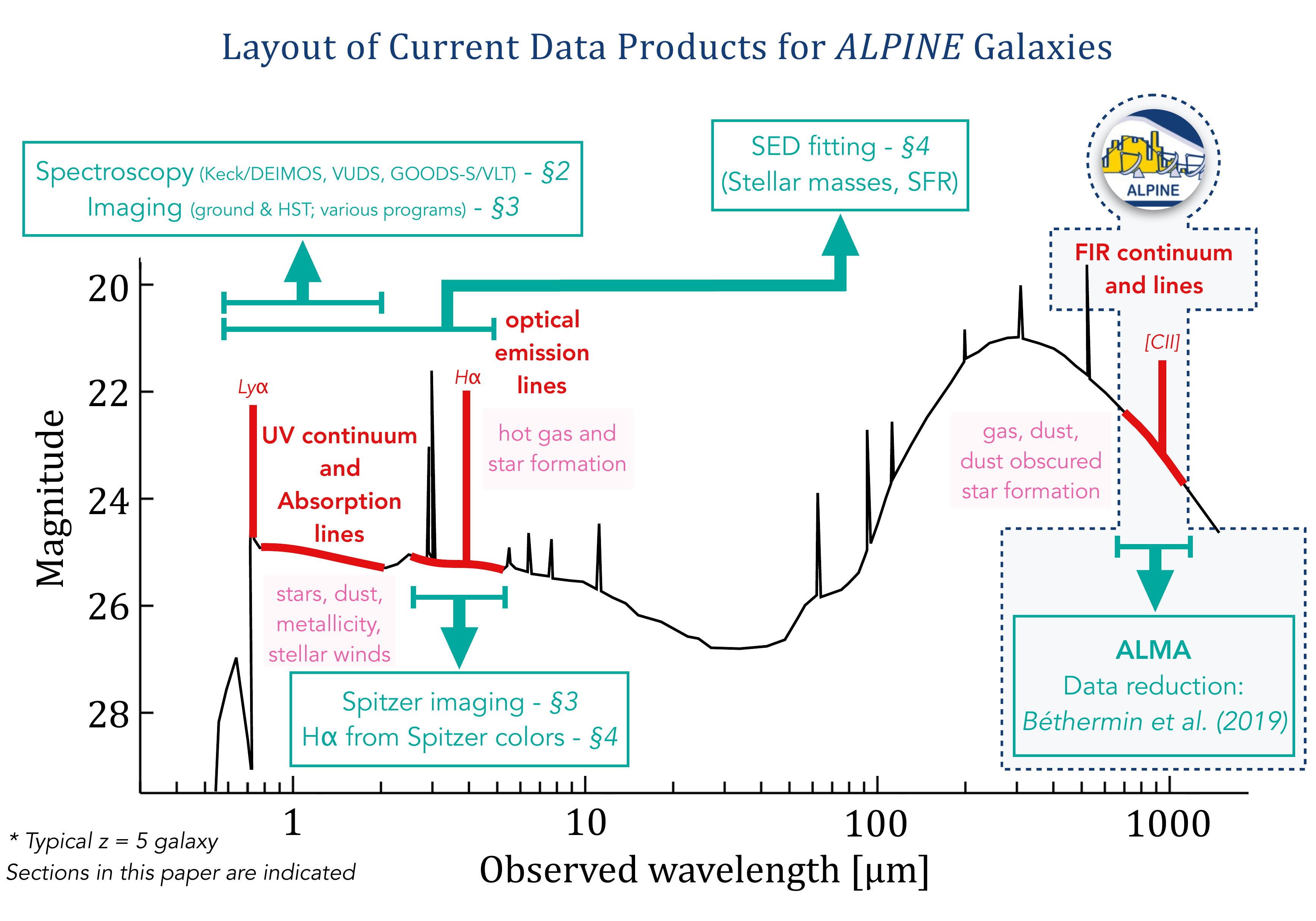}
\caption{\textit{ALPINE} builds the corner stone of a panchromatic survey at $z=4-6$. The diagram shows the multi-wavelength data products that are currently available for all the \textit{ALPINE} galaxies. The currently covered parts of the spectrum are indicated in red. The numbers link to sections in this paper where the data products and their analysis are explained in detail. The spectrum sketch is based on a typical $z=5$ galaxy \citep[adapted from][]{HARIKANE18}.  
\label{fig:ancillarydatasummary}}
\end{figure*}

\subsection{The Early Growth Phase in Galaxy Evolution}

Galaxy evolution undergoes several important phases such as the ionization of neutral Hydrogen at redshifts $z > 6$ (also known as the Epoch of Reionizaton) as well as a time of highest cosmic star-formation rate (SFR) density at $z\sim2-3$. The transition phase at $z=4-6$ (a time roughly $0.9$ to $1.5$ billion years after the Big Bang), often referred to as the \textit{early growth phase}, is currently in focus of many studies.
This time is of great interest for understanding galaxy evolution as it connects primordial galaxy formation during the epoch of reionization with mature galaxy growth at and after the peak of cosmic SFR density.
During a time of only $600\,{\rm Myrs}$, the cosmic stellar mass density in the universe increased by one order of magnitude \citep[][]{CAPUTI11,DAVIDZON17}, galaxies underwent a critical morphological transformation to build up their disk and bulge structures \citep{GNEDIN99,BOURNAUD07,AGERTZ09}, and their interstellar medium (ISM) became enriched with metal from sub-solar to solar amounts \citep{ANDO07,FAISST16b}, while at the same time the dust attenuation of the UV light significantly increased \citep{FINKELSTEIN12,BOUWENS15,FUDAMOTO17,POPPING17,CULLEN18,MA19,YAMANAKA19}. Furthermore, the most massive of these galaxies may become the first quiescent galaxies already at $z>4$ \citep{GLAZEBROOK17,VALENTINO19,TANAKA19,STOCKMANN19,FAISST19b}.
All this put together, makes the early growth phase an important puzzle piece to be studied in order to decipher how galaxies formed and evolved to become the galaxies (either star forming or quiescent) that we observe in the local universe.

It is evident from studies at lower redshift that multi-wavelength observations are crucial for us to be able to form a coherent picture of galaxy evolution. To capture several important properties of galaxies, a \textit{panchromatic} survey must comprise several spectroscopic and imaging datasets that cover a large fraction of the wavelength range of a galaxy's light emission, including
\textit{(i)} the rest-frame ultra-violet (UV) containing \lya~emission, as well as several absorption lines to study stellar winds and metallicity \citep{HECKMAN97,MARASTON09,STEIDEL10,FAISST16b}, 
\textit{(ii)} the rest-frame optical containing tracers of age (Balmer break) as well as important emission lines (e.g., \halpha) to quantify the star-formation and gas metal properties \citep{KENNICUTT98,KEWLEY08}, and
\textit{(iii)} the far-infrared (FIR) continuum and several FIR emission lines (e.g,. \Cii$\lambda158\,{\rm \mu m}$ or \nii$\lambda205\,{\rm \mu m}$) that provide insights into the gas and dust properties of galaxies \citep{DELOOZE14,PAVESI19}.

Fortunately, the early growth phase at redshifts $z~=~4~-~6$ is at the same time the highest redshift epoch at which, using current technologies, such a panchromatic study can be carried out.
The rest-frame UV part of the energy distribution at these redshifts has been probed in the past thanks to several large spectroscopic \citep{LEFEVRE15,HASINGER18} and imaging \citep{CAPAK07,MCCRACKEN12,AIHARA19} surveys from the ground as well as imaging surveys with the Hubble Space Telescope \citep[HST,][]{GROGIN11,KOEKEMOER11,SCOVILLE07b}. In addition, \halpha~has been accessed successfully through observations with the Spitzer Space Telescope \citep{STARK13,DEBARROS14,SMIT14,RASAPPU16,SMIT16,FAISST16b,FAISST19b,LAM19}. However, the FIR of $z>4$ galaxies has only been probed sparsely in the past in less than a dozen galaxies using the \textit{Atacama Large (Sub-) Millimeter Array} \citep[ALMA,][]{RIECHERS14,CAPAK15,WATSON15,WILLOTT15,STRANDET17,CARNIANI18,ZAVALA18a,ZAVALA18b,CASEY19,JIN19} as well as some as part of Herschel surveys in lensed and unlensed fields \citep[e.g.,][]{EGAMI10,COMBES12,CASEY12b,CASEY14}.
Commonly targeted by observations with ALMA is singly ionized Carbon (C$^{+}$) at $158\,{\rm \mu m}$, which is an important coolant for the gas in galaxies and is therefore broadly related to star formation activity and gas masses \citep{STACEY91,CARILLI13,DELOOZE14}.
The \Cii~emission line is one of the strongest in the FIR and is in addition conveniently located in the ALMA Band 7 at redshifts $z=4-6$ at one of the highest atmospheric transmissions compared to other FIR lines \citep[see, e.g., ][]{FAISST17b}. The origin of \Cii~emission is still debated. In addition to photo-dissociation regions (PDRs) and the cold neutral medium, a significant fraction can also origin from ionized gas regions or CO-dark molecular clouds \citep{PINEDA13,VALLINI15,PAVESI16}. Also, the increasing temperature of the Cosmic Microwave Background (CMB) has an effect on the relation between \Cii~and star formation \citep[][]{FERRARA19}. Both potentially complicates the interpretation of \Cii~as SFR indicator at high redshifts. Similar to \halpha, \Cii~traces the gas kinematics in a galaxy and is therefore an important component to quantify rotation- and dispersion-dominated systems as well as outflows \citep{JONES17,PAVESI18,KOHANDEL19,ALPINE_GINOLFI19}.

The FIR landscape has dramatically changed with the completion of the \textit{ALMA Large Program to Investigate C$^+$~at Early Times} (\textit{ALPINE}, \#2017.1.00428.L).
\textit{ALPINE} is laying the ground work for the exploration of gas and dust properties in $118$ \textit{main-sequence} star forming galaxies in the early growth phase at $4.4 < z < 5.9$ and herewith started the first panchromatic survey of its kind at these redshifts.

\subsection{ALPINE in a Nutshell}\label{sec:nutshell}

\begin{figure}[t!]
\includegraphics[width=1.0\columnwidth, angle=0]{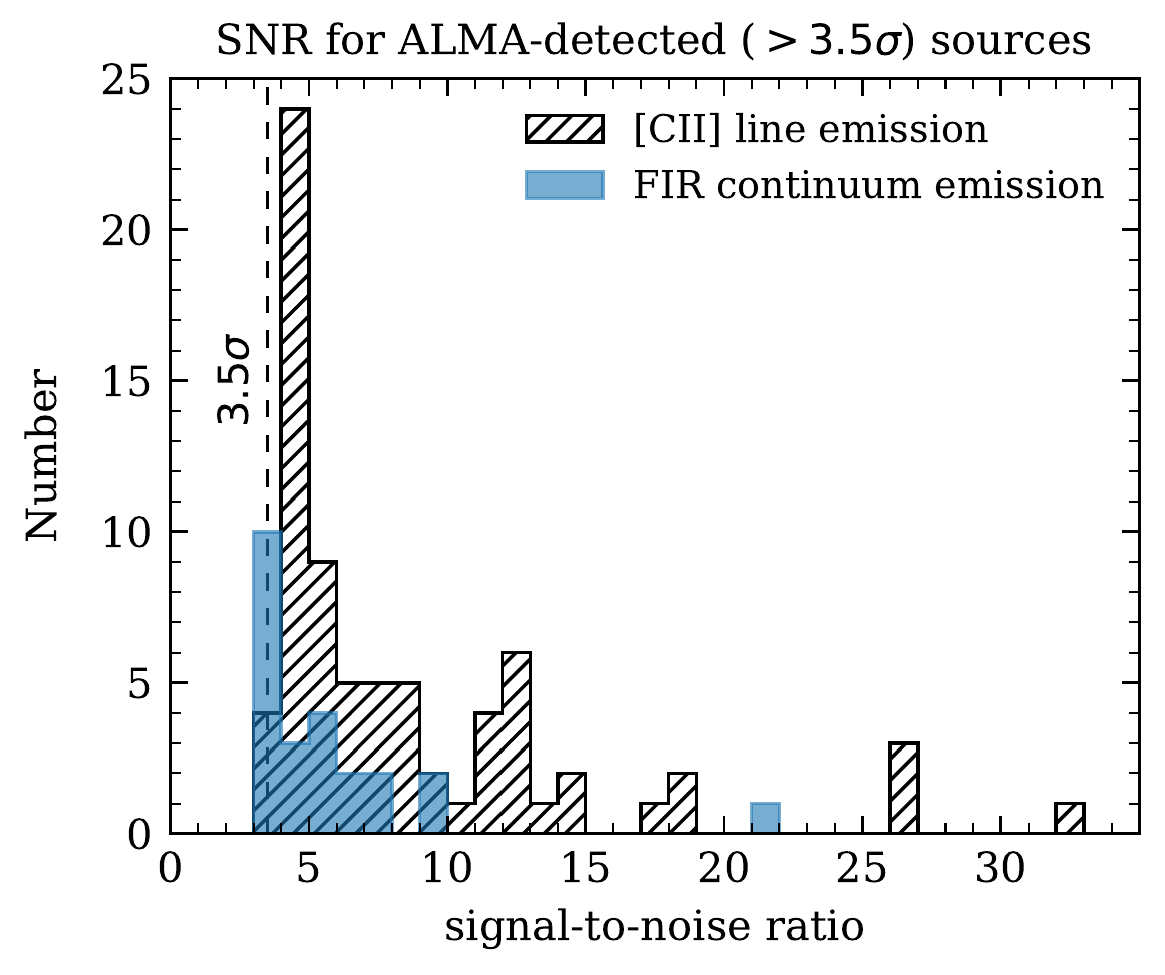}
\caption{Signal-to-Noise (SNR) of the ALMA-detected sources in the \textit{ALPINE} sample. The different histograms show the numbers for \Cii~and continuum detections above $3.5\sigma$. For more information, see \citet{ALPINE_LEFEVRE19} and \citet{ALPINE_BETHERMIN19}.
\label{fig:snr}}
\end{figure}

In the following, we summarize the scope of the \textit{ALPINE} survey, we refer to \citet{ALPINE_LEFEVRE19} for a broader overview of the program.
\textit{ALPINE} is a 69 hour large ALMA program started in Cycle 5 in May 2018 and completed during Cycle 6 in February 2019. In total, $118$ galaxies have been observed in Band 7 (covering \Cii~emission at $158\,{\rm \mu m}$ and its nearby continuum) at a spatial resolution of $<1.0\arcsec$ and with integration times $\sim30\,{\rm minutes}$ on-source depending on their predicted \Cii~flux. The galaxies origin from two fields, namely the \textit{Cosmic Evolution Survey} field \citep[COSMOS, 105 galaxies,][]{SCOVILLE07} and the \textit{Extended Chandra Deep Field South} \citep[ECDFS, 13 galaxies,][]{GIACCONI02}.
Due to gaps in the transition through the atmosphere, the galaxies are split in two different redshift ranges spanning $4.40 < z < 4.65$ and $5.05 < z < 5.90$ with medians of $\left< z \right> = 4.5$ and $5.5$ and galaxy numbers of $67$ and $51$, respectively.
All galaxies are spectroscopically confirmed by either \lya~emission or rest-UV absorption lines and are selected to be brighter than an absolute UV magnitude of $M_{\rm 1500} = -20.2$. This limit is roughly equivalent to a SFR cut at $10\,{\rm M_{\odot}\,yr^{-1}}$ and corresponds roughly to a limiting luminosity in \Cii~emission of $L_{\rm [CII]} = 1.2^{+1.9}_{-0.9} \times 10^8\,{\rm L_{\odot}}$ \citep[assuming the relation derived by][]{DELOOZE14}. Assuming a $3.5\sigma$ detection limit, the \Cii~detection rate is $64\%$ and continuum emission is detected in $19\%$ of the galaxies (see Figure~\ref{fig:snr}).

The main science goals enabled by \textit{ALPINE} are diverse and cover many crucial research topics at high redshifts:
\begin{itemize}
    \item[$-$] connecting \Cii~to star-formation at high redshifts,\vspace{-0.3cm}
    \item[$-$] coherent study of the total SFR density at $z>4$ including the contribution of dust-obscured star formation,\vspace{-0.3cm}
    \item[$-$] study of gas dynamics and merger statistics from \Cii~kinematics and quantification of UV-faint companion galaxies,\vspace{-0.3cm}
    \item[$-$] study of gas fractions and dust properties at $z>4$,\vspace{-0.3cm}
    \item[$-$] the first characterization of ISM properties using ${\rm L_{FIR}}/{\rm L_{UV}}$ and \Cii/FIR continuum diagnostics for a large sample at $z>4$,\vspace{-0.3cm}
    \item[$-$] quantifying outflows and feedback processes in $z>4$ galaxies from \Cii~line profiles.
\end{itemize}{}

Note that \textit{ALPINE} provides at the same time the equivalent of a blind-survey of approximately $25$ square-arcminutes. This enables us to estimate the obscured fraction of star-formation (mostly below $z=4$) by finding UV-faint galaxies with FIR continuum or \Cii~emission. The serendipitous continuum sources and \Cii~detections are discussed in detail in \citet{ALPINE_BETHERMIN19} and \citet{ALPINE_LOIACONO19}. A more detailed description of these science goals can be found in our survey overview paper \citep[][]{ALPINE_LEFEVRE19}.

\textit{ALPINE} is based on a rich set of ancillary data, which makes it the first panchromatic survey at these high redshifts including imaging and spectroscopic observations at FIR wavelengths (see Figure~\ref{fig:ancillarydatasummary}). The backbone for a successful selection of galaxies are rest-frame UV spectroscopic observations from the Keck telescope in Hawaii as well as the European \textit{Very Large Telescope} (VLT) in Chile. These are complemented by ground-based imaging observations from rest-frame UV to optical, HST observations in the rest-frame UV, and Spitzer coverage above the Balmer break at rest-frame $4000\,{\textrm \AA}$. The latter is crucial for the robust measurement of stellar masses at these redshifts \citep[e.g.,][]{FAISST16a}.

\vspace{0.5cm}
For a survey overview of \textit{ALPINE} see \citet[][]{ALPINE_LEFEVRE19} and for details on the data analysis see \citet[][]{ALPINE_BETHERMIN19}.
In this paper, we present these valuable ancillary data products and detail several basic measurements for the \textit{ALPINE} galaxies. The outline of the paper is sketched in Figure~\ref{fig:ancillarydatasummary}. Specifically, in Section~\ref{sec:spectroscopy}, we present the spectroscopic data and detail the spectroscopic selection of the \textit{ALPINE} galaxies. In the same section, we also present stacked spectra and touch on velocity offsets between \lya, \Cii, and absorption line redshifts.
Section~\ref{sec:photometry} is devoted to the photometric data products, which include ground- and space-based photometry. In Section~\ref{sec:sedfitting}, we detail the derivation of several galaxy properties from the observed photometry. These include stellar masses, SFRs, UV luminosities, UV continuum slopes, as well as \halpha~emission derived from Spitzer colors.
We conclude and summarize in Section~\ref{sec:end}.
All presented data products are available in the online printed version of this paper\footnote{\alpinedataurl}. The different catalogs and their columns are described in detail in the Appendix~\ref{app:dataproducts}. HST cutouts and rest-frame UV spectra for each of the \textit{ALPINE} galaxies are shown in Appendix~\ref{app:additionalfigures}.

Throughout the paper we assume the $\Lambda$CDM cosmology with $H_0 = 70\,{\rm km\,s^{-1}\,Mpc^{-1}}$, $\Omega_\Lambda = 0.70$, and $\Omega_{\rm m} = 0.30$. All magnitudes are given in the AB system \citep{OKE74} and stellar masses and SFRs are normalized to a \citet[][]{CHABRIER03} initial mass function (IMF).

\begin{deluxetable}{l l c c }
\tabletypesize{\scriptsize}
\tablecaption{Spectroscopy and selection of \textit{ALPINE} galaxies \label{tab:preselection}}
\tablewidth{0pt}
\tablehead{
\colhead{Survey} & \colhead{Selection} & \colhead{Number} & \colhead{Ref.}
}
\startdata
\multicolumn{4}{c}{\textbf{COSMOS field (105 galaxies)}  }\\[0.2cm]
\textit{Keck/DEIMOS}$^\dagger$ & & \textbf{84} & \textit{1}\\
& narrow-band ($z\sim4.5$)$^a$ & 6 & \\
& narrow-band ($z\sim5.7$)$^b$ & 23 & \\
& LBG (color)$^c$ & 41 & \\
& pure photo-\textit{z}$^d$ & 9 & \\
& $4.5\,{\rm \mu m}$ excess & 4 & \\
& X-ray (Chandra) & 1 &  \\[0.15cm]
 & with \lya~emission & 66 & \\
 & weak \lya~emission or absorption & 18 & \\[0.2cm]
%
\textit{VUDS} & & \textbf{21} & \textit{2} \\
& photo-$z$ + LBG & 21  & \\
& [narrow-band ($z\sim4.5$) & 3]$^\ddagger$ & \\
& [narrow-band ($z\sim5.7$) & 1]$^\ddagger$ & \\
& [LBG (color) & 1]$^\ddagger$ & \\
& [$4.5\,{\rm \mu m}$ excess & 1]$^\ddagger$ & \\[0.15cm]
 & with \lya~emission & 16 & \\
 & weak \lya~emission or absorption & 5 & \\[0.2cm]
\multicolumn{4}{c}{\textbf{ECDFS field (13 galaxies)}  }\\[0.2cm]
\textit{VLT GOODS-S} & & \textbf{11}  & \textit{3} \\
& primarily LBG (color) & 11  & \\[0.15cm]
 & total with \lya~emission & 6 & \\
 & total without \lya~emission & 5 & \\[0.2cm]
\textit{HST/GRAPES} & & \textbf{2}  & \textit{4}\\
& Grism (no \textit{a priori} selection) & 2  &\\[0.15cm]
 & with \lya~emission & 2 & \\
 & weak \lya~emission or absorption & 0 & \\[0.2cm]
\enddata
\tablenotetext{\dagger}{For a detailed description of the selection criteria, we refer to \citet{MALLERY12} and \citet{HASINGER18}.\\[-0.5cm]}
\tablenotetext{\ddagger}{Six of these galaxies are also observed as part of the Keck/DEIMOS survey (ref. 1). The corresponding number per selection from the Keck/DEIMOS program is given in square-brackets for those six galaxies.\\[-0.5cm]}
\tablenotetext{a}{\lya~emitters selected with \textit{NB711}.\\[-0.5cm]}
\tablenotetext{b}{\lya~emitters selected with \textit{NB814}.\\[-0.5cm]}
\tablenotetext{c}{Color-selected galaxies in $B$, $g^{+}$, $V$, $r^{+}$, and $z^{++}$ using the criteria from \citet{OUCHI04,CAPAK04,CAPAK11,IWATA03,HILDEBRANDT09}.\\[-0.5cm]}
\tablenotetext{d}{Galaxies with a photometric redshift $z>4$ with a probability of$>50\%$ based on the \citet{ILBERT10} photo-z catalog.\\[-0.5cm]}
\tablenotetext{}{References:
(1) \citet{CAPAK04,MALLERY12,HASINGER18},
(2) \citet{LEFEVRE15},
(3) \citet{VANZELLA07,VANZELLA08,BALESTRA10},
(4) \citet{MALHOTRA05,RHOADS09}
}\vspace{-1cm}
\end{deluxetable}

\begin{figure*}
\includegraphics[width=2.1\columnwidth, angle=0]{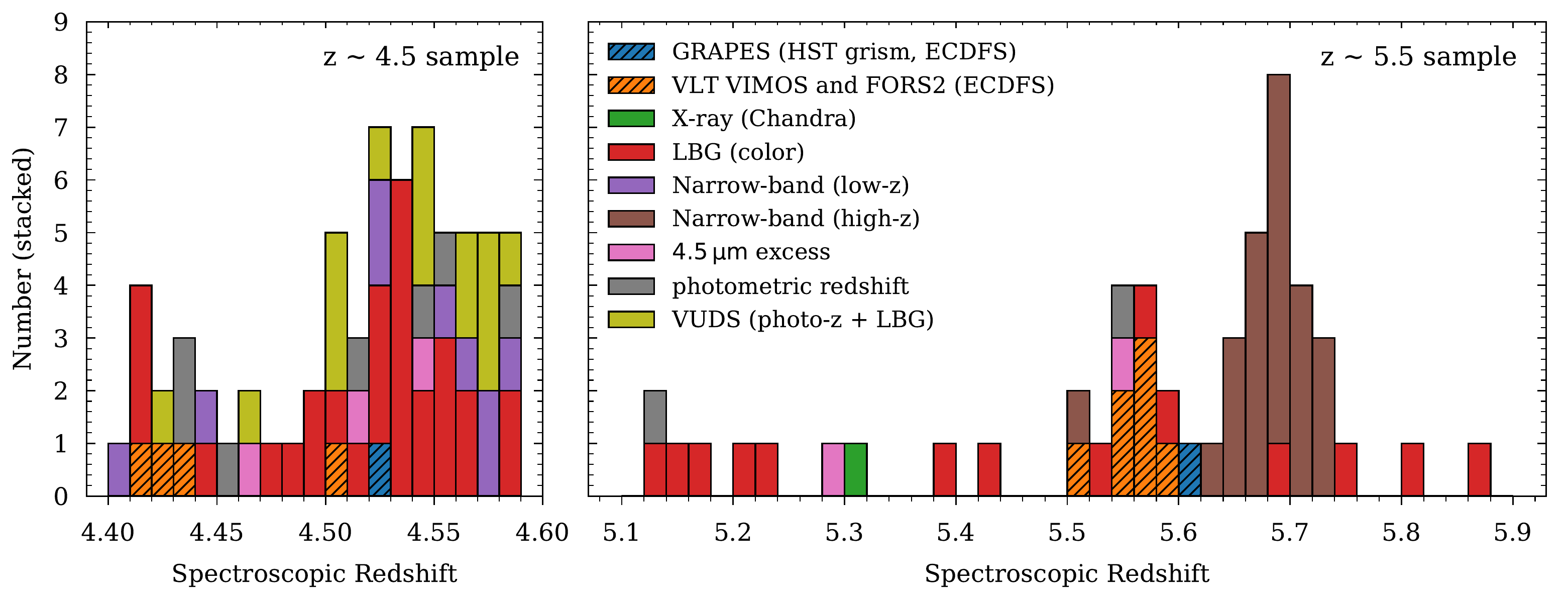}
\caption{Redshift distribution of \textit{ALPINE} galaxies. Each bar shows the stacked number of different selections per bin (see Table~\ref{tab:preselection} and description in text). The bins with galaxies from the ECDFS field are hatched. The left and right panels show galaxies in the two different redshift bins. 
\label{fig:redshiftdist}}
\end{figure*}

\begin{figure*}
\includegraphics[width=2.1\columnwidth, angle=0]{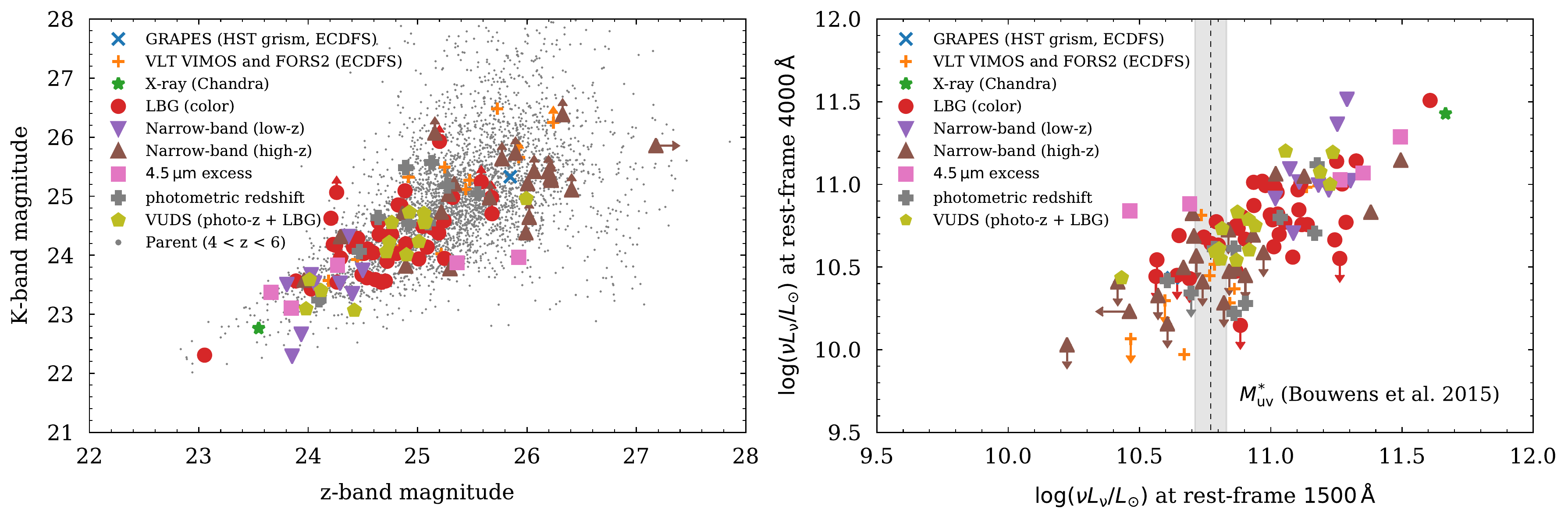}
\caption{Comparison of observed (i.e., not corrected for dust) $z$-band and $K$-band magnitudes (left) and luminosities (right) for different selections listed in Table~\ref{tab:preselection}. The measurements on the parent sample in COSMOS at $4 < z < 6$  is shown in light gray. The color-coding is the same as in Figure~\ref{fig:redshiftdist}. The arrows show $1\sigma$ upper limits. The gray area denotes the $M_{\rm UV}^{*}$, the knee of the UV luminosity function, which corresponds to $-21.1\pm0.15$ (or $\log(\nu L_{\nu}/{\rm L_{\odot}}) \sim 10.77$) at $z=5$ \citep[][]{BOUWENS15}. The derivation of the photometry is described in detail in Section~\ref{sec:photometry}. \label{fig:magcomparisonselections}}
\end{figure*}

\section{Spectroscopic Data and Selection} \label{sec:spectroscopy}

\subsection{Spectroscopic selection of ALPINE galaxies}
\label{sec:spectroscopyselection}

The \textit{ALPINE} survey is only possible due to a spectroscopic pre-selection of galaxies from large spectroscopic surveys on COSMOS and ECDFS. This is because the ALMA frequency bands are narrow ($\sim 1000\,{\rm km\,s^{-1}}$), and in order to observe \Cii~emission the redshift has to be known within a precision of $\sim1000\,{\rm km/s}$.
The galaxy selection is refined to optimize the efficiency of the ALMA observations by creating groups of galaxies in spectral dimensions. Our sample also includes $7$ galaxies that were previously observed with ALMA by \citet[][]{RIECHERS14} and \citet[][]{CAPAK15}. These are \textit{HZ1}, \textit{HZ2}, \textit{HZ3}, \textit{HZ4}, \textit{HZ5}, \textit{HZ6}/\textit{LBG-1}, and \textit{HZ8}, which correspond to the \textit{ALPINE} galaxies \textit{DC\_536534}, \textit{DC\_417567}, \textit{DC\_683613}, \textit{DC\_494057}, \textit{DC\_845652}, \textit{DC\_848185}, and \textit{DC\_873321}, respectively.
Furthermore, four galaxies from the VUDS survey (\textit{vc\_5101288969}, \textit{vc\_5100822662}, and \textit{vc\_510786441} in COSMOS and \textit{ve\_530029038} in ECDFS) are observed twice (resulting in a total number of $122$ observations). The duplicate observations are used for quality assessment. \citet[][]{ALPINE_BETHERMIN19} describes the combination of these observations.

The rest-frame UV spectroscopic data from which the \textit{ALPINE} sample is selected combine various large surveys on the COSMOS and ECDFS fields.
Out of the $105$ \textit{ALPINE} galaxies on the COSMOS field, $84$ are obtained by the large DEIMOS spectroscopic survey \citep{CAPAK04,MALLERY12,HASINGER18} at the Keck telescope in Hawaii. The remaining spectra on the COSMOS field are obtained from the VIMOS Ultra Deep Survey \citep[VUDS,][]{LEFEVRE15,TASCA17} at the VLT in Chile. In total $6$ of the VUDS spectra are independently also observed as part of the Keck/DEIMOS survey (\textit{vc\_5100559223}, \textit{vc\_5100822662}, \textit{vc\_5101218326}, \textit{vc\_5101244930}, \textit{vc\_5101288969}, \textit{vc\_510786441}). The redshifts are consistent within $280\,{\rm km\,s^{-1}}$ and we do not find any systematic offsets between the two observations (see also Section~\ref{sec:absorptionmeasurement}).
Out of the $13$ galaxies in the ECDFS field, $11$ are obtained from spectroscopic observations with VIMOS (9) and FORS2 (2\footnote{One of these galaxies, \textit{ve\_530029038}, has also been observed by the VUDS survey.}) at the VLT \citep{VANZELLA07,VANZELLA08,BALESTRA10}, and $2$ come from the HST grism survey \textit{GRAPES} \citep{MALHOTRA05,RHOADS09}.
The spectral resolution of the different dataset varies between $R\,\sim\,100$ (ECDFS/\textit{GRAPES} grism), $R\,\sim\,180$ (ECDFS/VIMOS), $R\,\sim\,230$ (COSMOS/VUDS), $R\,\sim\,660$ (ECDFS/FORS2), and $R\,\sim\,2500$ (COSMOS/DEIMOS).

Biases towards dust-poor star-forming galaxies with strong rest-frame UV emission lines (such as \lya) can be common in purely spectroscopically selected samples. To minimized such biases as much as possible, the spectroscopically observed galaxies have been pre-selected through a variety of different selection methods.
The largest fraction of galaxies in \textit{ALPINE} is drawn from the Keck/DEIMOS and VUDS surveys on the COSMOS field. Both surveys include galaxies preselected in various ways, resulting in the most representative and inclusive spectroscopic high-redshift galaxy sample.
Specifically, the VUDS survey combines predominantly a photometric redshift selection with a color-selected Lyman Break Galaxy (LBG) selection \citep{LEFEVRE15}, known as the Lyman-break drop-out technique \citep[see, e.g.,][]{STEIDEL96,DICKINSON98}. The Keck/DEIMOS survey (providing $71\%$ of the total \textit{ALPINE} sample) consists of galaxies that are selected by narrow-band surveys at $z\sim4.5$ ($7\%$) and $z\sim5.7$ ($27\%$), the drop-out technique (color selection) over the whole redshift range ($49\%$), as well as purely by photometric redshifts ($11\%$). In addition, $4$ galaxies are selected by a $4.5\,{\rm \mu m}$ excess and one galaxy was preselected through X-ray emission using the Chandra observatory.
On the ECDFS field, the galaxies are mostly color-selected.
Table~\ref{tab:preselection} summarizes the different selections and corresponding numbers of galaxies and provides a complete list of references. We also list the numbers of galaxies with \lya~emission ($76\%$) and weak \lya~emission or \lya~absorption ($\sim24\%$). Note that the Keck/DEIMOS and VUDS samples have similar \lya~emission properties. However, note that above $z=5$, the \textit{ALPINE} sample is strongly dominated by narrow-band selected galaxies.

Figure~\ref{fig:redshiftdist} shows the distribution of redshifts of the \textit{ALPINE} galaxies in the COSMOS and ECDFS fields. The colored histogram bars show stacked numbers of galaxies that are preselected by the different methods discussed above. The bins with galaxies in the ECDFS field are hatched. The narrow-band selected galaxies are prominent at $z\sim5.7$ and represent the largest fraction of galaxies at $z>5$ in \textit{ALPINE}. On the other hand, the $z<5$ sample consists mostly of color-selected galaxies.
The VUDS galaxies are most represented at $z < 5$, while the DEIMOS spectra and the galaxies in ECDFS cover the whole redshift range.

Figure~\ref{fig:magcomparisonselections} shows the distribution of observed magnitudes as well as rest-frame $1500\,{\textrm \AA}$ and $\sim4000\,{\textrm \AA}$ luminosity of galaxies selected by the different methods. The photometry that is used is explained in detail in Section~\ref{sec:photometry}. The $1500\,{\textrm \AA}$ rest-frame luminosity is derived from SED fitting (see Section~\ref{sec:uvmags} for details). The $4000\,{\textrm \AA}$ rest-frame luminosity is derived directly from the UltraVISTA $K_{\rm s}$ and VLT $K_{\rm s}^{\rm v}$ magnitude for galaxies in the COSMOS and GOODS-S field, respectively. The magnitudes and luminosities are not corrected for dust attenuation. Note that the $K$-band is rest-frame $3000\,{\rm \AA}$ at the highest redshifts ($z=5.9$), hence at these redshift older and dustier galaxies would be biased to lower luminosities. As expected for spectroscopically selected galaxies, the \textit{ALPINE} sample covers the brighter part of the galaxy magnitude and luminosity distribution. The different selection methods on their own are distributed differently in this parameter space. Most noticeably, the $z\sim5.7$ narrow-band selected galaxies reside at the faintest luminosities, while the $4.5\,{\rm \mu m}$ continuum excess selected galaxies are among the brightest. The X-ray Chandra detected galaxy \textit{DC\_845652} (green star) at $z=5.3$ outshines all of the galaxies in UV luminosity.

All in all, although naturally biased to the brightest galaxies, this diverse selection function makes \textit{ALPINE} an exemplary panchromatic survey that enables the study of a representative high-$z$ galaxy sample at UV, optical, and FIR wavelengths.

\begin{figure}[t!]
\includegraphics[width=1\columnwidth, angle=0]{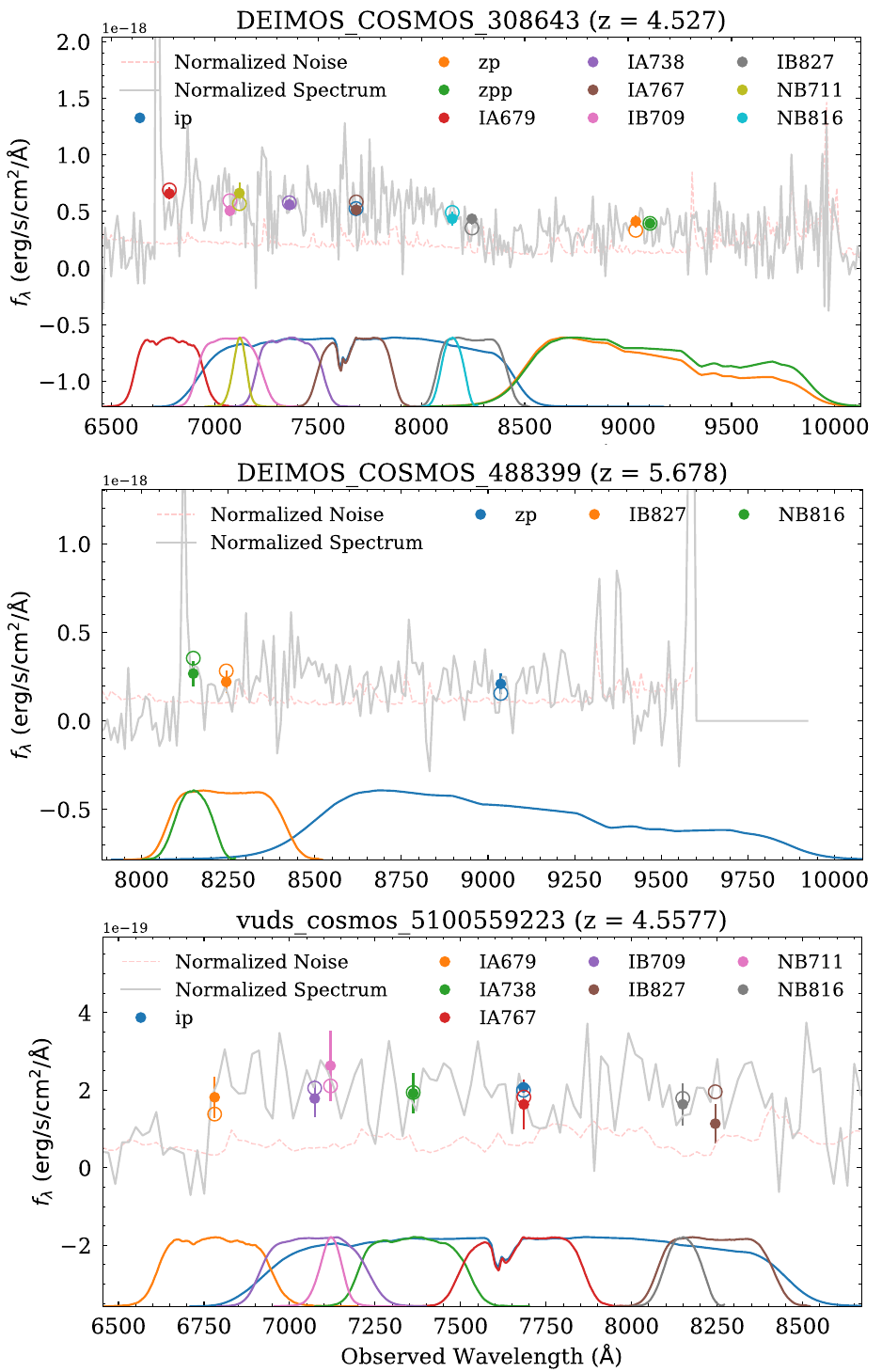}
\caption{Absolute calibration of rest-frame UV spectra. Shown are three examples at $z=4.53$, $z=4.56$, and $z=5.68$. The spectra are convolved by the filters and the photometry (open circles) is compared to the total and Galactic extinction corrected broad, intermediate, and narrow-band photometry from catalogs (filled circles) described in Section~\ref{sec:photometry}. \label{fig:spectrumcalib}}
\end{figure}

\begin{figure*}
\includegraphics[width=2.1\columnwidth, angle=0]{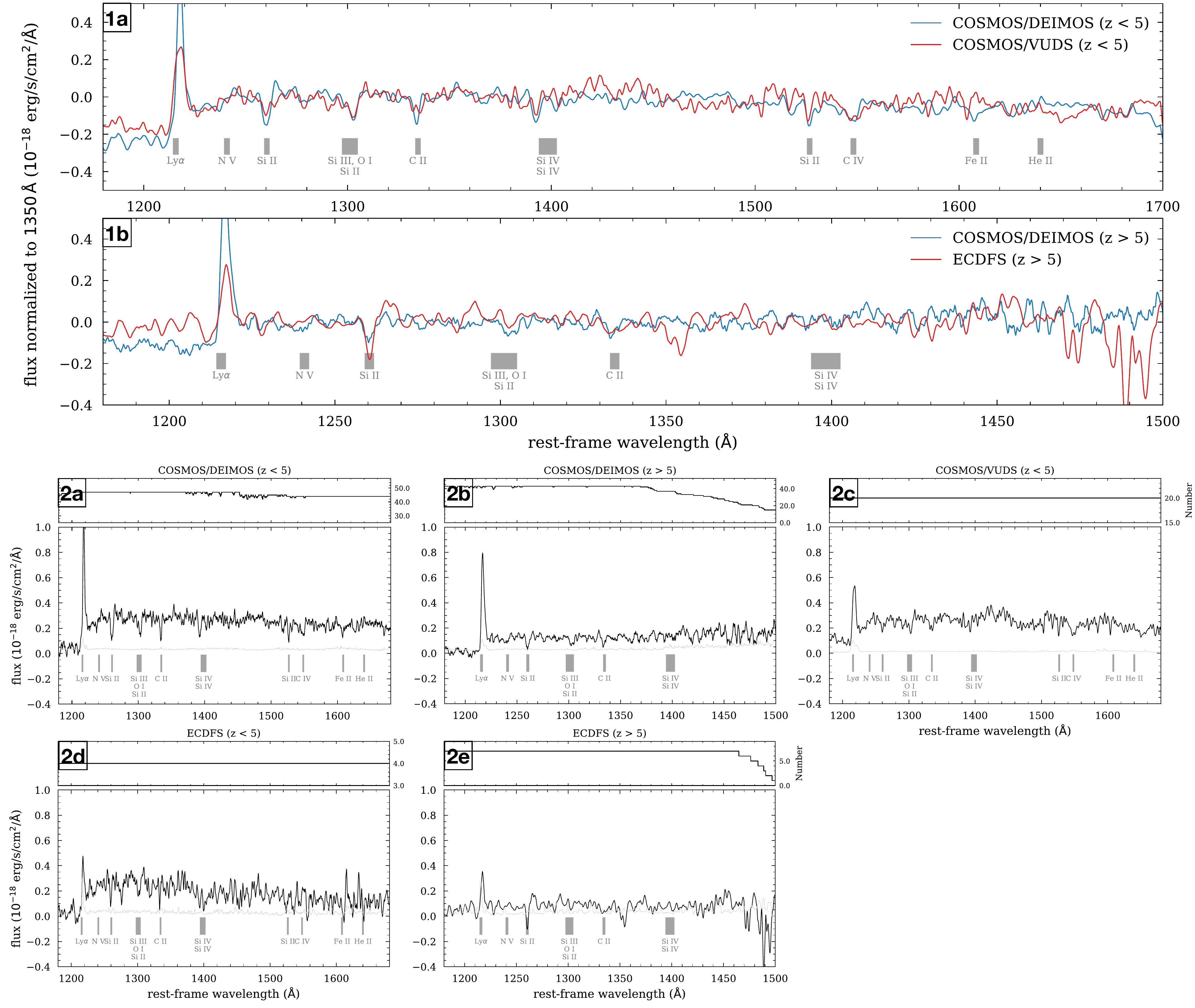}
\caption{Examples of stacked \textit{ALPINE} spectra. Panels \textit{1a} and \textit{1b} show stacked spectra at $z<5$ (in COSMOS from DEIMOS observations and as part of the VUDS survey) and $z>5$ (on COSMOS from DEIMOS and on ECDFS from VIMOS and FORS2 observations), respectively. The stacks are all normalized to the continuum between $1300\,{\textrm \AA}$ and $1400\,{\textrm \AA}$ and common emission and absorption features are indicated with gray bars. Note that the VUDS spectra have a lower native resolution ($R\sim230$) compared to the DEIMOS observations ($R\sim2500$), therefore the latter have been degraded in resolution using a 1-dimensional Gaussian window function for visual comparison. Panels \textit{2a} through \textit{2e} show stacks at $z<5$ and $z>5$ for the different datasets. The number of spectra included per wavelength is shown on the top of each panel. The uncertainty in flux is indicated by the light gray line. The $y$-axis scale is the same such that the continuum brightness can be compared. \label{fig:spectrastacks}}
\end{figure*}

\subsection{Uniform calibration of spectra}
\label{sec:spectroscopycalibration}

All of the rest-frame UV spectra discussed in Section~\ref{sec:spectroscopyselection} are relative flux corrected to remove sensitivity variations across the spectrograph as well as to correct atmospheric absorption features. However, not all of the spectra have been absolute flux calibrated, which is important to measure absolute quantities such as their \lya~emission. Hence, we recalibrate the spectra using the Galactic extinction corrected total broad, intermediate, and narrow-band photometry of the \textit{ALPINE} galaxies (see Section~\ref{sec:photometry} for details on the photometry). It turns out that the absolute flux calibrated spectra are in excellent agreement (within better than $5\%$ in flux) with our measured photometry and the recalibration is not necessary in these cases.
As the spectra come from different surveys, we convert them to a common format during the recalibration procedure.

To perform the absolute flux calibration, we convolve each of the spectra with the transmission functions of the various optical broad, intermediate, and narrow-band filters that exist on the COSMOS and ECDFS fields, respectively. On average, we use $4-9$ filters for galaxies at $z < 5$ and $2-4$ at $z>5$. If the filter extends further than the spectrum, we extrapolate the spectrum by its medium continuum value. If the filter extends significantly beyond the spectrum ($>50\%$), we do not consider the filter. We then compare the photometry obtained from the spectra to the total and Galactic extinction corrected photometry discussed in Section~\ref{sec:photometry}, which allows us to obtain an average correction factor for each spectrum.
We found that a single number for this correction per galaxy is enough for the calibration as the spectra already have been relative flux calibrated. Since the uncalibrated spectra are mostly in units of counts, this correction is on the order of $10^{-21}$ for most galaxies.

Our recalibration corrects for slit-losses and seeing variations. We also scale the variance in order to conserve the S/N of the spectrum.
The final precision of our calibration is around $5-10\%$ in flux, which corresponds to the $1\sigma$ uncertainty in the photometry.
Note that we do not consider undetected spectroscopic fluxes in this procedure, however, we use the constraints gained from the upper limits in the photometry for the calibration.
Figure~\ref{fig:spectrumcalib} shows three absolute calibrated spectra at $z\sim4.53$, $z\sim4.56$ (with weak \lya), and $z\sim5.68$ to visualize our method. The filters that were used for the calibration are indicated in colors.

\begin{figure*}[t!]
\includegraphics[width=2\columnwidth, angle=0]{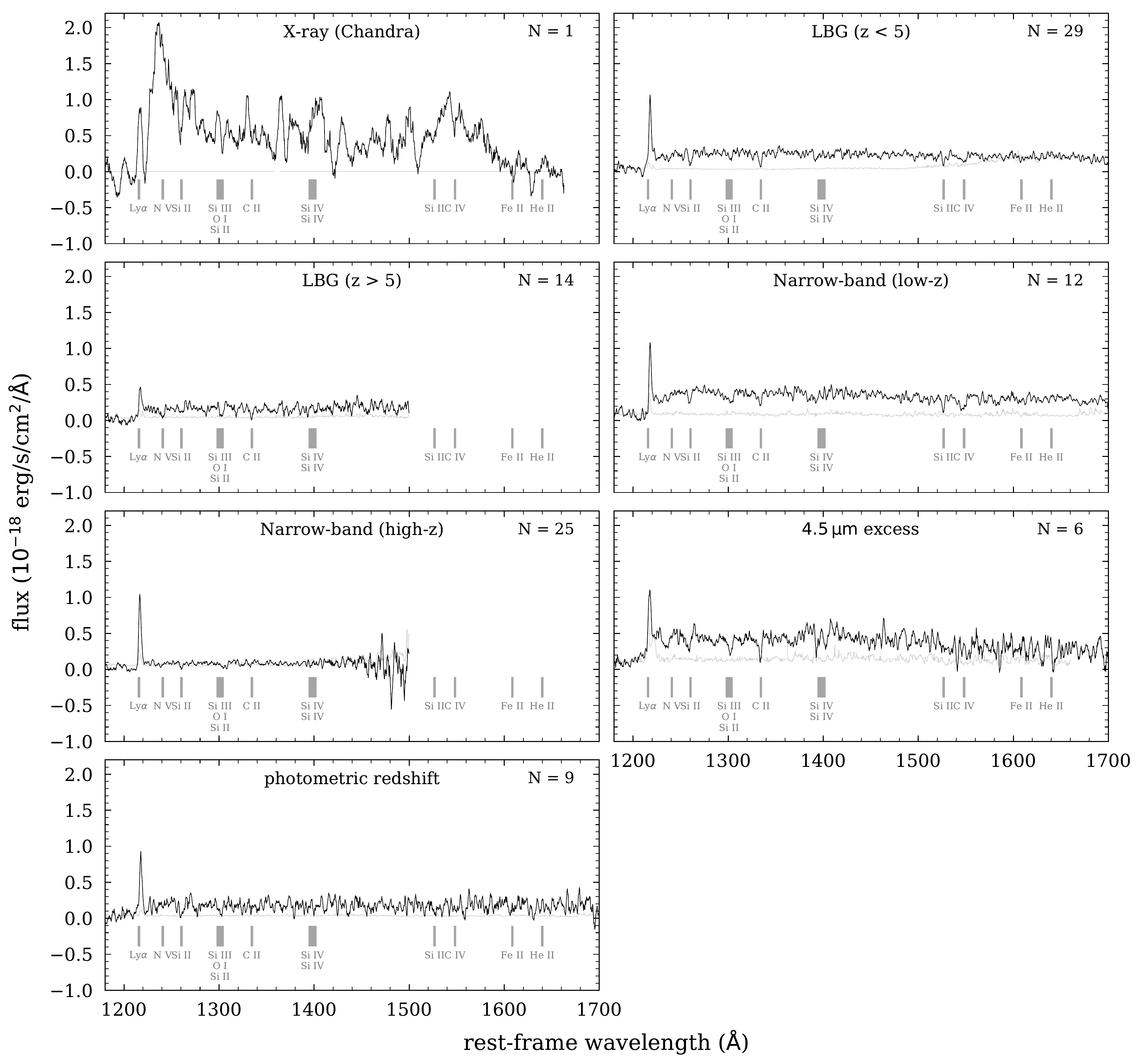}
\caption{Stacked spectra in COSMOS for each of the selections discussed in Section~\ref{sec:spectroscopyselection} and listed in Table~\ref{tab:preselection}. Emission and absorption features are indicated by gray bars and the number of spectra in the stack is shown on the upper right corner. We also show the X-ray detected galaxy (\textit{DC\_845652}) at $z=5.3$, which shows strong and broad \nv~and \civ~emission. The uncertainty in flux is indicated by the light gray line. \label{fig:spectrastacksselection}}
\end{figure*}

\subsection{Stacked Spectra: Overview over Rest-Frame UV Emission and Absorption Lines in \textit{ALPINE} Galaxies}
Figures~\ref{fig:spectrastacks} and~\ref{fig:spectrastacksselection}, show stacks of different spectra. In order to create median-stacks of the spectra, we resample the spectra to a common wavelength grid and stack them in rest-frame using their respective redshifts derived from \Cii~or, if not available, from rest-UV absorption lines or \lya~emission. All stacks are subsequently binned to a resolution of $2\,{\textrm \AA}$ for visual purposes to emphasize the UV absorption features. To obtain a per-pixel uncertainty from the sky background for each stack (visualize by the gray line), we simply combine the inverse variances in the individual spectra in quadrature. The latter are the original inverse variance that we adjusted to the new normalization described in Section~\ref{sec:spectroscopycalibration}.

Panels \textit{1a} and \textit{1b} of Figure~\ref{fig:spectrastacks} compare the full stacked spectra of galaxies at $z<5$ in COSMOS from observations with DEIMOS and as part of VUDS, as well as at $z>5$ from observations in COSMOS from DEIMOS and in ECDFS from VIMOS and FORS2. In the former case, we adjust the resolution of the DEIMOS spectra ($R\sim2500$) to that of the VUDS observations ($R\sim230$) by applying a 1-dimensional Gaussian smoothing. The spectra are normalized to the median flux in the rest-frame wavelength range between $1300$ and $1400\,{\textrm \AA}$ before stacking. For stacks of galaxies at $z>5$, the rest-frame wavelength reaches up to $1500\,{\textrm \AA}$, while for lower redshift stacks we show wavelengths up to rest-frame $1700\,{\textrm \AA}$.
Several prominent spectral features are visible in the stacks in both redshift bins (indicated by gray bars). These include the \lya~emission line and \nv~at $1241\,{\textrm \AA}$ and in addition UV absorption lines such as \siii~at $1260\,{\textrm \AA}$, the \siiii-\oi-\siii~complex at $1301\,{\textrm \AA}$, the two \siiv~lines at $1398\,{\textrm \AA}$, as well as \siii, \civ, and \heii~at $1527\,{\textrm \AA}$, $1548\,{\textrm \AA}$, and $1640\,{\textrm \AA}$, respectively. Furthermore, we see indication of \feii~absorption at $1608\,{\textrm \AA}$ in the COSMOS/DEIMOS spectra stack at $z<5$.
The depth of the UV absorption features are comparable for the different observations with the different instruments, verifying similar quality and little biases. However, note that the features in the ECDFS spectra are less pronounced due to the factor $\sim6$ smaller number of spectra contributing to the stacks compared to the DEIMOS stacks.
Panels \textit{2a} through \textit{2e} show the stacks for variously selected datasets below and above $z=5$. The spectra are not normalized before stacking in these cases to provide a comparison of the absolute flux values for the different redshifts and samples to the reader. The number of spectra per wavelength are shown on the top right for each panel. Note again that the number of high-redshift spectra drops towards redder wavelengths. This has to be kept in mind when analyzing the spectral features in the stacks. Emission and absorption lines are indicated as in the other panels.
As expected, the stacked spectra at higher redshifts are fainter, but still significant UV absorption features are present \citep[see also][]{FAISST16b,PAHL19,KHUSANOVA19}.

Figure~\ref{fig:spectrastacksselection} shows stacked spectra in COSMOS observed with DEIMOS for the different selection categories (see also Table~\ref{tab:preselection} and Figure~\ref{fig:redshiftdist}). We split the LBG category in galaxies below and above $z=5$. All the spectra are smoothed with a  Savitzky-Golay filter with size of $2\,{\textrm \AA}$ for visualization purposes. The total number of spectra per stack is indicated in the upper left corner. All panels are scaled the same way to emphasize differences in brightness.
The X-ray detected galaxy at $z=5.3$ is UV bright compared to the other stacks and shows strong \nv~emission with overlaid \siii~absorption as well as broad \civ~emission. LBGs (i.e., color-selected galaxies) are preferentially fainter but of similar continuum brightness as narrow-band selected galaxies at $z\sim4.5$. The latter show significant \cii, \siiv, and \civ~absorption. As expected, narrow-band selected galaxies at $z\sim5.7$ show strong \lya~emission and a faint continuum such that the S/N is too low to detect UV absorption features at great significance. The stack of galaxies selected by photometric redshifts shows to first order similar properties as the LBGs. The $4.5\,{\rm \mu m}$-excess continuum selected galaxies are on average the continuum brightest galaxies and show significant \lya~emission as well as absorption features.

\subsection{Rest-UV Emission and Absorption Lines and Velocity Offsets}\label{sec:uvspectrameasurements}

\subsubsection{Measurements}\label{sec:absorptionmeasurement}

We measure basic quantities from the individual rest-frame UV spectra. These include the redshift and equivalent width of \lya~emission as well as redshifts from various absorption lines.

The \lya~redshift ($\zlya$) is based on the peak of the (asymmetric) \lya~emission to allow a direct comparison with models of \lya~radiative transfer \citep[see, e.g., ][]{HASHIMOTO15}. The \lya~flux is measured by fitting a Gaussian to the line and for measuring the equivalent width ($\sim f^{\rm tot}_{\rm line}/f_{\rm continuum}$) the continuum redward of the \lya~line is used. These measurements are explained in more detail in \citet[][]{ALPINE_CASSATA19}.

The absorption redshifts are measured for each individual spectrum, if possible, using the lines \siii~($1260.4\,{\textrm \AA}$), \oi~($1302.2\,{\textrm \AA}$)\footnote{Here we refer to the \oi~absorption line complex consisting of \siiii, \ciii, \oi, and \siii.}, \cii~($1334.5\,{\textrm \AA}$), \siiv~($1393.8\,{\textrm \AA}$) and \siiv~($1402.8\,{\textrm \AA}$), \siii~($1526.7\,{\textrm \AA}$), and \civ~($1549.5\,{\textrm \AA}$)\footnote{This absorption consists of two lines ($1548.2\,{\textrm \AA}$ and $1550.8\,{\textrm \AA}$) and here we give the average wavelength.}. The first four are covered by observations in all galaxies, while the coverage of the latter depends on the redshift of the galaxy.
Note that some of the above lines are predominantly formed in the ISM (low-ionization interstellar [IS] lines; \siii, \oi, \cii, \siii), while others are formed in stellar winds (high-ionization wind lines; \siiv~or \civ) and therefore can display strong velocity shifts \citep[e.g.,][]{CASTOR79,LEITHERER11}. To increase the S/N of our measurements, we use all the above lines to derive an absorption line redshift (referred to as $\ziswind$), but we compare the individual redshift from the IS ($\zis$) and wind ($\zwind$) lines to investigate potential systematic differences.
Before performing any measurements, we subtract a continuum model from each individual spectrum. The model is derived by fitting a 4$^{\rm th}$-order polynomial to the spectrum, which is smoothed by a $5\,{\textrm \AA}$ box kernel.
We then fit the above absorption lines in five different rest-frame wavelength windows
[$1240\,{\textrm \AA}$,$1280\,{\textrm \AA}$],
[$1280\,{\textrm \AA}$,$1320\,{\textrm \AA}$],
[$1320\,{\textrm \AA}$,$1350\,{\textrm \AA}$],
[$1370\,{\textrm \AA}$,$1420\,{\textrm \AA}$], and
[$1500\,{\textrm \AA}$,$1570\,{\textrm \AA}$].
For the separate fit of the IS and wind lines, we split the last window into two ranges, namely [$1510\,{\textrm \AA}$,$1540\,{\textrm \AA}$] and
[$1530\,{\textrm \AA}$,$1570\,{\textrm \AA}$] to separate the IS line \siii~and the wind line \civ, respectively.
The absorption lines can be significantly asymmetric due to stellar winds and the effect of optical depth. Fitting a single Gaussian to them could therefore bias the redshift measurements. Instead, we use the stacked spectrum of LBGs at $z\sim3$ from \citet[][]{SHAPLEY06} as a template, which we cross-correlate to the observed data within the wavelength range of a given window by $\chi^2$ minimization.
We let the redshift vary within a velocity range of $\pm600\,{\rm km\,s^{-1}}$ (corresponding to roughly $0.01$ in redshift) around a prior absorption redshift, which is obtained by a manual cross-correlation of the same template to all possible absorption lines at once using the interacting redshift-fitting tool \texttt{SpecPro}\footnote{\url{http://specpro.caltech.edu}} \citep{MASTERS11}. We found that this approach significantly removes degeneracies in the fit and at the same time allows a visual inspection of all the spectra to flag the ones with low S/N where no reasonable fit can be obtained\footnote{The value of this visual flag is $-99$ if the S/N is too low to obtain a redshift, and $1$ and $2$ for reliable and very reliable redshift measurements, respectively.}.
For each galaxy, the so obtained $\chi^2(z)$ distribution is then converted into a probability density function $p(z)$ for each of the windows. These are combined, by choosing the necessary absorption lines, to a total probability $P(z)$ from which the final absorption line redshifts ($\ziswind$, $\zwind$, or $\zis$) are derived. The errors on these redshifts are derived by repeating this measurement $200$ times, thereby perturbing the fluxes according to a Gaussian error distribution with $\sigma$ defined by the average flux noise of the continuum. Typical uncertainties are on the order of $\pm100\,{\rm km\,s^{-1}}$.

\begin{figure}[t!]
\includegraphics[width=1.0\columnwidth, angle=0]{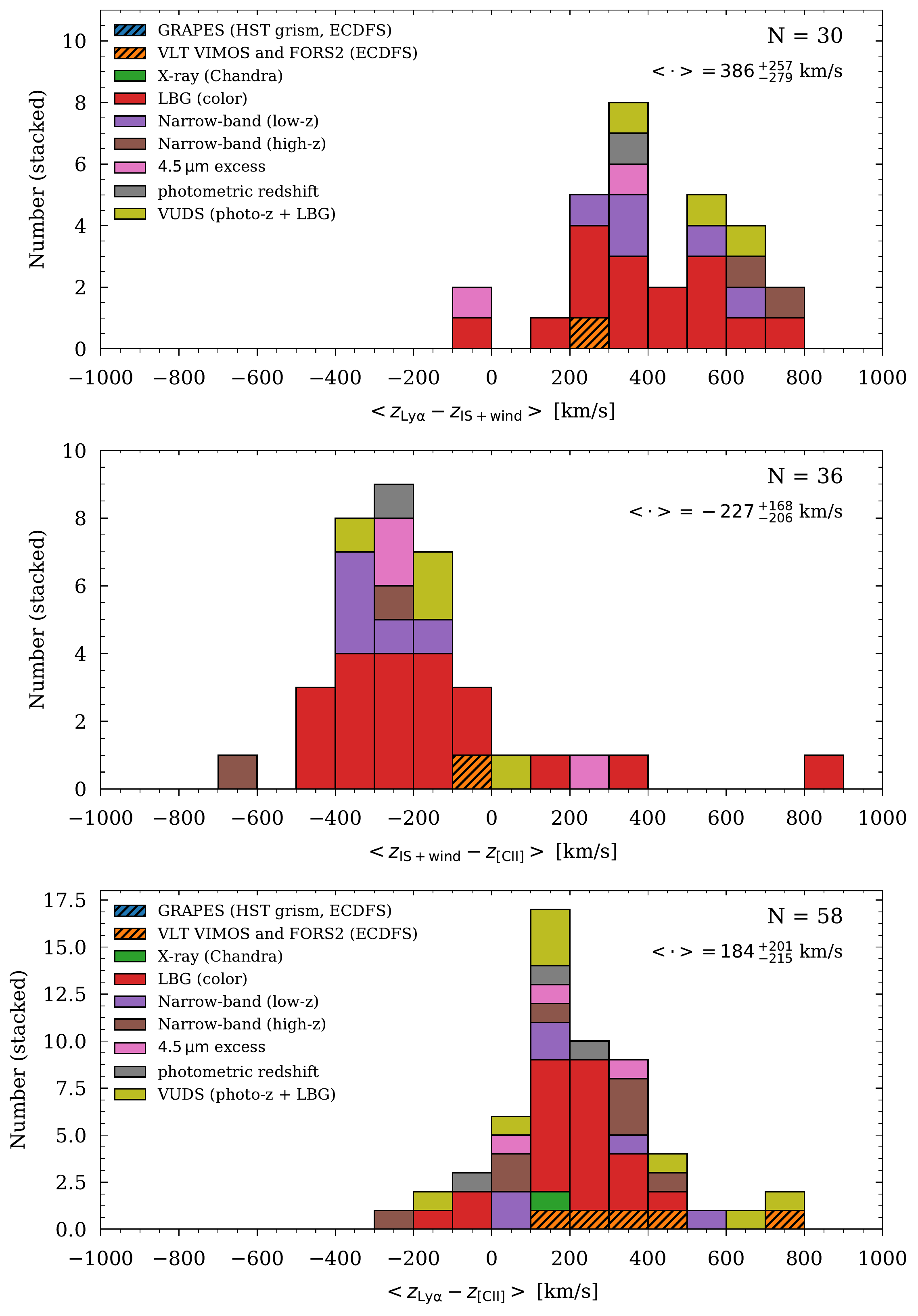}
\caption{Stacked histograms of velocity offsets between redshifts derived from different spectral features. The number of galaxies and median of the distribution (including scatter) are indicated. Shown are the velocity offsets between \lya~emission and IS$+$wind absorption lines (\textit{top panel}), as well as between \lya, IS$+$wind, and systemic redshift (\textit{middle} and \textit{bottom} panel). The latter two are in detail discussed in a forthcoming paper \citep{ALPINE_CASSATA19}. The average errors are on the order of $\pm100\,{\rm km\,s^{-1}}$, which corresponds to the size of the bins.
We do not find any significant biases introduced by the different selection methods (color coded as in previous figures).\vspace{-0.2cm}
\label{fig:velocity}}
\end{figure}

\begin{figure}[t!]
\includegraphics[width=1.0\columnwidth, angle=0]{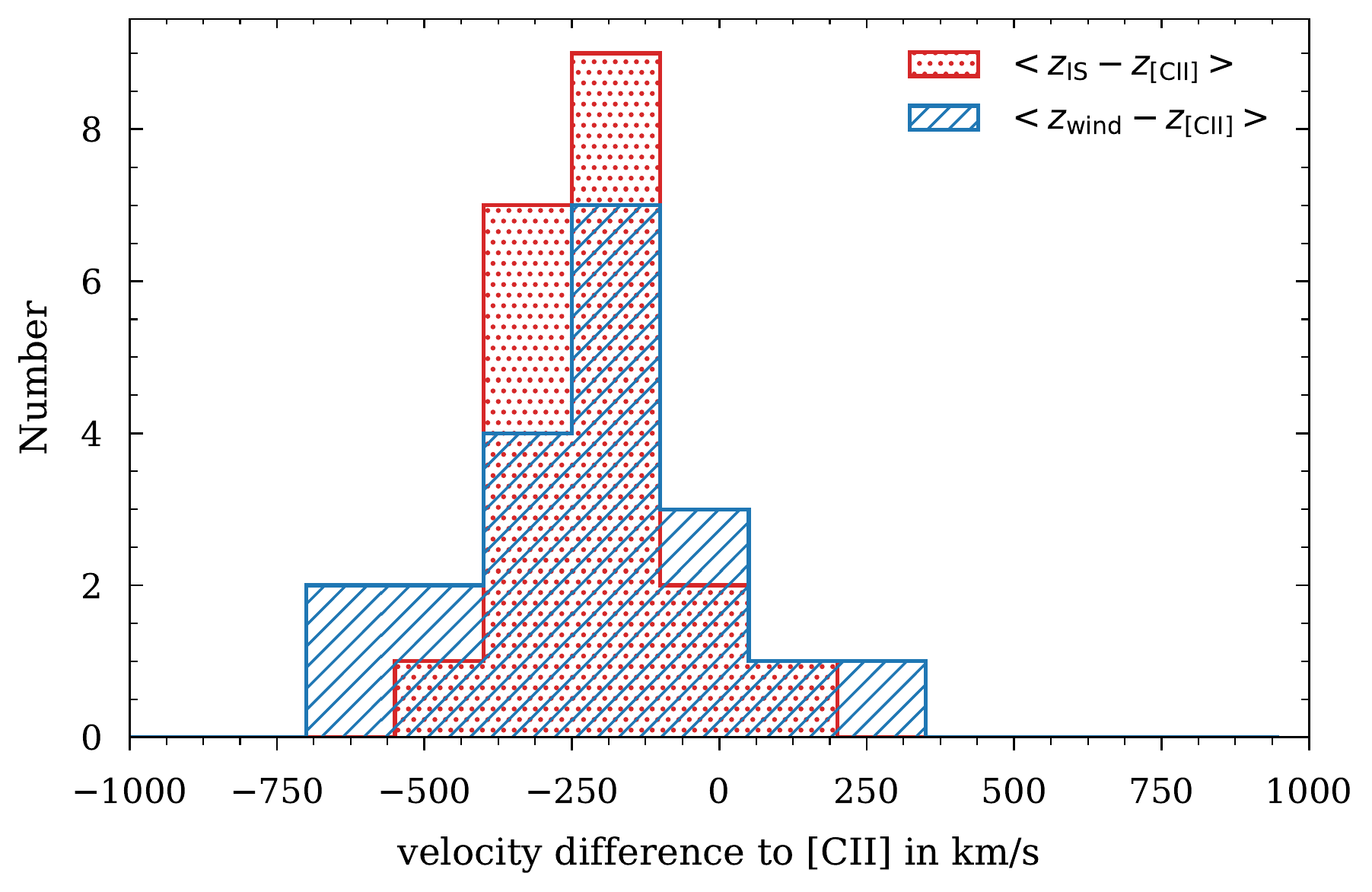}
\caption{Histogram of velocity offset with respect to systemic (defined by the \Cii$\lambda158{\rm \mu m}$~redshift) for IS (red; \oi, \cii, and \siii) and stellar wind affected absorption lines (blue; \siiv~and \civ). The average errors are on the order of $\sim200\,{\rm km\,s^{-1}}$, which corresponds to the size of the bins. \label{fig:velocity_iswind}\vspace{-0.2cm}}
\end{figure}

\begin{figure*}[t!]
\includegraphics[width=2.1\columnwidth, angle=0]{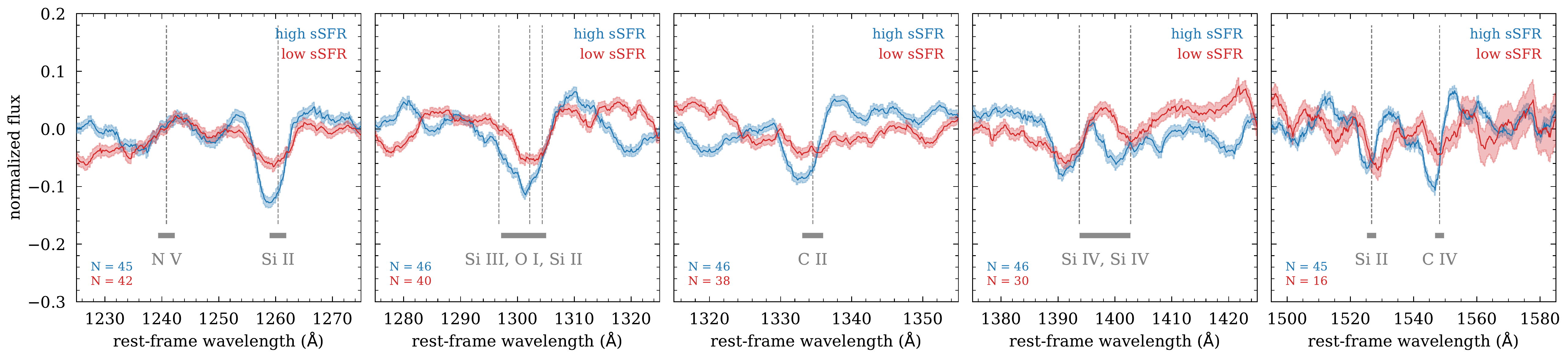}
\caption{Stacked spectra (in \cii~systemic redshift) in two bins of sSFR (red: $<4\,{\rm Gyr^{-1}}$, blue: $>5\,{\rm Gyr^{-1}}$) for five wavelength regions covering prominent rest-UV absorption lines. The derivation of the sSFR for the \textit{ALPINE} galaxies is detailed in Section~\ref{sec:sedfitting}.
The average number of spectra in each bin is indicated together with the prominent absorption and emission lines. We note systematically stronger blue shifts of all absorption lines for the high sSFR stack. Particularly, note the strong blue-shift of the high-ionization wind lines. The \civ~lines in the high sSFR bin also show indication of a more pronounced P-Cygni profile, indicative of strong stellar winds and outflows in high sSFR galaxies. The 1$\sigma$ uncertainties of the stacked spectra is indicated by the shaded regions. \label{fig:spectra_ssfr_stack}}
\end{figure*}

As mentioned in Section~\ref{sec:spectroscopyselection}, $6$ galaxies in COSMOS have been observed by the Keck/DEIMOS and VUDS spectroscopic surveys. Therefore there are two measurements for each of these galaxies.
Specifically, for \textit{vc\_5100559223}, \textit{vc\_5100822662}, \textit{vc\_5101218326}, and \textit{vc\_5101244930}, the IS$+$wind redshift measurements agree within $200\,{\rm km\,s^{-1}}$, $280\,{\rm km\,s^{-1}}$, $70\,{\rm km\,s^{-1}}$, and $110\,{\rm km\,s^{-1}}$. These values are on the order of the measurement uncertainties. Note that while the VUDS slits are oriented east-west, the DEIMOS slits can be oriented north-south or in any other angle. This different orientation could also be responsible for the differences in velocity offsets. On the other hand, for \textit{vc\_5101288969} and \textit{vc\_510786441}, we find significant differences of $1290\,{\rm km\,s^{-1}}$ and $1010\,{\rm km\,s^{-1}}$. A close inspection of the spectra shows that these are very low in S/N. Also, both have low visual quality flags ($-99$ and $1$, indicating not robust measurements are possible) and their redshifts are fit with less than $3$ lines, hence should not be trusted. For all $6$ spectra we decided to prefer the VUDS observations because of their slightly better S/N at a cost of lower resolution.

\subsubsection{Velocity Offsets with respect to \Cii~FIR Redshifts}\label{sec:velocity}

The detection of \Cii~by ALMA provides the systemic redshift of a galaxy. This enables us to study velocity offsets of rest-frame UV absorption lines and \lya~emission that will inform further about the properties of the ISM in these galaxies similarly to studies at lower redshifts using \halpha~and \cii]$\lambda1909$ \citep[e.g.,][]{STEIDEL10,MARCHI19}. Here we give an overview of the velocity properties and compare them for galaxies with high and low specific SFRs.

In the following, we define the velocity difference for two redshifts ($z_1$ and $z_2$) as $\left<z_1 - z_2\right> \equiv \Delta v_{\rm 12} = c \times (\frac{1+z_1}{1+z_2}-1)$ where $c=2.998\times10^5\,{\rm km\,s^{-1}}$.
The measurement of the \Cii~redshifts are detailed in \citet{ALPINE_BETHERMIN19}. They are defined as the peak of a Gaussian fit to the \Cii~line with spectral resolution of $25\,{\rm km\,s^{-1}}$. The uncertainty of the redshift measurements was estimated by a Monte Carlo simulations with perturbed fluxes according to the error per spectral bin. The average uncertainty is roughly $50-60\,{\rm km\,s^{-1}}$.
For the absorption lines, we require that $\ziswind$ is measured from at least three absorption lines and we only show galaxies that have not been flagged by our visual inspection with \texttt{SpecPro} as unreliable (flag $-99$). The average intrinsic measurement error per galaxy is $\pm100\,{\rm km\,s^{-1}}$. In relation to that, a systematic uncertainty of $0.5\,{\textrm \AA}$ in the rest-frame wavelength of the absorption lines (e.g,. due to calibration issues) turns into a velocity shift of $\sim120\,{\rm km\,s^{-1}}$.

Figure~\ref{fig:velocity} shows stacked histograms of velocity differences. The number of galaxies used as well as the median of the distribution with scatter (\textit{not} error on the median) are indicated as well.
The upper panel compares the velocities measured from \lya~and the IS$+$wind absorption lines. We find a median offset on the order of $386^{+257}_{-279}\,{\rm km\,s^{-1}}$, which is consistent with other measurements at the same redshifts \citep[see, e.g.,][]{FAISST16b,PAHL19} as well as at $z\sim2-3$ \citep{STEIDEL10}.
The center and bottom panels compare the IS$+$wind and \lya~redshifts to the systemic redshift \citep[here defined as the \Cii$\lambda 158\,{\rm \mu m}$ redshift,][]{ALPINE_BETHERMIN19}.
For the former we find an offset of $-227^{+168}_{-206}\,{\rm km\,s^{-1}}$ and for the latter we find $184^{+201}_{-215}\,{\rm km\,s^{-1}}$. These negative and positive velocity offsets can be related in a simple physical model involving the resonant scattering of \lya~photons and outflowing gas in the outskirts of galaxies \citep[see detailed discussion in][]{STEIDEL10}. The redshifted \lya~emission line (with respect to systemic) can be explained by resonant scattering of the \lya~photons. Preferentially, red-shifted \lya~photons scattered from the back of the galaxy can make it unscattered through the intervening gas inside the galaxy along the line of sight. The blueshift of IS absorption may depend on the outflow velocity of the absorbing gas as well as its covering fraction (or optical depth) inside the galaxy along the line of sight towards the observer. 
For a more in-depth discussion we refer to a companion paper by \citet{ALPINE_CASSATA19}.
Overall, we do not see a significant dependence of the velocity differences on the various selection techniques (color-coded in the figure).

Figure~\ref{fig:velocity_iswind} compares the velocity offsets between IS (\siii, \oi, \cii, \siii) and wind (\siiv, \civ) lines. We require that at least three IS lines and one wind line is measured. In addition, only galaxies that pass our visual classification (i.e., have flags other than $-99$, see above) are used. Overall, we do not see any statistical difference between IS and wind lines, although there is a tail towards higher blueshifts in the case of wind lines.
However, wind and outflows may be increased in galaxies with high and spatially dense star formation and young stellar populations. Therefore we would expect different velocity shifts for the absorption lines with respect to the systemic redshift for highly star-forming galaxies. 
In Figure~\ref{fig:spectra_ssfr_stack}, we investigate this picture by stacking galaxies at the extreme ends of the sSFR distribution (we refer to Section~\ref{sec:physicalproperties} for details on the measurement of the physical properties of our galaxies), namely low ($<4\,{\rm Gyr^{-1}}$) and high ($>5\,{\rm Gyr^{-1}}$) sSFR, in their corresponding rest-frames defined by the systemic redshift (i.e., \Cii$\lambda158{\rm \mu m}$ redshift). The sSFR is a good proxy of the star-formation density in a galaxy as well as the age of the current stellar population \citep[see, e.g., ][]{COWIE11}. We show the stacked spectra in five wavelength regions covering prominent absorption lines for each sSFR bin. The vertical dashed lines show the different absorption lines in the \Cii~rest-frame. First, we verify that the shifts between IS and wind lines are very similar for each sSFR bin (in concordance with Figure~\ref{fig:velocity}).
However, intriguing is that in the low sSFR stack, all absorption lines agree well with the \Cii~redshift, while in the high sSFR stack the lines are significantly blue shifted by $300-400\,{\rm km\,s^{-1}}$. We also note that in the high sSFR stack, the \civ~line shows a noticeable P-Cygni profile indicative of strong winds and outflows \citep{CASTOR79}. These findings fit well into a picture of strong winds and outflows produced by the high star-formation in these galaxies, which is also in line with recent results obtained through the stacking of \textit{ALPINE} \Cii~spectra \citep{ALPINE_GINOLFI19}.

\begin{figure}[t!]
\includegraphics[width=1.0\columnwidth, angle=0]{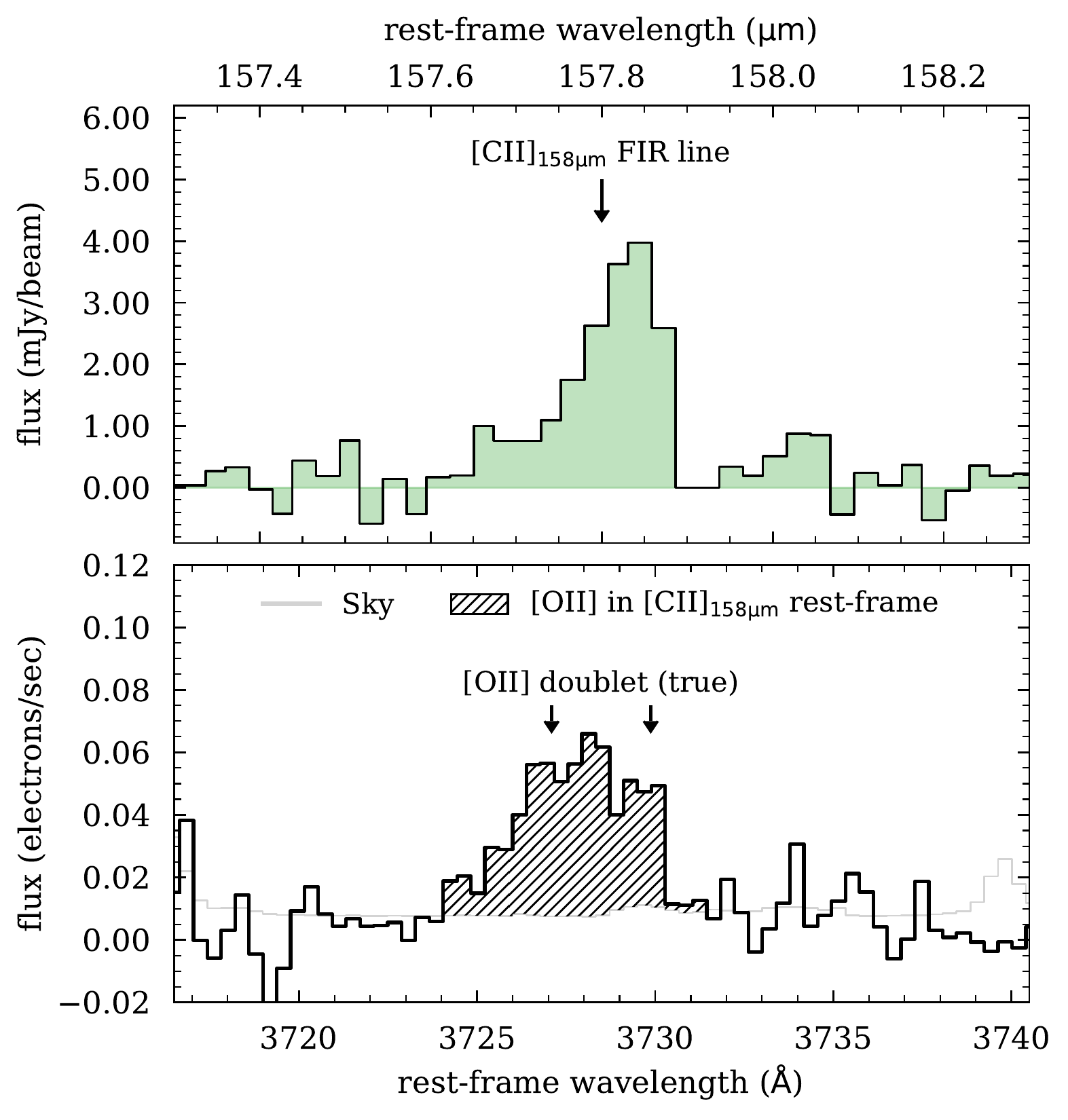}
\caption{ Comparison of optical and FIR emission lines in the case of galaxy \textit{DC\_881725}. \textit{Top:} \Cii~emission line at $158\,{\rm \mu m}$ observed with ALMA. \textit{Bottom:} The black line shows the optical \oii~doublet at $3727\,{\rm \AA}$ observed with the MOSFIRE spectrograph on Keck in the rest-frame of the \Cii~emission. The theoretical rest-frame wavelength of the doublet is indicate by the two arrows. This shows that there is no significant velocity offset between the \Cii~FIR and optical \oii~emission, hence verifies the validity of \Cii~as tracer of the systemic redshift of this galaxy.
\label{fig:mosfireoii}}
\end{figure}

\subsubsection{How well does \Cii~trace systemic redshift? Comparison to optical \oii~emission}
\label{sec:mosfireoii}

The extended nature of \Cii~may be indicative of its origin in the diffuse interstellar medium in addition to PDRs \citep{STACEY91,GULLBERG15,VALLINI15,FAISST17b}. Moreover, recent work by \citet[][]{ALPINE_GINOLFI19} shows that \Cii~emission is significantly affected by large-scale outflows caused by high star-formation in these galaxies. However, as shown by the same study, the outflows seem to be symmetric and therefore we do not expect them to significantly change the centroid of the \Cii~emission line.

During January 13-15, 2019, we were able to obtain a near-IR spectrum of one of our \textit{ALPINE} galaxies (\textit{DC\_881725} at $\zcii=4.5777$) using the \textit{Multi-Object Spectrometer For Infra-Red Exploration} \citep[MOSFIRE,][]{MCLEAN10,MCLEAN12} at the $10$-meter Keck I telescope on Mauna Kea in Hawaii. The observations of a total on-source integration time of $24\times 3\,{\rm min}$  in $K$ band ($1.92-2.40\,{\rm \mu m}$) were carried out under clear weather conditions with an excellent average seeing FWHM of $0.3\arcsec-0.4\arcsec$.
We performed a standard data reduction using the MOSFIRE data reduction pipeline\footnote{\url{https://keck-datareductionpipelines.github.io/MosfireDRP/}} (Version 2018). From the produced 2-dimensional spectrum and variance map, we extract the 1-dimensional spectrum at the spatial location of the galaxy using a weighted mean across $\pm3.5$ spatial pixels ($0.18\arcsec/{\rm px}$).

We are able to detect the optical \oii~doublet ($3727.09\,{\textrm \AA}$ and $3729.88\,{\textrm \AA}$) at the spatial position of the galaxy at a level of $>5\sigma$.
Note that this is the first detection of optical \oii~in a galaxy with \Cii~measurement from ALMA, which allows us for the first time to compare this two lines at these redshifts.
In bottom panel of Figure~\ref{fig:mosfireoii}, we show the final spectrum in the rest-frame of the \Cii~emission. The width of $\sim4\,{\textrm \AA}$ includes both \oii~lines. The theoretical rest-frame wavelength of the doublet is indicated by the black arrows. The position of the line agrees perfectly with the \Cii~redshift derived from ALMA, indicating that FIR \Cii~and optical \oii~trace the same systemic redshift. In addition, the top panel of the Figure shows the \Cii~FIR line for comparison of the central wavelength of the lines. Note that the actual width of the \Cii~line is more than $100$ times larger than the the one of the optical \oii~line.

\begin{figure*}
\includegraphics[width=2.1\columnwidth, angle=0]{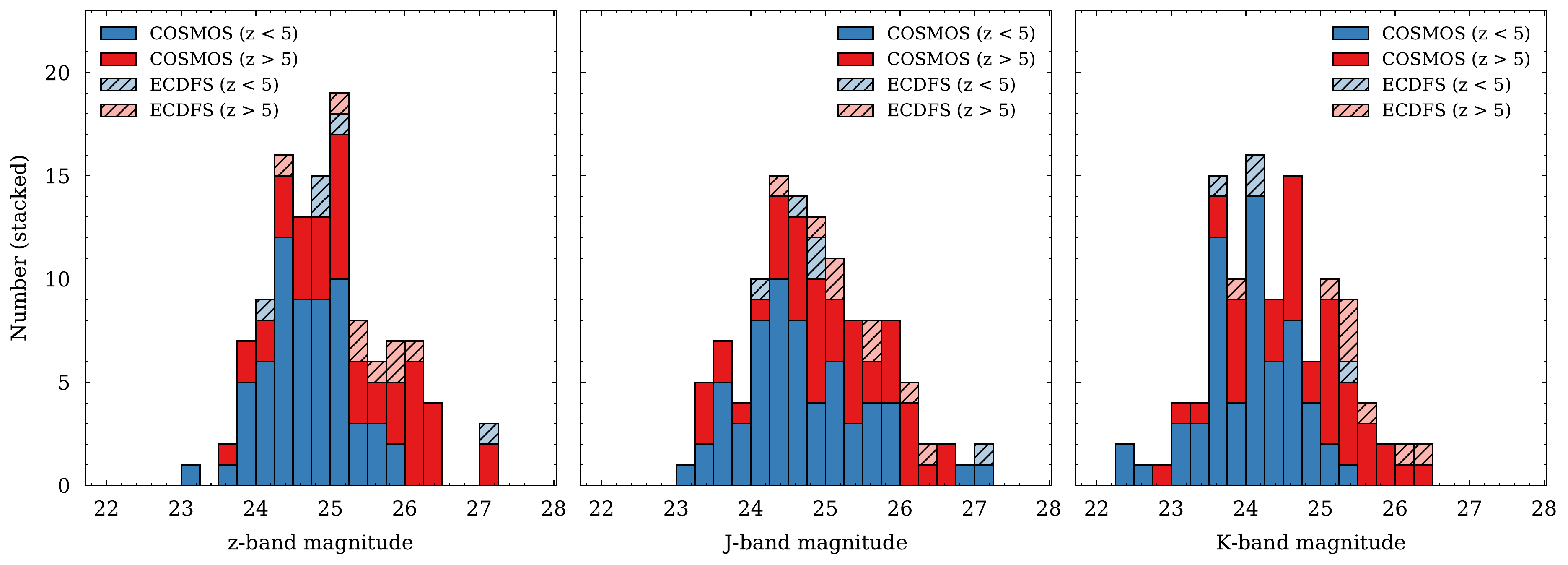}
\caption{Stacked histograms of the magnitude distribution for the \textit{ALPINE} galaxies in the COSMOS (solid) and ECDFS (hatched) fields. The blue and red color-coding indicates galaxies at $z<5$ and $z>5$, respectively. The magnitudes (from left to right) correspond to $z^{++}$, $J$, and $K_{\rm s}$ bands for COSMOS and $F850LP$, $J^{\rm v}$, and $K_{\rm s}^{\rm v}$ bands for ECDFS (see Tables~\ref{tab:photogoods} and~\ref{tab:photocosmos} for more information on the filters).  \label{fig:magnitudecomparison}}
\end{figure*}

\begin{figure*}
\includegraphics[width=2.1\columnwidth, angle=0]{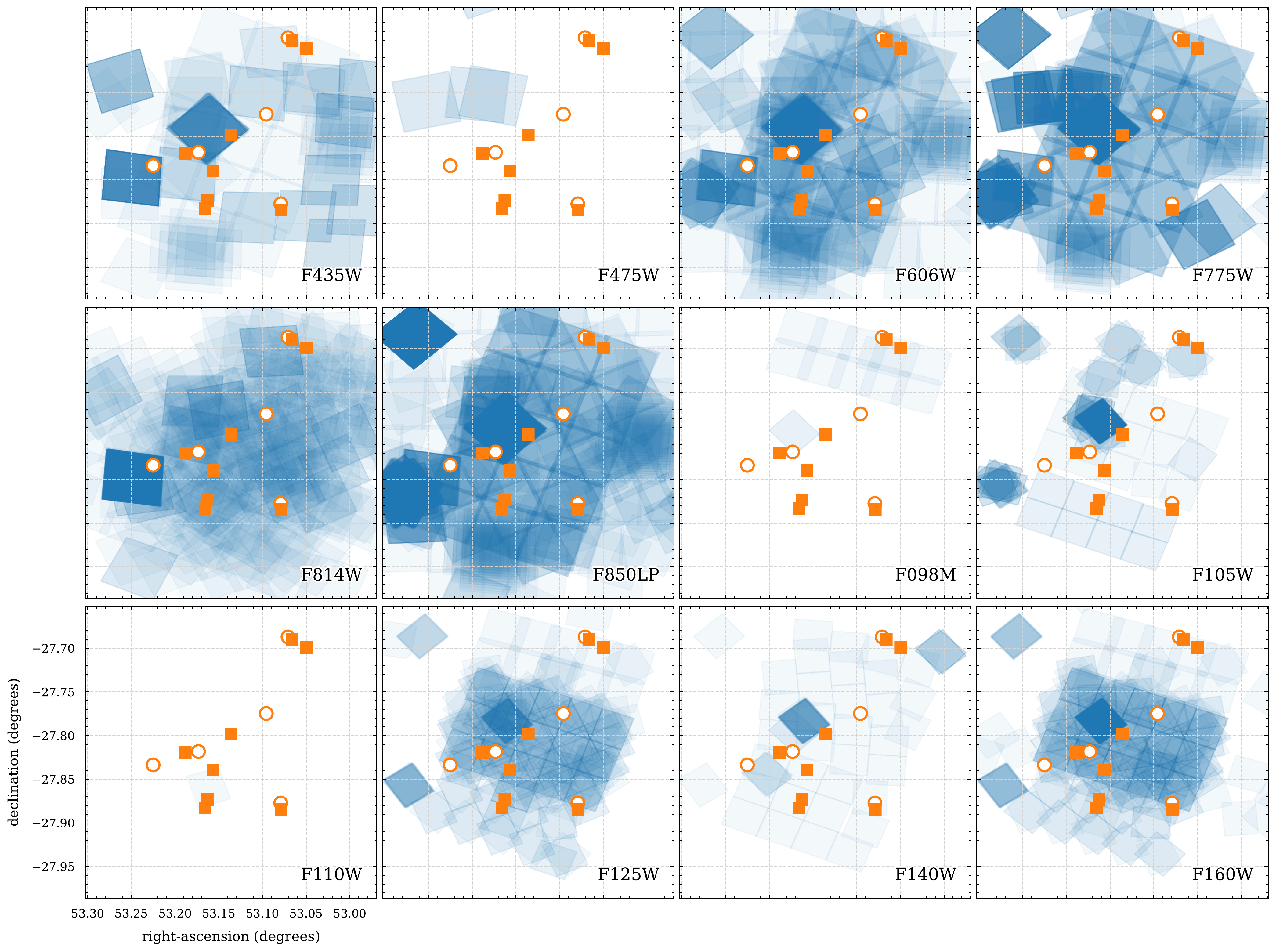}
\caption{HST pointing footprints on the ECDFS field (blue) for different HST ACS and WFC3/IR filters. Darker colors mean more observations. The \textit{ALPINE} galaxies are indicated by orange circles and squares for $z<5$ and $z>5$, respectively. \label{fig:goodsfootprints}}
\end{figure*}

\section{Photometry from Ground and Space} \label{sec:photometry}

In this section, we summarize the ground- and space-based photometric data that are available for the \ALPINE~galaxies in the COSMOS (105 galaxies) and ECDFS (13 galaxies) fields. Although these fields differ in survey depth, reduction methods, and number and type of photometric filters used, we find that their overall photometric measurements are comparable within $1\sigma$ limits after their conversion to total magnitudes and the correction for the specific biases of each survey. Therefore, we can treat them separately to first order for the matter of measuring various physical properties of the galaxies. 
The basis catalogs to which we match the \textit{ALPINE} galaxies in the COSMOS and ECDFS field are the \textit{COSMOS2015}\footnote{\url{http://cosmos.astro.caltech.edu/page/photom}} \citep{LAIGLE16} and the \textit{3D-HST}\footnote{\url{https://3dhst.research.yale.edu/Data.php}} \citep{BRAMMER12,SKELTON14} catalog, respectively.
A summary of the different data available on the two fields including filter names, wavelengths, $3\sigma$ depths, and references to the measurements are given in Tables~\ref{tab:photogoods} and~\ref{tab:photocosmos}, respectively.
In the following, we describe these data in more detail.

\subsection{Photometry on the ECDFS field}\label{sec:photometrygs}

The photometry for the galaxies on the ECDFS field is taken directly from the \textit{3D-HST} catalog, which provides ground-based observations as well as a wealth of data from HST imaging. The photometry (total fluxes and magnitudes) is corrected for Galactic extinction, PSF size as well as other biases, therefore no further correction are applied. The \textit{ALPINE} galaxies are matched visually to the spatially closest \textit{3D-HST} counterpart using the HST WFC3/IR $F160W$ image as reference. The spectroscopic redshifts match the photometric redshifts within their uncertainty ($\sim0.1-0.2$), ensuring that we identified the correct counterpart.

The ground-based photometry available in ECDFS (including references) is listed in Table~\ref{tab:photogoods}. Summarizing, this includes the $U38$, $b$, $v$, $R_{\rm c}$, and $I$ broad-band filters from the Wide Field Imager on the 2.2 meter MPG/ESO telescope, the $U$ and $R$ bands from VIMOS on the VLT, the near-IR filters $J^{\rm v}$, $H^{\rm v}$, and $K_{\rm s}^{\rm v}$ from ISAAC on the VLT, $J_{\rm w}$ and $K_{\rm s}^{\rm w}$ data taken by WIRCam on the CFHT, as well as 14 intermediate-band filter from the Suprime-Cam on the Subaru telescope. For galaxies at $z=4.5$ and $5.5$, the Lyman-break falls roughly in the $v$ and the $R_{\rm c}$-band and therefore the galaxies are expected to be only faintly (or not at all) visible in these and blue-ward filters. On the other hand, the galaxies are bright at observed near-IR wavelengths, i.e., filters red-ward of $z$-band (corresponding to roughly the $F850LP$ filter. Figure~\ref{fig:magnitudecomparison} shows the stacked $F850LP$ ($z$), $J^{\rm v}$ ($J$) and $K_{\rm s}^{\rm v}$ ($K$) magnitude distributions of the ECDFS \textit{ALPINE} galaxies split in $z<5$ (hatched blue) and $z>5$ (hatched red). As expected, the latter sample occupies slightly fainter magnitudes.

The space-based photometry includes the four Spitzer bands at $3.6\,{\rm \mu m}$, $4.5\,{\rm \mu m}$, $5.8\,{\rm \mu m}$, and $8.0\,{\rm \mu m}$. In addition, the public \textit{3D-HST} catalog includes a wealth of HST photometry. Specifically, it contains measurements in the ACS bands $F435W$, $F606W$, $F775W$, $F814W$, and $F850LP$ as well as in the WFC3/IR bands $F125W$, and $F160W$ bands for all $13$ \textit{ALPINE} galaxies. Only 10 galaxies have measurements in the WFC3/IR band $F140W$. The HST photometry is measured on PSF-matched images. As described in \citet{SKELTON14}, the Spitzer and ground-based photometry are measured using the \texttt{MOPHONGO} \citep[][]{LABBE06,WUYTS07,WHITAKER11}, which uses a high-resolution image (here the HST imaging) as spatial prior to estimate the contributions from neighboring blended sources in the lower resolution image.
The different depths of these observations as well as references are listed in Table~\ref{tab:photogoods}.
A query of the Barbara A. Mikulski Archive for Space Telescopes (MAST\footnote{\url{https://mast.stsci.edu}}) using the \texttt{mastquery} Python package\footnote{\url{https://github.com/gbrammer/mastquery}} shows that in addition to the HST measurements contained in the \textit{3D-HST} catalog, four, ten, and two galaxies have coverage in the WFC3/IR bands $F098M$, $F105W$, and $F110W$, respectively. None of the galaxies has ACS $F475W$ coverage.
These additional data that are not published in the \textit{3D-HST} catalog come from various other observation programs in and around the ECDFS field. We subsequently measure this additional photometry for all \textit{ALPINE} galaxies in ECDFS using \texttt{SExtractor} \citep[version 2.19.5,][]{BERTIN96} in different aperture sizes ($0.7\arcsec$ and $3\arcsec$) as well as auto magnitudes.
For this, we first create a mosaic of all the HST pointings that overlap with the \textit{ALPINE} galaxies using the \texttt{AWS-drizzler}\footnote{\url{https://github.com/grizli-project/grizli-aws}} tool that is part of the \texttt{grizli} \footnote{\url{https://github.com/gbrammer/grizli}} Python package (Brammer, in prep.). We use a $0.06\arcsec$ pixel scale and all HST images are registered to Gaia (see Section~\ref{sec:astrometry}). \texttt{SExtractor} is run with relative \texttt{THRESH\_TYPE}, and we set \texttt{DETECT\_MINAREA}, \texttt{DETECT\_THRESH}, \texttt{DEBLEND\_MINCONT}, and \texttt{DEBLEND\_NTHRESH} to $3$, $1.5$, $1.5$, $0.001$, and $64$, respectively.
If no object is detected above the threshold ($1.5\sigma$) within $0.7\arcsec$ (roughly the ground-based seeing) of the original \textit{ALPINE} coordinates, we consider the galaxy as undetected in a given band and replace its flux by a $1\sigma$ limit that is computed from the RMS noise at the position of the galaxy.
The photometry measured by \texttt{SExtractor} is subsequently corrected for galactic foreground extinction, which we assume to be constant for all galaxies in ECDSF at $\ebmv = 0.0069\,{\rm mag}$.
Figure~\ref{fig:goodsfootprints} summarizes the HST data available for the galaxies in the ECDFS field. The blue squares show the layout of all the HST pointings as of October 2019, with darker shades of blue indicating more observations. The \textit{ALPINE} galaxies at $z<5$ and $z>5$ are indicated with orange circles and squares, respectively.

\begin{figure*}
\includegraphics[width=2.1\columnwidth, angle=0]{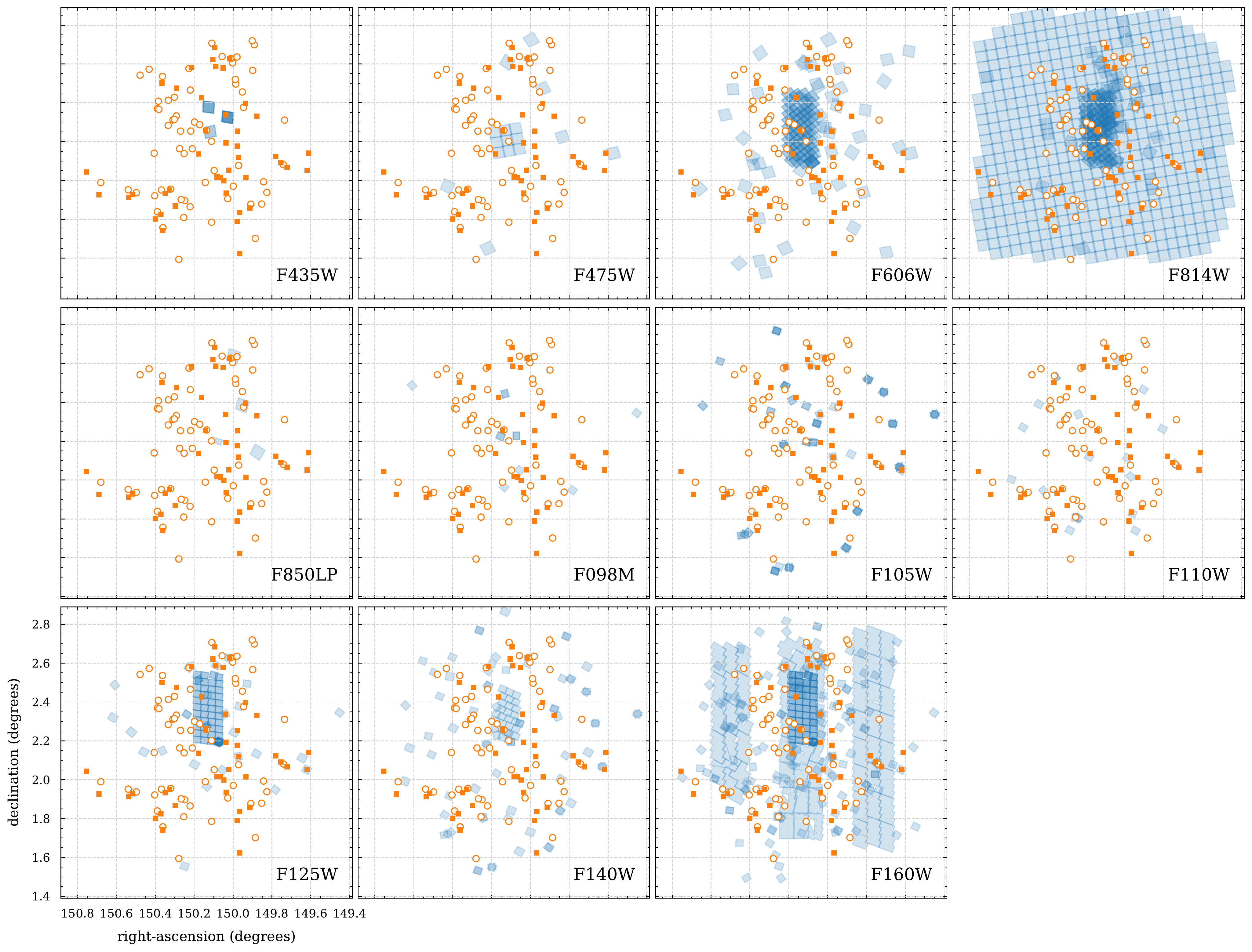}
\caption{HST pointing footprints on the COSMOS field (blue) for different HST ACS and WFC3/IR filters. Darker colors mean more observations. The \textit{ALPINE} galaxies are indicated by orange circles and squares for $z<5$ and $z>5$, respectively. \label{fig:cosmosfootprints}}
\end{figure*}

\subsection{Photometry on the COSMOS field}\label{sec:photometrycosmos}

Most of the \textit{ALPINE} galaxies (105 out of 118) reside in the COSMOS field and we match them to the latest photometric measurements presented in the \textit{COSMOS2015} catalog. The matching is again done on a galaxy-by-galaxy basis using the HST/ACS $F814W$ as well as UltraVISTA $K_{\rm s}$ images as references. We also match against the photometric redshifts given in the catalog in order to identify the correct counterpart\footnote{The photometric redshifts given in the \textit{COSMOS2015} catalog are consistent within 0.2 with our spectroscopic redshifts for more than $95\%$ of all cases.}.

In Table~\ref{tab:photocosmos}, we list all the photometric measurements available for the \textit{ALPINE} galaxies on the COSMOS field. Summarizing, these include $u^{*}$-band observations from MegaCam on CFHT, the $B$, $V$, $r^{+}$, $i^{+}$, $z^{++}$ as well as 12 intermediate-band and 2 narrow-band filters from the Suprime-Cam on Subaru, the $Y_{\rm HSC}$-band from the Hyper Suprime-Cam on Subaru as well as near-IR bands $H_{\rm w}$ and $K_{\rm s}^{\rm w}$ from WIRCam on CFHT and $Y$, $J$, $H$, and $K_{\rm s}$ from VIRCAM on the VISTA telescope. In addition, the galaxies are covered by the four Spitzer channels from $3.6\,{\rm \mu m}$ to $8.0\,{\rm \mu m}$ from the SPLASH survey\footnote{\url{https://splash.caltech.edu}} \citep{CAPAK-SPITZER12,STEINHARDT14,LAIGLE16}. As described in \citep[][]{LAIGLE16}, the Spitzer photometry is measured using \texttt{IRACLEAN} \citep[][]{HSIEH12}, which uses positional priors from higher resolution imaging (in this case the $zYJKHK_{\rm s}$ detection $\chi^2-$image) to deblend the photometry.

Contrary to the \textit{3D-HST} photometry catalog, the fluxes and magnitudes in the \textit{COSMOS2015} catalog are not total and not corrected for systematic biases and Galactic extinction. To perform these corrections, we follow the steps outlined in the appendix of \citet{LAIGLE16}. 
Specifically, we use the $3\arcsec$ diameter aperture magnitudes ($\mathcal{M}^3$), which we correct for photometric ($o_i$, see their equation 4) and systematic offsets ($s_f$, see their table 3) by applying
\begin{equation}
\mathcal{M}^{tot, uncorr}_{i,f} = \mathcal{M}^{3}_{i,f} + o_i - s_f,
\end{equation}
where $i$ is the object identifier and $f$ denotes the different filters. The total magnitudes are subsequently corrected for Galactic extinction by applying 
\begin{equation}\label{eq:extinctioncorr}
\mathcal{M}^{tot}_{i,f} = \mathcal{M}^{tot, uncorr}_{i,f} - {\rm EBV}_i \times F_f,
\end{equation}
where ${\rm EBV}_i$ is the Galactic extinction from the \citet{SCHLEGEL98} maps on COSMOS for each object as given in the catalog and $F_f$ are the extinction factors per filter given in table 3 of \citet{LAIGLE16}.

Figure~\ref{fig:magnitudecomparison} shows the stacked magnitude distribution in $z^{++}$ ($z$), $J$, and $K_{\rm s}$ ($K$) bands for the $z<5$ (blue) and $z>5$ (red) sub-samples. As expected, the high-redshift galaxies are fainter in all bands. In addition, we find the magnitude distributions between the galaxies in ECDFS and COSMOS to be similar. This indicates no major discrepancies in photometric, hence physical properties between the two samples.

In terms of HST imaging, all galaxies except one are observed in ACS $F814W$ \citep{SCOVILLE07b,KOEKEMOER07}. In addition to this, a MAST-search shows that several galaxies are covered by other observing programs in the ACS bands $F435W$ (3), $F475W$ (5), $F606W$ (21), and $F850LP$ (5) as well as in the WFC3/IR bands, $F105W$ (11), $F110W$ (5), $F125W$ (16), $F140W$ (13), and $F160W$ (53)\footnote{There is also data in the  $F098W$ filter, however, unfortunately no \textit{ALPINE} galaxies are covered.}. Note that the observations in $F160W$ primarily come from the CANDELS survey \citep[covering the central part of COSMOS,][]{GROGIN11,KOEKEMOER11} as well as the ``drift and shift'' \citep[DASH,][]{MOMCHEVA17} survey. While the CANDELS imaging is deep ($>27.5$ magnitudes at $3\sigma$), the data from the DASH survey is much shallower ($25.0$ magnitudes at $3\sigma$) and therefore only half of the galaxies are detected. Furthermore, the spatial sampling in the latter does not allow a detailed study of the structure of the galaxies.
Figure~\ref{fig:cosmosfootprints} summarizes the HST pointings on the COSMOS field (blue). The CANDELS area in the center of the COSMOS field as well as the three DASH stripes are evident. The location of the \textit{ALPINE} galaxies is indicated with circles ($z<5$) and squares ($z>5$). The photometry available is summarized in Table~\ref{tab:photocosmos} including depths (where applicable) and references.

In order to measure the  photometry in these HST bands, we use \texttt{SExtractor} in the same way as described in Section~\ref{sec:photometrygs}.
We first create a mosaic ($0.06\arcsec$ pixel sizes and registered to Gaia) using the \texttt{AWS-drizzler}.
We subsequently measure the HST photometry of on each of the images for all \textit{ALPINE} galaxies with coverage using \texttt{SExtractor} in apertures ($0.7\arcsec$ and $3\arcsec$) as well as auto magnitudes. If no object is detected above the set threshold level within $0.7\arcsec$ of the original \textit{ALPINE} coordinates, we consider the galaxy as undetected in a given band. Similar to Equation~\ref{eq:extinctioncorr}, we correct the HST photometry for galactic foreground extinction using the \citet{SCHLEGEL98} extinction map.
We  compare these measurements to the ground-based photometry, by first performing a PSF matching by smoothing the original HST images with a Gaussian kernel with FWHM of $0.7\arcsec$ and measure the photometry in a $3\arcsec$ aperture (as used in the \textit{COSMOS2015} catalog). 
We also compare the ground-based ($i^+$, $Y$, $J$, $H$) with the HST ($F814W$, $F105W$, $F125W$, $F160W$) photometry in approximately matching filter bands. We find an overall agreement on average of $0.2$ magnitudes (dark gray region) with an expected increase of scatter at fainter magnitudes. For the four brightest sources in $H$-band ($<23.4\,{\rm AB}$), the ground-based measured photometry is systematically up to $0.5$ magnitudes brighter in three out of four cases. Due to the low number of galaxies, it is difficult to investigate this statistically.
A more detailed measurement of the HST photometry including deblending of specific sources such as merging or clumpy galaxies, will be provided in a forthcoming paper.

\begin{figure*}
\includegraphics[width=1.05\columnwidth, angle=0]{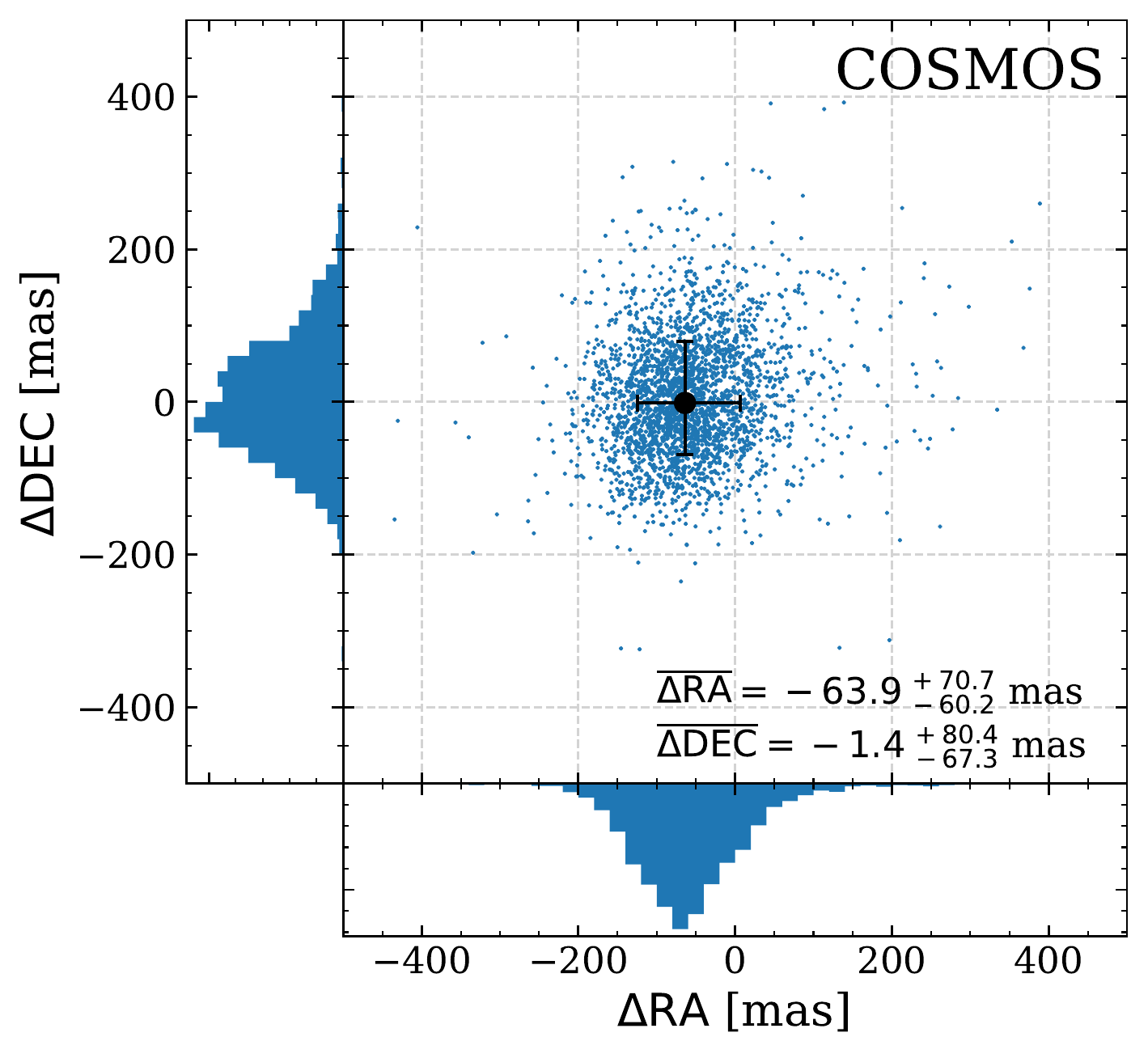}
\includegraphics[width=1.05\columnwidth, angle=0]{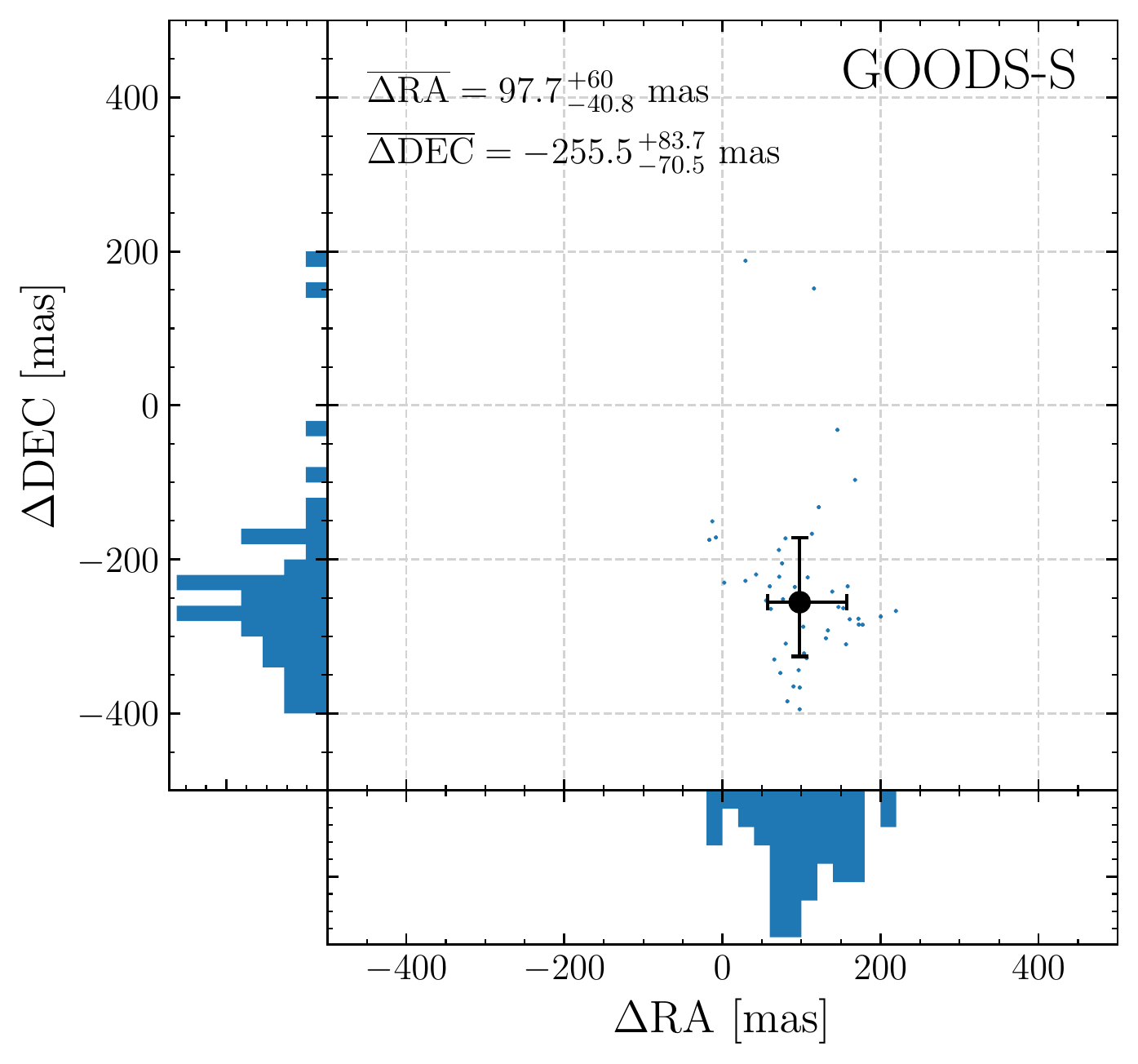}
\caption{Scatter diagrams and histograms of offsets between the Gaia reference frame and the \textit{COSMOS2015} (left) and \textit{3D-HST} (ECDFS, right) catalogs. The offsets are in the sense ``Gaia $-$ COSMOS'' and ``Gaia $-$ ECDFS'', respectively. For COSMOS we find only a systematic offsets in right-ascension direction of $-64\,{\rm mas}$. For ECDFS, the offsets are large in both directions. In addition to this, we measure a scatter in the astrometry of $\sim100\,{\rm mas}$ for both fields in both coordinates. \label{fig:gaiaoffsets}}
\end{figure*}

\subsection{Astrometric offsets between ALMA and ancillary data} \label{sec:astrometry}

Astrometric accuracy is crucial in many cases.
First, accurate spatial offsets of light emission from rest-frame UV, optical, and sub-mm wavelengths reveal the properties of the interstellar medium such as the location and interplay of stars, dust, and gas.
Second, the robustness of the identification of sub-mm counterparts needs a high astrometric accuracy of all involved datasets.
As described in Sections~\ref{sec:photometrygs} and~\ref{sec:photometrycosmos}, the HST images produced for the \textit{ALPINE} galaxies are all aligned to Gaia. Our tests show that the average offsets are less than 15 milli-arcseconds (mas) in both right-ascension and declination with a scatter of no more than $30\,{\rm mas}$ (G. Brammer, private communication).
Unfortunately, the positional accuracy for current catalogs (such as \textit{3D-HST} or \textit{COSMOS2015} that are used here) are lower and we expect significant offsets between those astrometric solution and ALMA data products. In the following, we characterize these offsets.

According to the technical handbook, ALMA observations are currently registered to the \textit{International Celestial Reference Frame} (ICRF) to an accuracy better than $\sim5\,{\rm mas}$. The ICRF is based on hundreds of extragalactic radio sources such as quasars distributed over the whole sky. The positional accuracy $\Delta p$ of single observations can be estimated by 
\begin{equation}
\Delta p = \frac{70000}{\nu \cdot B \cdot \sigma},
\end{equation}
where $\nu$ is the observed frequency in GHz, $B$ is the maximum baseline length in kilometers, and $\sigma$ is the S/N at the peak of emission.
For \ALPINE~($\nu = 330\,{\rm GHz}$ and $B=0.2\,{\rm km}$ for C43-1) this leads to $\Delta p = 1060/\sigma$. The calibrators are detected well above $50\sigma$, which leads to an absolute positional accuracy of $\sim20\,{\rm mas}$ or better.

To check the astrometric alignment of the photometric catalogs used here, we make use of the Gaia DR2 catalog \citep[][]{MIGNARD18a}, which provides currently the most accurate absolute astrometry. As shown in \citet{MIGNARD18b}, there are no significant offsets between this reference frame and the ICRF frame used by ALMA, hence this test directly reveals potential differences in astrometry between the \textit{3D-HST} and \textit{COSMOS2015} catalogs and our ALMA observations.
Using the proper motion information of the stars from Gaia, we project their positions back in time to the year of calibration of the data products by to using the equations
\begin{equation}
\begin{aligned}
\Delta\alpha &= \cos{(\delta)} \cdot \mathcal{P}_{\alpha} \cdot (t_{\rm ref} - t_{\rm gaia}), \\
\Delta\delta &= \mathcal{P}_{\delta} \cdot (t_{\rm ref} - t_{\rm gaia}) ,
\end{aligned}
\end{equation}
where $\alpha$ and $\delta$ denote the right-ascension and declination (and $\mathcal{P}_{\alpha}$ and $\mathcal{P}_{\delta}$ their proper motion), $t_{\rm gaia}$ is the Gaia reference frame in years (here 2015.5), and $t_{\rm ref}$ is the reference frame of the calibration of the catalogs.
To increase accuracy, we only include stars with a proper motion in both coordinates of less than $5\,{\rm mas\,yr^{-1}}$ in the following. Note that no parallax motion is included in the above formulae, which would result in less than $5\,{\rm mas\,yr^{-1}}$ astrometric shifts.

The \textit{COSMOS2015} catalog is calibrated to the MegaCam $i-$band data that was taken in 2004 \citep[][]{LAIGLE16}.
For ECDFS, the catalog (\textit{3D-HST}) is calibrated to the same reference system as the CANDELS HST images, namely ground-based $R-$band data taken in 2001 \citep[][]{KOEKEMOER11,SKELTON14}. However, the exact year is not important as we have selected stars with a relatively slow proper motion.
In order to select stars on the COSMOS field, we use the ACS/F814W \texttt{SExtractor} catalog \citep{SCOVILLE07b} and select sources with a magnitude brighter than $23\,{\rm AB}$ and \textit{SExtractor} star/galaxy classifier value of $>0.8$. These stars are then matched to the \textit{COSMOS2015} catalog to obtain their position in that catalog.
For ECDFS we extract stars directly from the \textit{3D-HST} catalog by selecting sources brighter than $23\,{\rm AB}$ in F160W with \texttt{star\_flag} $ = 1$ and a \texttt{SExtractor} star/galaxy classifier value larger than $0.8$.
 Other selections (e.g., different magnitude cuts) do not affect the following results.
The Gaia stars are then matched to the star catalogs in COSMOS and ECDFS to obtain the astrometric offsets. In total, $47$ and $2724$ Gaia stars are used in ECDFS and COSMOS, respectively.

Figure~\ref{fig:gaiaoffsets} shows scatter plots and histograms comparing the position of the Gaia stars to the positions in the catalogs.
We find significant systematic offsets in the astrometry in ECDFS of $98\,{\rm mas}$ in right-ascension and $-256\,{\rm mas}$ in declination. These offsets are consistent with what was found in earlier studies \citep[][]{DUNLOP17,FRANCO18,WHITAKER19}. For COSMOS we only find a significant offset in right-ascension of $-64\,{\rm mas}$. In addition to that, there is a significant scatter in the astrometry on the order of $100\,{\rm mas}$ in both coordinates in both fields.

To compute the astrometric offset of individual \textit{ALPINE} galaxies, we make use of the fact that the HST images are already aligned to the Gaia reference frame (see details in Sections~\ref{sec:photometrygs} and~\ref{sec:photometrycosmos}). Specifically, we compute the offsets between the coordinates measured on the ACS/F814W images and the original coordinates from the \textit{3D-HST} or \textit{COSMOS2015} catalog. If no ACS $F814W$ image is available or if the galaxy is not detected (which happens for redshifts $z>5$), we use the deepest image red of $F814W$. If none is available (in four cases), we report the average offset as shown in Figure~\ref{fig:gaiaoffsets}.

Note that the coordinates given in the final ancillary data catalog are not corrected for these offsets. However, we give the offsets for each galaxy in the columns \texttt{delta\_RA} and \texttt{delta\_DEC}, which can be \textit{added} to the original coordinates to obtain Gaia-corrected right-ascensions and declinations (see Appendix~\ref{app:dataproducts}).

\section{Physical properties}\label{sec:physicalproperties}

In this section, we detail measurements of various basic physical properties of the \textit{ALPINE} galaxies that are based on their total, extinction corrected photometry described in Section~\ref{sec:photometry}.
These include physical quantities from SED fitting such as stellar masses, SFRs, ages, and dust attenuation (\S\ref{sec:sedfitting}), and UV continuum slopes (\S\ref{sec:uvslopes}), as well as quantities directly derived from the photometry such as UV magnitudes and luminosities (\S\ref{sec:uvmags}) and estimates of the \halpha~luminosity and equivalent width from Spitzer colors (\S\ref{sec:halphaemission}).

\FloatBarrier
\begin{deluxetable}{l c  c  c  c}[h!]
\tabletypesize{\scriptsize}
\tablecaption{Photometry available for galaxies on the ECDFS field.\label{tab:photogoods}}
\tablewidth{0pt}
\tablehead{
\colhead{Observatory/Instrument} & \colhead{Filter} & \colhead{Central $\lambda$} & \colhead{$3\sigma$ depth} & \colhead{Ref.}\\[-0.3cm]
\colhead{} & \colhead{} & \colhead{($\rm \AA$)} & \colhead{(mag)} & \colhead{}
}
\startdata
\\[0cm]
\multicolumn{5}{c}{\textbf{Ground-based} }\\[0.2cm]
MPG-ESO/WFI & $U38$ & 3633.3 & 27.3 & 1 \\
 & $b$ & 4571.2 & 26.6 & 1 \\
 & $v$ & 5377.0 & 26.6 & 1 \\
 & $R_{\rm c}$ & 6536.3 & 26.9 & 1 \\
 & $I$ & 9920.2 & 26.5 & 1 \\
VLT/VIMOS & $U$ & 3720.5 & 28.6 & 2 \\
 & $R$ & 6449.7 & 27.8 & 2 \\
 VLT/ISAAC & $J^{\rm v}$ & 12492.2 & 25.6 & 3 \\
& $H^{\rm v}$ & 16519.9 & 25.1 & 3 \\
 & $K_{\rm s}^{\rm v}$ & 21638.3 & 25.0 & 3 \\
CFHT/WIRCam & $J_{\rm w}$ & 12544.6  & 25.1 & 4 \\
 & $K_{\rm s}^{\rm w}$ & 21590.4 & 24.5 & 4 \\
Subaru/Suprime-Cam  & $IA427$ & 4263.4 & 25.7 & 5  \\
& $IA445$ & 4456.0 & 25.7 & 5 \\
& $IA505$ & 5062.5 & 25.8 & 5 \\
& $IA527$ & 5261.1 & 26.7 & 5 \\
& $IA550$ & 5512.0 & 26.0 & 5 \\
& $IA574$ & 5764.8 & 25.7 & 5 \\
& $IA598$ & 6000.0 & 26.6 & 5 \\
& $IA624$ & 6233.1 & 26.5 & 5  \\
& $IA651$ & 6502.0  & 26.7 & 5 \\
& $IA679$ & 6781.1  & 26.6 & 5 \\
& $IA738$ & 7371.0 & 26.5 & 5 \\
& $IA767$ & 7684.9 & 25.5 & 5 \\
& $IA797$ & 7981.0 & 25.2 & 5 \\
& $IA856$ & 8566.0 & 25.0 & 5 \\[0.2cm]
 \multicolumn{5}{c}{\textbf{Space-based} }\\[0.2cm]
 HST/ACS  & F435W  & 4328.7  & 29.1  & 6\\
 & $F606W$ & 5924.8  & 29.1 (29.0) & 6,7 \\ 
 & $F775W$ & 7704.8  & 28.4 & 6 \\
 & $F814W$ & 8058.2  & 28.9 & 7 \\
 & $F850LP$ & 9181.2  & 27.9 & 6 \\ 
HST/WFC3 & $F125W$ & 12516.3 & 27.6 & 8 \\
 & $F140W$ & 13969.4 & 26.7 & 9 \\
 & $F160W$ & 15391.1 & 27.7 & 8 \\
\textit{Spitzer}/IRAC & ch$_1$ & 35634.3 & $\sim26.0$$^{a}$ & 10\\ 
& ch$_2$ & 45110.1 & $\sim26.0$$^{a}$ & 10 \\ 
& ch$_3$ & 57593.4 & 24.4 & 11\\
 & ch$_4$ & 79594.9 & 24.3 & 11 \\
\enddata
\tablenotetext{a}{The exposure time varies between $10\,{\rm ks}$ and $300\,{\rm ks}$ (corresponding to roughly 1.8 in magnitudes).\\[-0.5cm]}
\tablenotetext{}{References:
(1) \citet{HILDEBRANDT06,ERBEN05},
(2) \citet{NONINO09},
(3) \citet{WUYTS08,RETZLAFF10},
(4) \citet{HSIEH12},
(5) \citet{CARDAMONE10},
(6) \citet{GIAVALISCO04},
(7) \citet{KOEKEMOER11},
(8) \citet{GROGIN11,KOEKEMOER11},
(9) \citet{BRAMMER12,VANDOKKUM13},
(10) \citet{ASHBY13,GUO13},
(11) \citet{DICKINSON03}}
\end{deluxetable}
\FloatBarrier
\nopagebreak

\FloatBarrier
\begin{deluxetable}{l c  c  c  c}[h!]
\tabletypesize{\scriptsize}
\tablecaption{Photometry available on the COSMOS field.\label{tab:photocosmos}}
\tablewidth{0pt}
\tablehead{
\colhead{Observatory/Instrument} & \colhead{Filter} & \colhead{Central $\lambda$} & \colhead{$3\sigma$ depth} & \colhead{Ref.}\\[-0.3cm]
\colhead{} & \colhead{} & \colhead{($\rm \AA$)} & \colhead{(mag)$^{\dagger}$} & \colhead{}
}
\startdata
\\[0cm]
\multicolumn{5}{c}{\textbf{Ground-based} }\\[0.2cm]
CFHT/MegaCam & $u^{*}$ & 3823.3 & 26.6 & 1 \\
Subaru/Suprime-Cam & $B$ & 4458.3 & 27.0 & 1,2\\
 & $V$ & 5477.8 & 26.2 & 1,2\\ 
& $r^{+}$ & 6288.7 & 26.5 & 1,2 \\
 & $i^{+}$ & 7683.9 & 26.2 & 1,2 \\
 & $z^{++}$ & 9105.7  & 25.9 & 1,2 \\
 & $IA427$ & 4263.4 & 25.9 & 1,2 \\
 & $IA464$ & 4635.1 & 25.9 & 1,2  \\
 & $IA484$ & 4849.2 & 25.9 & 1,2 \\
 & $IA505$ & 5062.5 &  25.7 & 1,2 \\
& $IA527$ & 5261.1 & 26.1 & 1,2 \\
& $IA574$ & 5764.8 & 25.5 & 1,2 \\
& $IA624$ & 6233.1 & 25.9 & 1,2  \\
& $IA679$ & 6781.1  & 25.4 & 1,2 \\
& $IA709$ & 7073.6 & 25.7 & 1,2 \\
& $IA738$ & 7361.6 & 25.6 & 1,2 \\
 & $IA767$ & 7684.9 & 25.3 & 1,2 \\
 & $IA827$ & 8244.5 & 25.2 & 1,2 \\
 & $NB711$ & 7119.9 & 25.1 & 1,2  \\
 & $NB816$ & 8149.4 & 25.2 & 1,2  \\
Subaru/HSC & $Y_{\rm HSC}$ & 9791.4  & 24.4 & 1 \\
CFHT/WIRCam & $H_{\rm w}$ & 16311.4  & 23.5 & 1 \\
 & $K_{\rm s}^{\rm w}$ & 21590.4  & 23.4 & 1 \\
VISTA/VIRCAM & $Y$ & 10214.2 & 24.8 (25.3)$^a$ & 1   \\ 
& $J$ & 12534.6 & 24.7 (24.9)$^a$ & 1 \\
& $H$ & 16453.4  & 24.3 (24.6)$^a$ & 1 \\
& $K_{\rm s}$ & 21539.9 & 24.0 (24.7)$^a$ & 1 \\[0.2cm]
 \multicolumn{5}{c}{\textbf{Space-based} }\\[0.2cm]
HST/ACS  & F435W  & 4328.7   & $-^{b}$  & $-$\\
& $F475W$ & 4792.3  & $-^{b}$ & $-$ \\
& $F606W$ & 5924.8  & $-^{b}$ & $-$ \\
& $F814W$ & 8058.2  & 29.2 & 3 \\
& $F850LP$ & 9181.2  & $-^{b}$ & $-$ \\
HST/WFC3 & $F098M$ & 9877.4  & $-^{b}$ & $-$ \\
 & $F105W$ & 10584.9 & $-^{b}$ & $-$ \\
& $F110W$ & 11623.8  & $-^{b}$ & $-$ \\
& $F125W$$^c$ & 12516.3 & $27.6^{b,c}$ & 4 \\
& $F140W$ & 13969.4 & $-^{b}$ & $-$ \\
& $F160W$$^c$ & 15391.1 & $27.5$ $(25.0)^{b,c,d}$ & 4,5 \\
\textit{Spitzer}/IRAC & ch$_1$ & 35634.3 & 25.5 & 1,6,7\\
& ch$_2$ & 45110.1 & 25.5 & 1,6,7 \\
& ch$_3$ & 57593.4 & 23.0 & 1,7\\
 & ch$_4$ & 79594.9 & 22.9 & 1,7 \\
\enddata
\tablenotetext{\dagger}{Depth is measured in $3\arcsec$ aperture for ground-based photometry.\\[-0.5cm]}
\tablenotetext{a}{Depths of the ultra-deep area are given in parenthesis.\\[-0.5cm]}
\tablenotetext{b}{Ancillary pointings, therefore depth varies depending on the specific observations.\\[-0.5cm]}
\tablenotetext{c}{Most of these data come from the CANDELS survey and their depth is indicated.\\[-0.5cm]}
\tablenotetext{d}{Some of the F160W data comes from the \textit{DASH} survey (depth given in parenthesis).\\[-0.5cm]}

\tablenotetext{}{References:
(1) see \citet{LAIGLE16},
(2) \citet{TANIGUCHI07,TANIGUCHI15}
(3) \citet{SCOVILLE07,KOEKEMOER07},
(4) \citet{GROGIN11,KOEKEMOER11}, 
(5) \citet{MOMCHEVA17}, 
(6) \citet{CAPAK-SPITZER12,STEINHARDT14}, 
(7) \citet{SANDERS07}
}
\end{deluxetable}
\FloatBarrier

\subsection{Stellar mass and SFRs from SED fitting}\label{sec:sedfitting}

\subsubsection{Fitting Method}\label{sec:sedfittingmethod}

For consistency and comparability with other studies on the COSMOS field, we choose the \texttt{LePhare} SED fitting code\footnote{\url{http://www.cfht.hawaii.edu/~arnouts/lephare.html}} \citep{ARNOUTS99,ILBERT06} to derive stellar masses, SFRs, light-weighted stellar population ages, absolute magnitudes, optical dust reddening, and UV continuum slopes of the \textit{ALPINE} galaxies.

Importantly, stellar masses, SFR, and sSFR values are computed from the marginalized probability distribution functions over all the models and uncertainties are given in $\pm1\sigma$ of this distribution.

We use a set of synthetic templates based on the \citet[][]{BRUZUALCHARLOT03} stellar population library, which we tune to represent best galaxies at redshifts between $4 < z < 6$.
In detail, we use a series of different star-formation histories (SFH) based on exponentially declining (with $\tau=$ 0.1, 0.3, 1.0, and 3.0 Gyrs), delayed\footnote{The delayed SFR is parameterzed as $\psi(t) \propto \tau^{-2}\,t\,e^{-t/\tau}$ such that $\psi(t)$ is maximal at $t=\tau$.} (with $\tau=$ 0.1, 0.5, 1.0, and 3.0 Gyrs), and constant star formation. 
We add dust attenuation corresponding to a stellar $\ebmvs$ from $0$ to $0.5$ spaced in steps of $0.05$ assuming a \citet{CALZETTI00} dust attenuation law. To account for metallicity dependence, we use a solar ($\Zsol$) and $0.2\,\Zsol$ metallicity. We also adopt a \citet{CHABRIER03} IMF in the following. The model SEDs are generated for logarithmically spaced ages starting from $50\,{\rm Myrs}$ to the age of the universe at the redshift of each galaxy.
To each SED, various rest-frame UV and optical emission lines are added following the description in \citet[][]{ILBERT09} by using common conversions outlined in \citet[][]{KENNICUTT98}. Specifically, the UV luminosity at $2300\,{\textrm \AA}$ is converted to a SFR using the relation ${\rm SFR\,(\Msol\,yr^{-1})} = 1.4 \times 10^{-28}\,L_{\rm \nu}\,{\rm (erg\,s^{-1}\,Hz^{-1})}$ and subsequently translated to an \oii~emission line flux using the relation ${\rm SFR\,(\Msol\,yr^{-1})} = (1.4\pm0.4) \times 10^{-41}\,L_{\rm [OII]}\,({\rm erg\,s^{-1}})$. Other emission lines (\lya, \oiii, \hbeta, \halpha) are derived by assuming specific ratios to \oii~that are calibrated by observations \citep[see detailed description with references in][]{ILBERT09}.

The models are fit to the photometry described in Section~\ref{sec:photometry} (and listed in Tables~\ref{tab:photogoods} and~\ref{tab:photocosmos}). Specifically, for the galaxies in the ECDFS field we use the ground-based observations in $U$, $R$, $J^{\rm v}$, $H^{\rm v}$, and $K_{\rm s}^{\rm v}$, as well as the intermediate-bands $IA427$, $IA505$, $IA527$, $IA574$, $IA624$, $IA679$, $IA738$, $IA767$, and $IA856$, and all four Spitzer channels ch$_1$, ch$_2$, ch$_3$, and ch$_4$. We also include observations in the HST filters $F435W$, $F606W$, $F775W$, $F814W$, $F850LP$, $F125W$,  $F140W$, and $F160W$ that are properly combined with the ground-based and Spitzer measurements in the \textit{3D-HST} catalog.
For galaxies in the COSMOS field, we include the ground-based observations in $u^*$, $B$, $V$, $r^{+}$, $i^+$, $z^{++}$, $Y$, Y$_{\rm HSC}$, $J$, $H$, $H_w$, $K_{\rm s}$, and $K_{\rm s}^{\rm w}$, as well as the intermediate-bands $IA427$, $IA464$, $IA484$, $IA505$, $IA527$, $IA574$, $IA624$, $IA679$, $IA709$, $IA738$, $IA767$, and $IA827$. We use all four Spitzer channels but no HST observations as only the one filter ($F814W$) exists for all galaxies.

\begin{figure*}
\includegraphics[width=2.1\columnwidth, angle=0]{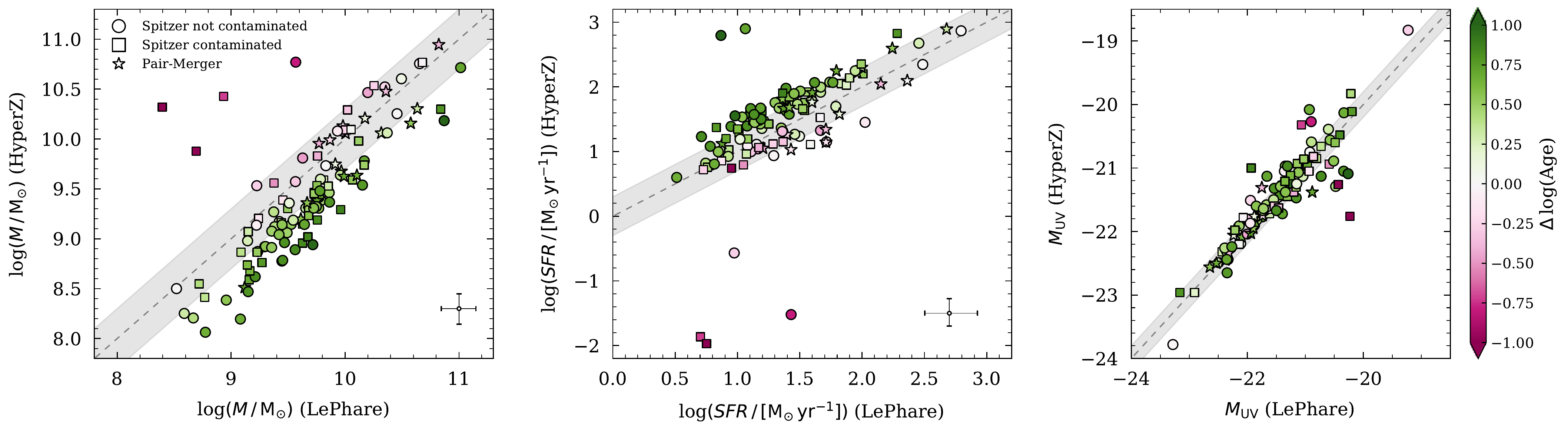}
\caption{Comparison of stellar mass (left), SED-derived SFR (middle), and absolute rest-frame UV magnitude (right) measured by \texttt{LePhare} and \texttt{HyperZ}. The symbols are color-coded by the logarithmic difference in stellar population age (note that the scale is the same for all panels). The squares denote galaxies whose Spitzer photometry is blended with a nearby bright galaxy or star. The star-symbols denote mergers based on the classification in \citet{ALPINE_LEFEVRE19}. The 1-to-1 relation is shown as dashed line and a $\pm0.3$ margin is shown by the gray band. The systematic differences in stellar masses derived by \texttt{HyperZ} are likely caused by different implementations of rest-frame optical emission lines, which affect most the measurement of stellar masses (based on rest-frame optical light).\label{fig:sedfitcomparison}}
\end{figure*}

The fits are performed in flux density space ($f_\nu$ in Jansky), which has several advantages compared to magnitude space.
Specifically, it allows a proper statistical treatment of limits in the data of all bands. While in the case of magnitude limits an arbitrary significance level (e.g., $1\sigma$, $3\sigma$) has to be defined in order to use them in SED fitting, in flux density space no arbitrary choice needs to be made by the user as the limits manifest themselves only in the error bars of those bands where no flux is measured. This is important as slightly different levels of significance set for the limits can have profound effects on the output parameters, especially in cases where limits are imposed in astrophysically important parts of the rest-frame spectra (e.g., the Balmer break).
In addition, some SED codes, including \texttt{Le Phare}, have difficulties dealing with limits (in the case of magnitudes) in a statistically consistent manner, and, instead, remove from consideration any models that slightly exceed these limits. Moving to flux density space for all fits alleviates these concerns.

Several studies have found that the photometric errors are generally underestimated \citep[e.g.,][]{ILBERT06,ILBERT09,ILBERT13,SKELTON14,LAIGLE16}. Artificially increasing the errors has been found to mitigate issues associated with poorly measured bands, poorly estimated zero points, or inhomogeneous methods of measuring photometry.
To avoid artificially small errors that would dominate the $\chi^2$ budget of a fit of a given galaxy, we follow the prescription of \citet{LAIGLE16} and rescale the official flux density errors by a factor of 1.1. In addition, following \citet{ILBERT09}, we correct for underestimated photometric errors due to varying PSF sizes between optical and Spitzer imaging by adding in quadrature the following systematic errors to the photometric error budget prior to fitting: $0.01\,{\rm mag}$ for all ground-based UV/optical broad-band measurements; $0.05\,{\rm mag}$ for all ground-based UV/optical intermediate-bands and near-IR broad-band measurements; $0.1\,{\rm mag}$ for Spitzer $3.6\,{\rm \mu m}$ and $4.5\,{\rm \mu m}$ and $0.3\,{\rm mag}$ for Spitzer $5.8\,{\rm \mu m}$ and $8.0\,{\rm \mu m}$ measurements.

The large PSF sizes of the Spitzer observations result in a large risk of blended photometry. This can lead to an overestimation of flux and hence to an overestimation of stellar masses\footnote{Note that the stellar masses are primarily constrained through the Spitzer photometry that covers rest-frame wavelengths redward of the Balmer break.}. A cleaner and manual deblending of the Spitzer photometry (e.g., using a position prior from HST imaging) is possible and will be pursued in a forthcoming paper.
In the following, we flag galaxies that have a bright companion galaxy within a Spitzer $4.5\,{\rm \mu m}$ PSF FWHM ($2.5\arcsec$) based on the ground-based and HST imaging data. In total, one-third of the galaxies have contaminated Spitzer photometry to some degree. For $19\%$ of the galaxies, the Spitzer photometry is severely contaminated and should not be trusted. We checked our simple flagging scheme against the contamination flags given in the \textit{3D-HST} catalog, which are based on the flux fraction that is overlapping with the galaxy for which the photometry is measured. For the $13$ \textit{ALPINE} galaxies in ECDFS, we find excellent agreement between both classifications. The Spitzer contamination flags are included in the ancillary data catalog (column \texttt{spitzer\_cont}).

\begin{figure*}
\includegraphics[width=2.1\columnwidth, angle=0]{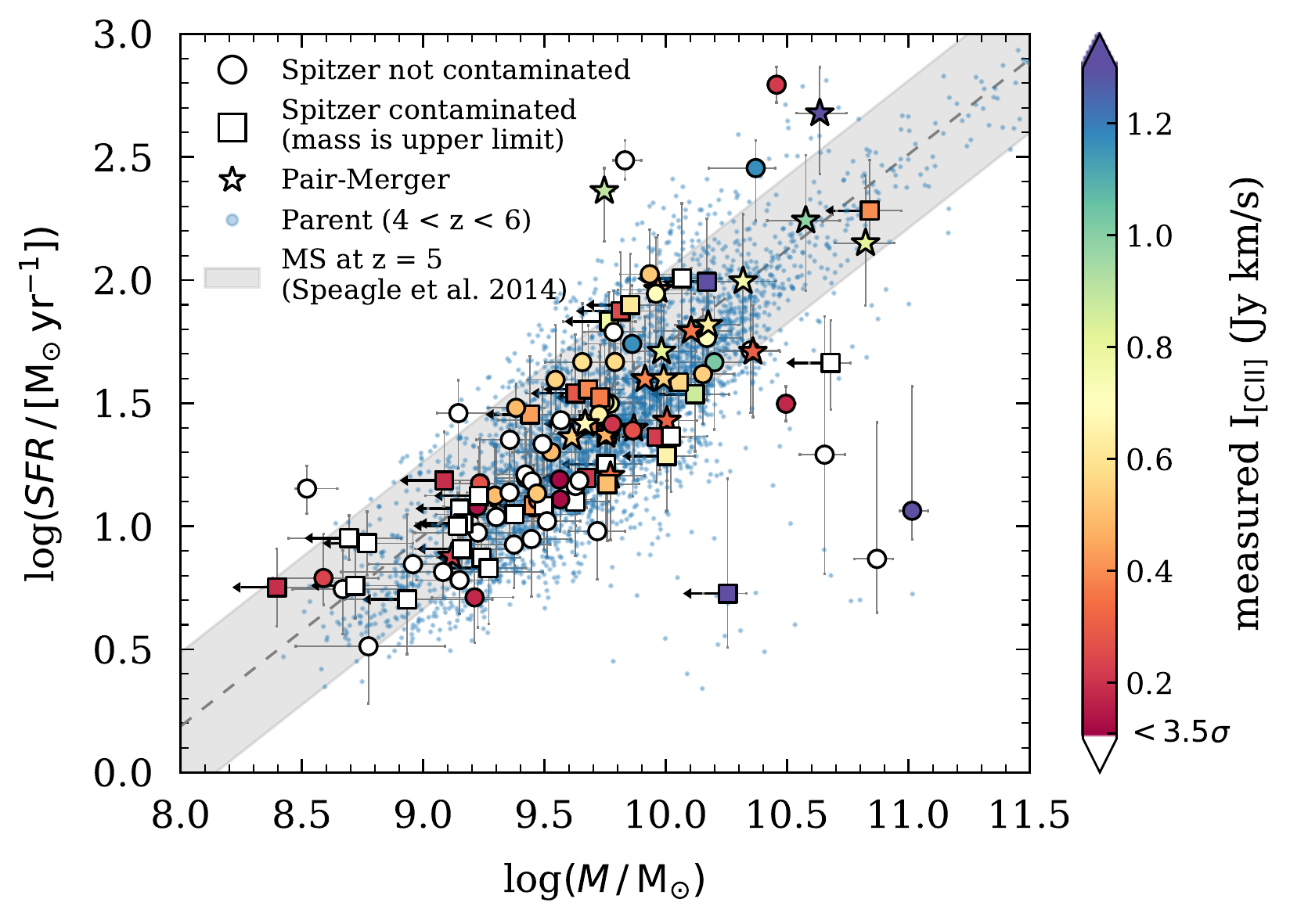}
\caption{Relation between stellar mass and SFR (main-sequence) of our \textit{ALPINE} galaxies compared to all COSMOS galaxies at $4 < z < 6$ (blue points) and the main-sequence parameterization at $z=5$ by \citet[][]{SPEAGLE14} (gray band with $\pm0.3\,{\rm dex}$ width). Galaxies with contaminated Spitzer photometry are  marked with squares (their stellar mass is likely an upper limit) and mergers \citep[classification by][]{ALPINE_LEFEVRE19} are shown as stars. The color denotes the \Cii~flux in ${\rm Jy\,km\,s^{-1}}$ measured by \textit{ALPINE}. Galaxies that are not detected at the $3.5\sigma$ level are shown with white face color. \label{fig:sedmainsequence}}
\end{figure*}

\subsubsection{Systematic Uncertainties in Physical Properties from Modeling Assumptions} \label{sec:sedfittingcomparison}

Depending on their exact methods, different SED fitting codes may measure different physical properties even with the same photometry provided as input. 
In addition to more physical reasons (such as different assumptions on the stellar population models and implementation of emission lines), the different treatment of undetected fluxes (or fitting in magnitude space), varying minimization techniques and weightings, and different scaling of the error of the input photometry can contribute to these discrepancies.

To investigate the amplitude of such differences, we compare the measurements from \texttt{LePhare} to a modified version of the \texttt{HyperZ} code \citep{BOLZONELLA00} that includes the effects of nebular emission \citep{SCHAERER09}, to estimate such systematic uncertainties in the fitted parameters. To minimize the degeneracy with other assumptions, we run \texttt{HyperZ} on the exact same photometry and with the exact same model SEDs (i.e., same metallicity, age, SFH, and dust attenuation law) as described in Section~\ref{sec:sedfittingmethod}.
Figure~\ref{fig:sedfitcomparison} compares the stellar mass (left), SED-derived SFR (middle), and absolute UV magnitude (right, see also Section~\ref{sec:uvmags}) derived by the two codes. The symbols are color-coded by difference in stellar population age (positive values indicate younger ages derived by \texttt{HyperZ}).
We find that the stellar masses derived by \texttt{HyperZ} are systematically smaller by $0.3-0.4\,{\rm dex}$. Similarly, the SFRs are systematically larger by $\sim0.3\,{\rm dex}$.

As expected, the absolute UV magnitudes are largely in agreement, as they are to first order independent of the physical parameters, and just represent a translation of the fitted UV flux.

The differences in stellar mass and SFR are due to the effect of nebular emission, which results in younger ages and larger emission line corrections of the intrinsic rest-frame optical continuum (observed by the Spitzer broad-bands) in the \texttt{HyperZ} models, and therefore directly affects the stellar mass measurements \citep[see, e.g.,][]{DEBARROS14}. This is confirmed by the fact that the stellar masses and SFR measured by the two codes agree well (within a factor of two) if the emission lines are turned off. The effect of nebular emission found by \texttt{HyperZ} may be overestimated for the \textit{ALPINE} galaxies, e.g., if the emission lines are more strongly attenuated than the continuum (see Section~\ref{sec:halphaemission}).
For consistency and comparability with other studies on the COSMOS field, we choose the \texttt{LePhare} fitting results as the default.

Next to the systematic offset discussed above, we also find four galaxies (\textit{DC\_472215}, \textit{DC\_503575}, \textit{DC\_722679}, \textit{DC\_790930}) whose stellar mass measurements are significantly discrepant, by more than one order of magnitude, between the two codes. Specifically, they are fitted with an old, low-SFR, and massive galaxy template with \texttt{HyperZ}, while a young, high-SFR, and low-mass galaxy template is preferred by \texttt{LePhare}. Three of these outliers have significantly contaminated Spitzer photometry (indicated by the squares). As a consequence, the apparent Spitzer fluxes and stellar masses are overestimated and the optical colors are artificially reddened, which makes their stellar masses largely unreliable. Furthermore, we artificially increased the errors of the Spitzer photometry in our \texttt{LePhare} measurements (see Section~\ref{sec:sedfittingmethod}), hence they have smaller weights, which might reduce the effect of photometric contamination on the fit.

In addition, we investigate the effect of different dust reddening laws on the \texttt{LePhare} measurements. For this, we compare a \citeauthor{CALZETTI00} reddening with a Small Magellanic Cloud \citep[SMC,][]{PREVOT84} reddening. The latter might be more suitable for metal poor low-mass galaxies. Running \texttt{LePhare} with the same settings but adopting a SMC reddening curve, we find only small changes in the stellar mass and SFR measurements. Specifically, for the former we find an average offset towards lower stellar masses in the case of SMC dust of $0.05\,{\rm dex}$ and a maximal offset of $0.2\,{\rm dex}$. For latter we find similar offsets towards lower SFRs in the case of SMC dust.

Finally, we note that the galaxy properties derived here are consistent within a factor of two with the ones published in the \textit{COSMOS2015} catalog (based on \textit{photometric} redshifts). We conclude this from comparing galaxies with the same spectroscopic and photometric redshift within $0.1$.

\begin{figure*}
\includegraphics[width=0.7\columnwidth, angle=0]{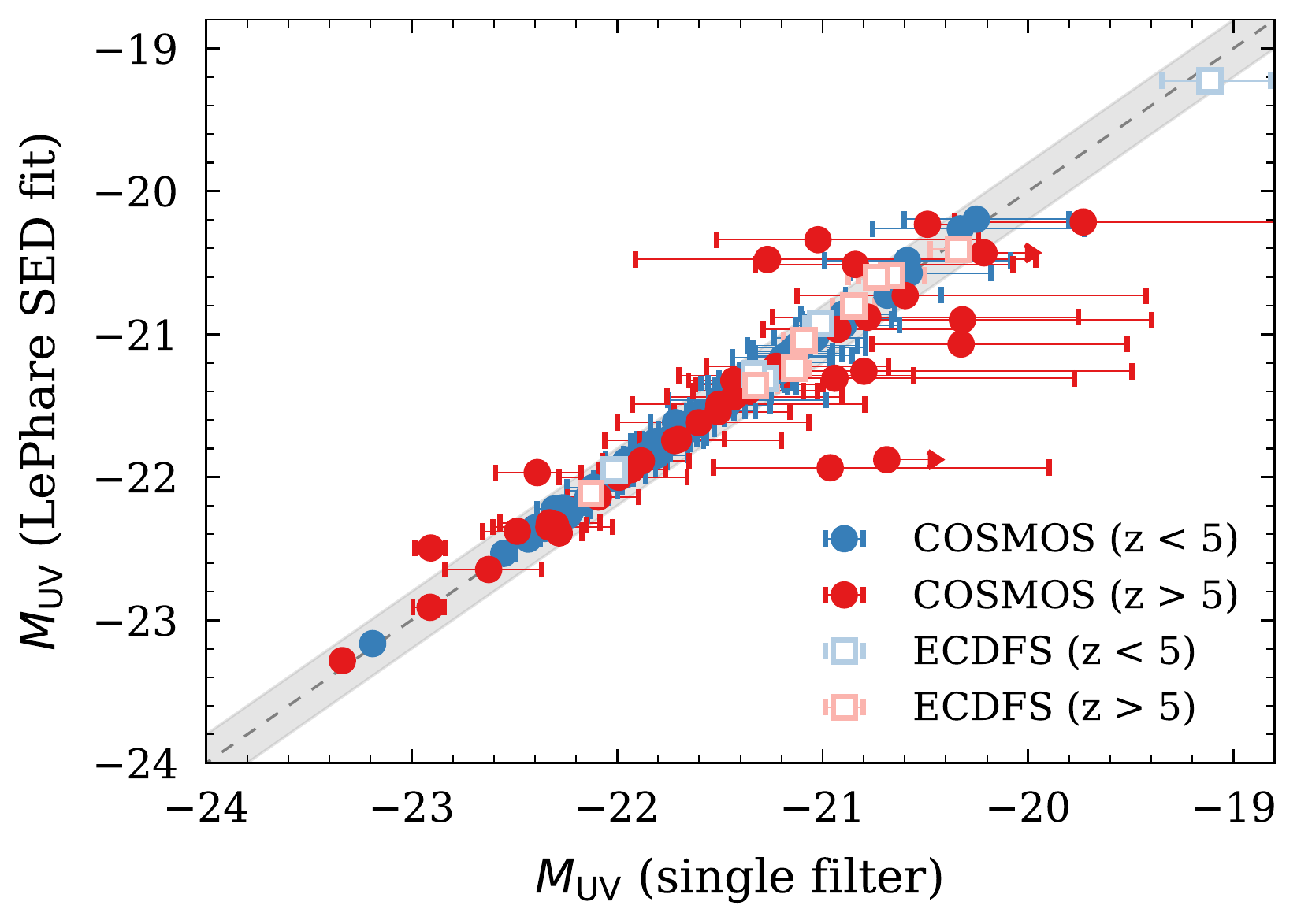}
\includegraphics[width=0.7\columnwidth, angle=0]{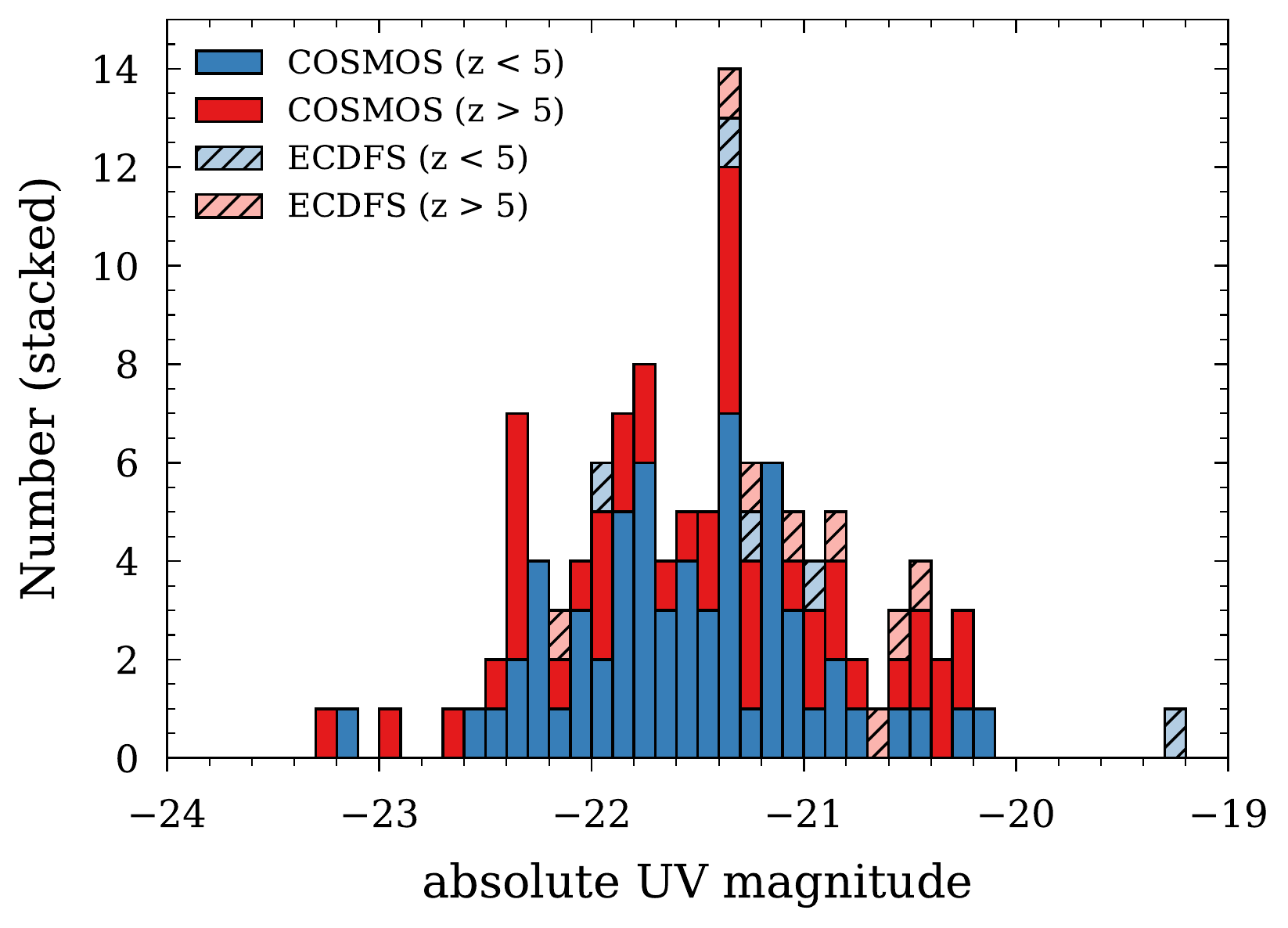}
\includegraphics[width=0.7\columnwidth, angle=0]{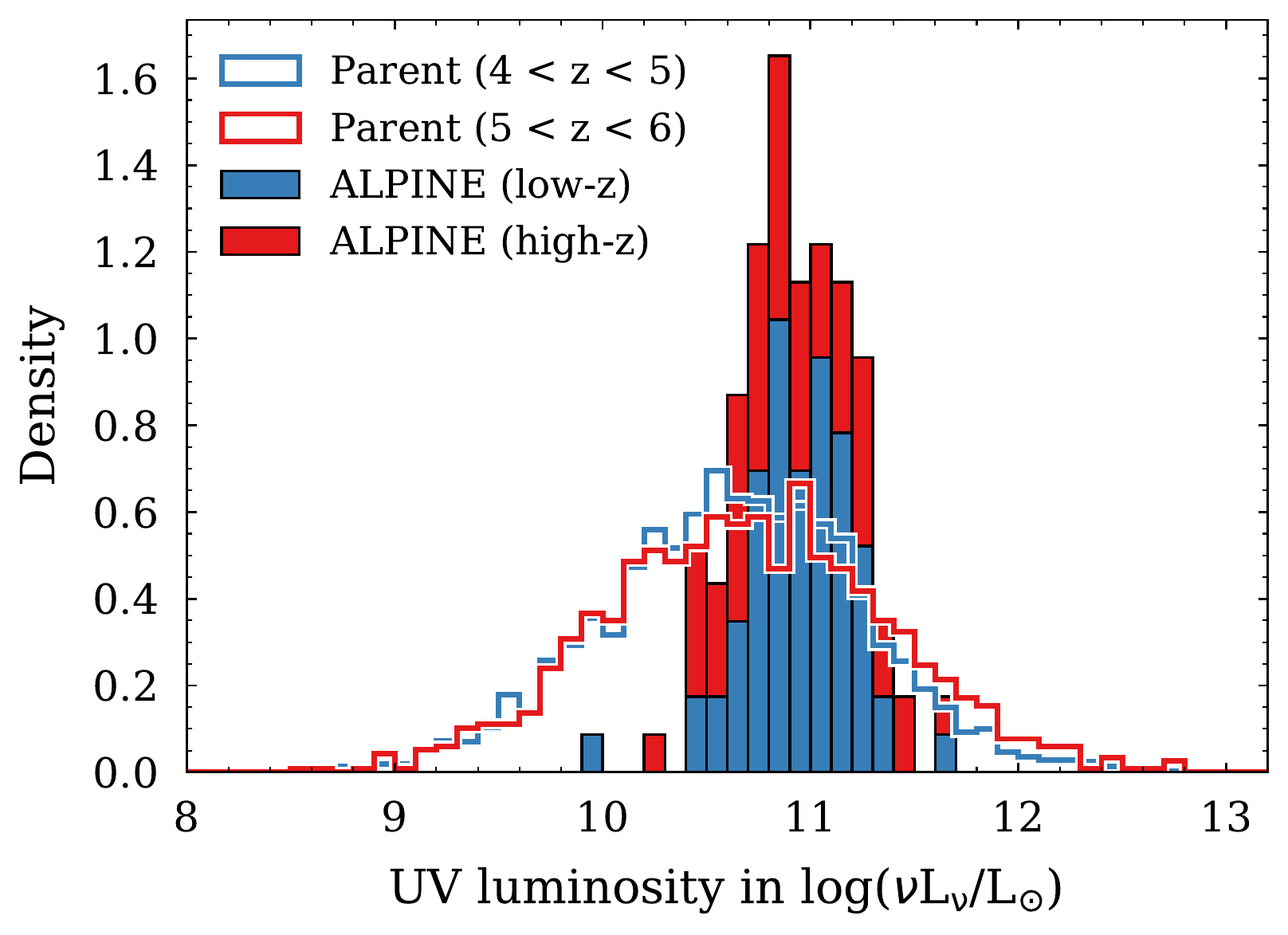}
\caption{\textit{Left:} Comparison of absolute UV magnitudes derived from \texttt{LePhare} and from one single filter close to rest-frame $1500\,{\textrm \AA}$. The scatter at fainter magnitudes is a S/N effect in the latter measurement. The gray stripe shows $\pm0.3$ magnitudes around the 1-to-1 line.
\textit{Middle:} Distribution of absolute UV magnitudes of the whole \textit{ALPINE} sample.
\textit{Right:} The UV luminosity distribution of \textit{ALPINE} galaxies in relation to a parent sample from the \textit{COSMOS2015} catalog selected in the same redshift range. \textit{ALPINE} galaxies occupy brighter luminosity, caused by the selection in $M_{\rm UV}$ (see Section~\ref{sec:spectroscopyselection}). \label{fig:muvdistributions}}
\end{figure*}

\subsubsection{The \textit{ALPINE} galaxies on the $z=5$ main-sequence and \Cii~fluxes}\label{sec:ms}

Figure~\ref{fig:sedmainsequence} summarizes the results of this section by showing the relation between stellar mass and SFR (the main-sequence) of our \textit{ALPINE} galaxies together with the \Cii~emission (in ${\rm Jy\,km\,s^{-1}}$) measured by ALMA (in color). The measurements are compared to all galaxies (with photometric redshifts) at $4 < z < 6$ in the \textit{COSMOS2015} catalog (blue points) as well as the main-sequence parameterization by \citet[][]{SPEAGLE14} at $z=5$ (gray band with $\pm0.3\,{\rm dex}$ width).

The comparison to the photometrically selected COSMOS parent sample indicates that the \textit{ALPINE} sample is a fair representation of the overall population of star-forming $z>4$ galaxies. The sample also includes at higher stellar masses galaxies that lie $2-3\sigma$ below the main-sequence. Note that two of these galaxies at $\logm\sim10.3$ and $10.7$ have contaminated Spitzer photometry, and therefore their stellar masses are upper limits. Other two galaxies at $\logm\sim10.6$ and $10.9$ do not show \Cii~emission, which is expected if they are systems of low SFR below the main-sequence. The galaxy with \Cii~detection at $\logm\sim10.1$ is classified as ``extended dispersion dominated'' by our \Cii~morpho-kinematic classification \citep[see][]{ALPINE_LEFEVRE19} and the HST/ACS imaging suggest a clumpy morphology. Although significantly below the main-sequence, \Cii~is still detected in that galaxy, perhaps indicative of dust-obscured star-formation.

The color-coding of the points suggests a \Cii~emission increase along the star-forming main-sequence. Furthermore, the fraction of \Cii~detected galaxies significantly drops below $\logm\sim9.3$ or a SFR of less than $\sim10\,{\rm M_{\odot}\,yr^{-1}}$. This could be due to the effect of metallicity on the \Cii~line strength by either indirectly a higher ionization state or directly through lower Carbon abundance \citep[e.g.,][]{NARAYANAN17}. The relationship between star formation and \Cii~emission will be studied in more detail in a forthcoming paper \citep[][]{ALPINE_SCHAERER19}.

Note that there are galaxies with a SFR of less than $\sim10\,{\rm M_{\odot}\,yr^{-1}}$ in our sample, contrary to our initial selection. We emphasize that the initial selection was based on the observed absolute UV magnitude and not on any property derived from SED fitting (such as SFR). This discrepancy is therefore expected within the uncertainty of measuring SFRs from SED fitting.

\subsection{Measurement of UV magnitudes and luminosities}\label{sec:uvmags}

The UV luminosities and absolute UV magnitudes at rest-frame $1500\,{\textrm \AA}$ (not dust corrected) are measured during the SED-fitting process with \texttt{LePhare} and are defined by the transmission curve of the \textit{GALEX} FUV filter ($\sim1500\,{\textrm \AA}$). As shown in Section~\ref{sec:sedfittingcomparison}, the absolute UV magnitudes measured by \texttt{LePhare} and \texttt{HyperZ} are in very good agreement.

We also compare these measurements to a more direct method by using the observed flux in a single filter that is closest to rest-frame $1500\,{\textrm \AA}$ (Subaru $z^{++}$ filter at $z<5$ and UltraVISTA $Y$ or $Y_{\rm HSC}$ band at $z>5$ for galaxies in COSMOS and the HST filter $F850LP$ for all the galaxies in ECDFS).
The left panel in Figure~\ref{fig:muvdistributions} shows a very good agreement between the two methods. The scatter is mainly due to the low S/N of the single-filter measurements of the second method (indicated by the large error bars). The scatter is enhanced for galaxies at $z>5$ also due to the fact that the UltraVISTA $Y$-band observations are less deep than the $z^{++}$ observations used for $z < 5$ galaxies.
In the following, we will use the more robust absolute UV magnitude from \texttt{LePhare} as they depend less on the S/N of single observations.

The middle panel in Figure~\ref{fig:muvdistributions} shows the distribution of absolute UV magnitudes for galaxies in the COSMOS and ECDFS fields in two redshift bins as in the previous figures. The bulk of galaxies are between $M_{\rm UV}$ of $-22.7$ and $-20.2$ (consistent with the faint absolute UV magnitude limit of the survey, see Section~\ref{sec:nutshell}). One of the galaxies in ECDFS (\textit{CANDELS\_GOODSS\_37}) is significantly fainter ($M_{\rm UV} = -19.2$). This galaxy has been added to the sample to fill in an empty frequency window. The measurement of the absolute UV magnitude from single band and SED fit agree and the galaxy is compact and isolated (i.e., no contamination in the photometry). The fit to its photometry with \texttt{LePhare} suggests a dust-free, low-mass ($\logm = 9.22$) galaxy that is forming stars at a rate typical for the main-sequence ($\logsfr = 0.97$).

The right panel of Figure~\ref{fig:muvdistributions} compares the distribution of the UV luminosity of the \textit{ALPINE} galaxies (stacked filled histogram for low and high redshift) to the same parent sample selected from the \textit{COSMOS2015} catalog and used in Figure~\ref{fig:sedmainsequence} split in two redshift bins (empty histogram). As expected, the absolute UV magnitude cut applied for the selection of the \textit{ALPINE} sample (see Section~\ref{sec:spectroscopyselection}) causes a bias towards the brighter end of the parent distribution.

\subsection{UV continuum slopes}\label{sec:uvslopes}

\subsubsection{Method}\label{sec:uvslopesmethod}

The UV continuum slope ($\beta$, defined as $f_{\rm \lambda} \propto \lambda^{\beta}$) generally correlates with the attenuation of stellar light by dust and is therefore an important tool to study the dust properties of galaxies especially at high redshfits \citep[e.g.,][]{MEURER99,BOUWENS09,FINKELSTEIN12}.
The UV continuum slope of a galaxy can be derived by various methods. Here, we compute $\beta$ from the best-fit SEDs derived by \texttt{LePhare}.
Compared to deriving the slopes directly from the observed photometry by a linear fit, this approach results in less biased $\beta$ measurements in the case of low S/N observations \citep[e.g.,][]{FINKELSTEIN12,BARISIC17}. This is particularly true for the \textit{ALPINE} galaxies, whose rest-frame UV continuum is predominantly covered by relatively shallow ground-based imaging. Deep HST coverage by a sufficient number of bands of this wavelength range is only available for a small fraction of the galaxies.

The $\beta$ slopes are derived by a robust linear fit  (to avoid the fit being affected by any absorption or emission lines) to the logarithmic slope of the best-fit \texttt{LePhare} SED in the wavelength range between $1300\,{\textrm \AA}$ and $2300\,{\textrm \AA}$. To quantify uncertainties, we perturb the fluxes of each filter according to their individual errors assuming a Gaussian error distribution, refit the galaxies, and re-measure $\beta$ from the resultant best-fit SED. For each galaxy we repeat this procedure $1000$ times to produce a probability density function. The uncertainties for $\beta$ quoted here are the $1\sigma$ percentiles of this distribution and are on average on the order of $\Delta\beta = 0.2-0.3$.

\begin{figure}
\includegraphics[width=1\columnwidth, angle=0]{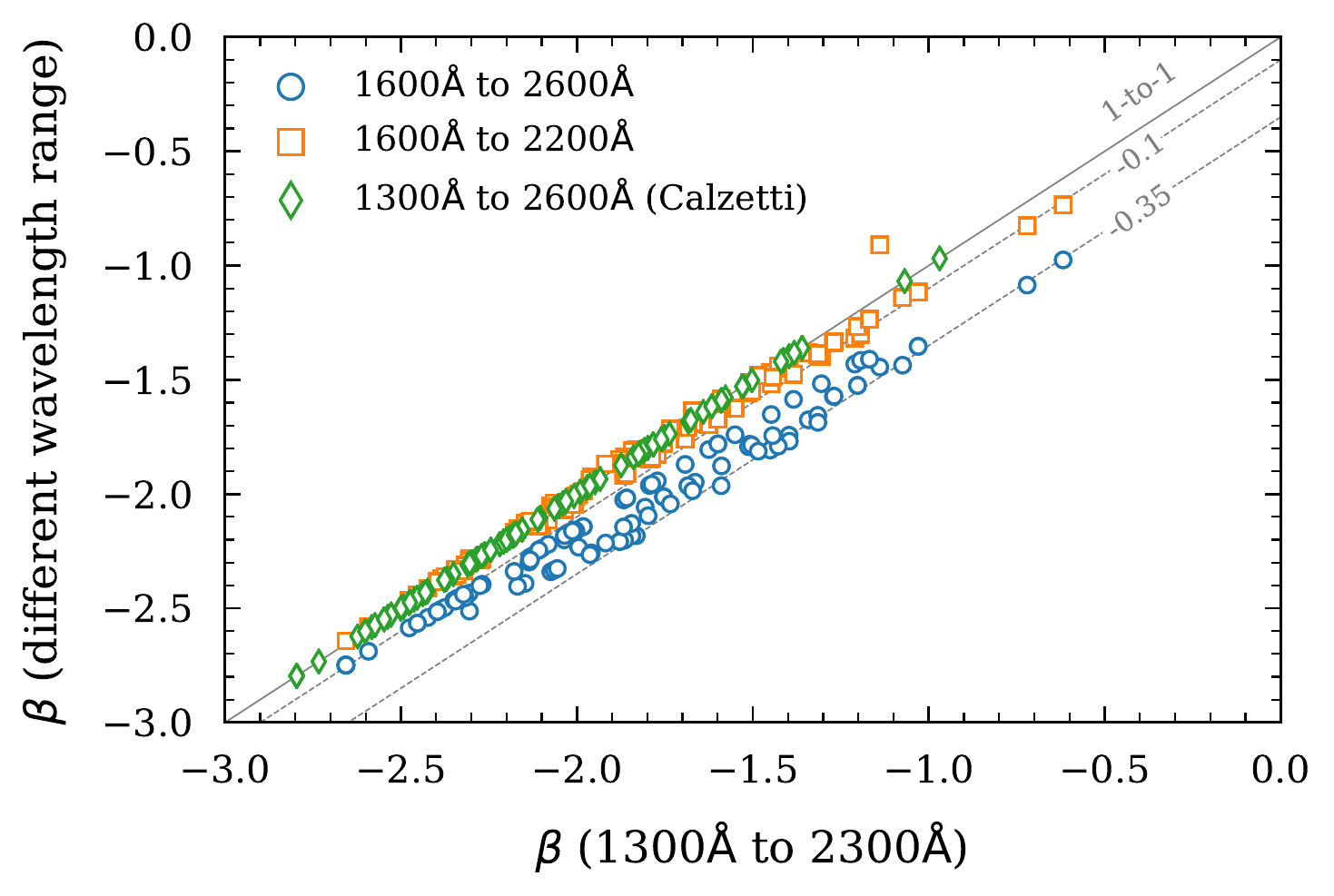}\\
\includegraphics[width=1\columnwidth, angle=0]{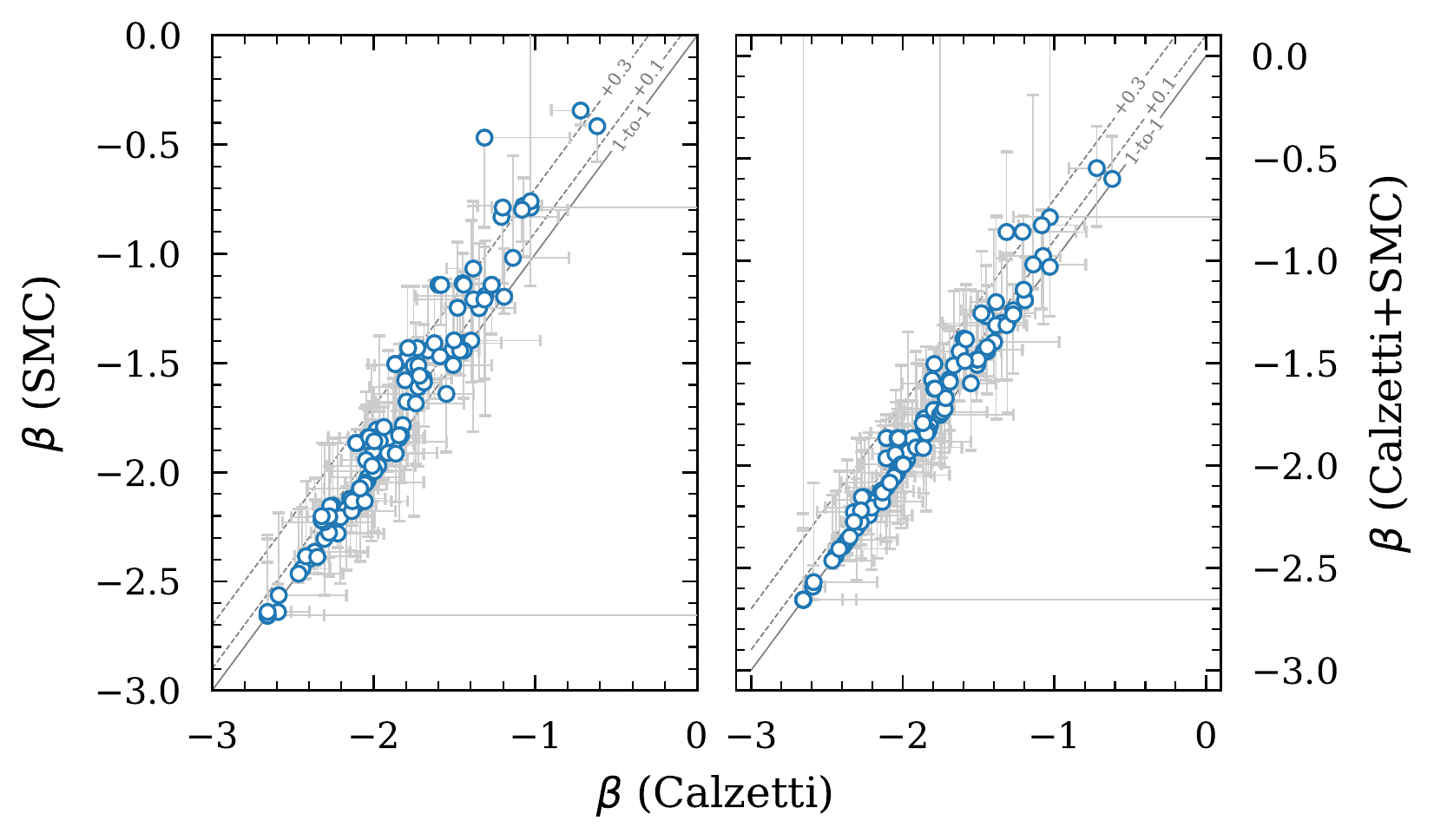}
\caption{Dependence of UV continuum slope $\beta$ on the definition of the wavelength regions and assumption of dust attenuation law. The solid line shows the 1-to-1 relation and the dashed lines show different offsets.
\textit{Top:} Dependence of $\beta$ on the adopted wavelength window (with respect to our choice, $1300\,{\textrm \AA}$ to $2300\,{\textrm \AA}$).
\textit{Bottom:} Dependence of $\beta$ on the assumed dust attenuation law. Using an SMC dust attenuation results in redder slopes compared to a Calzetti reddening law.
\label{fig:uvslopetests}\vspace{-0.5cm}}
\end{figure}

\begin{figure}
\includegraphics[width=1.0\columnwidth, angle=0]{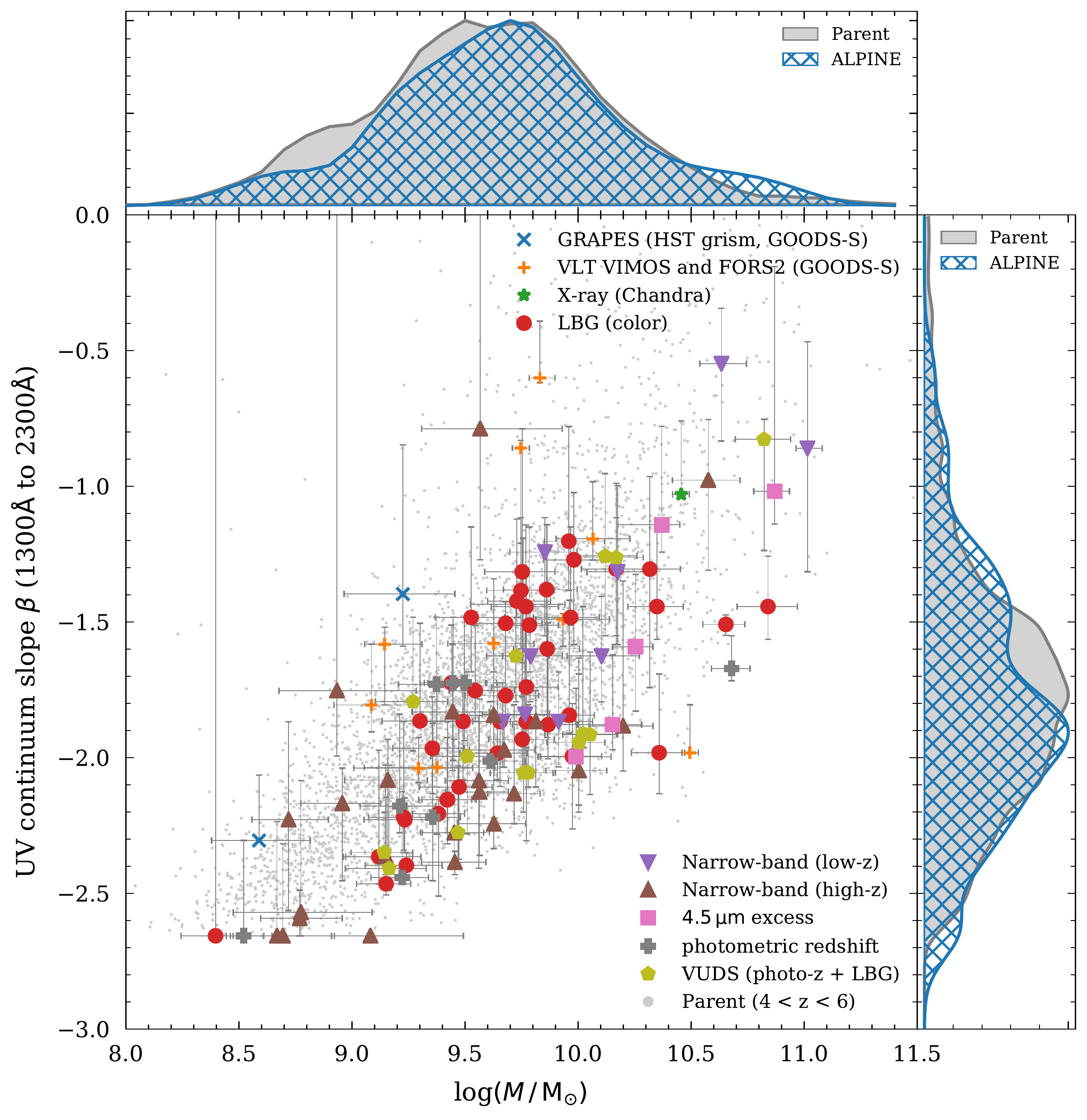}
\caption{Comparison of UV slopes and stellar mass. The offset panels show kernel density estimates of the $\beta$ and stellar mass distribution. The \textit{ALPINE} galaxies (large symbols) are split into their method of selection (see Section~\ref{sec:spectroscopyselection}). We also show the data from our parent sample at $4 < z < 5$ (gray dots). Statistically, our \textit{ALPINE} peaks at $\sim0.2\,{\rm dex}$ higher stellar masses and $\sim 0.3$ bluer $\beta$.  \label{fig:uvslopesmass}}
\end{figure}

\subsubsection{Systematic Uncertainties and Dependencies on Dust Attenuation}\label{sec:uvslopesuncertainties}

In addition to photometric uncertainties, several model assumptions affect the measurement of $\beta$. We found that the two most important ones are the wavelength range over which $\beta$ is fit and the assumed dust attenuation law. 

We choose the wavelength range over which $\beta$ is measured to be consistent with the definition of several other studies \citep{CALZETTI94,MEURER99,FINKELSTEIN12,BOUWENS14,BARISIC17,FUDAMOTO17}.
In the upper panel of Figure~\ref{fig:uvslopetests}, we demonstrate how the $\beta$ measured for our \textit{ALPINE} galaxies would change if different wavelength ranges are used.
First, we do not find any differences in our measurements compared to the definition by \citet[][]{CALZETTI94}, who use $10$ discrete fitting windows between $1300\,{\textrm \AA}$ and $2600\,{\textrm \AA}$ to avoid strong absorption and emission lines (green diamonds).
The other symbols show the comparison to different wavelength ranges and we notice significant offsets from our measurements. For example, defining $\beta$ between $1600\,{\textrm \AA}$ and $2200\,{\textrm \AA}$ results in up to $\Delta\beta = 0.1$ bluer slopes (orange squares). Using a significantly redder wavelength range, $1600\,{\textrm \AA}$ to $2600\,{\textrm \AA}$, leads to $0.1-0.35$ bluer slopes (blue circles) compared to our definition. Note that the offset varies as a function of $\beta$ itself $-$ specifically, differences are enhanced towards redder slopes.

The second, more physically driven, quantity that affects the measurement of $\beta$ is the assumed dust attenuation law. As described in Section~\ref{sec:sedfittingcomparison}, the choice of the dust attenuation law has a negligible affect on the stellar masses and SFRs. This is not the case for the $\beta$ slopes as shown in the lower panels of Figure~\ref{fig:uvslopetests}. The left panel compares $\beta$ derived using Calzetti and SMC dust attenuation. We notice a consistent positive offset of up to $\Delta\beta = 0.3$ for the reddest slopes.
We compared the reduced $\chi^2$ values output by \texttt{LePhare} for fits using a Calzetti and SMC dust attenuation in order to derive a preference for either of the dust attenuations. We find that the $\chi^2$ values show insignificant differences, which lets us conclude that we are not able to distinguish between the different dust attenuations based on our SED fitting. Hence, we decided that the best way is to be agnostic about the dust attenuation and combine for each galaxy the two probability density functions $P_{\rm Calzetti}(\beta)$ and $P_{\rm SMC}(\beta)$ derived from our Monte Carlo approach (Section~\ref{sec:uvslopesmethod}) assuming equal weighting to derive the median $\beta$ and its $1\sigma$ uncertainties. In the lower left panel of Figure~\ref{fig:uvslopetests}, the final combined $\beta$ slopes are compared to the $\beta$ derived assuming a Calzetti attenuation. The offset towards redder $\beta$ is significantly reduced due to a narrower probability density functions assuming Calzetti dust (hence the average $\beta$ are drawn to the Calzetti solution in most cases).

\subsubsection{The $\beta$ Slopes of the \textit{ALPINE} Galaxies in Context}\label{sec:uvslopesmass}

Figure~\ref{fig:uvslopesmass} shows our $\beta$ measurement (marginalized over both Calzetti and SMC dust) as a function of stellar mass split in the different methods of selection (Section~\ref{sec:spectroscopyselection}). As the UV slope is mostly affected by the dust attenuation, the strong correlation between $\beta$ and stellar mass is not surprising as more massive galaxies are expected to be more dusty. The $z\sim5.5$ narrow-band selected galaxies have statistically the bluest slopes, indicating their dust-poor nature. The other galaxies are spread out over the whole parameter space.
We also show the data from our parent sample at $4 < z < 6$ in gray and compare their  $\beta$ slope and stellar mass distribution to the \textit{ALPINE} sample in the kernel density estimate plots.
Note that the $\beta$ slope distribution of \textit{ALPINE} galaxies peaks at $\Delta\beta \sim 0.2$ bluer values than the parent sample at the same redshift. This is a minor bias (likely caused by our spectroscopic selection) that has to be kept in mind for future analyses.

\begin{deluxetable}{l c  c c}
\tabletypesize{\scriptsize}
\tablecaption{List of basis stellar population models for the parameterization of the rest-frame optical continuum at $4 < z < 5$ to derive \halpha~emission from Spitzer colors.\label{tab:fitinput}}
\tablewidth{0pt}
\tablehead{
\colhead{Model} & \colhead{SFH} & \colhead{Metallicity} & \colhead{Dust attenuation}\\[-0.2cm]
\colhead{} & \colhead{} & \colhead{($Z_{\odot} = 0.02$)} & \colhead{}
}
\startdata
A & constant & 0.02 & Calzetti\\
B & constant & 0.004 & Calzetti\\
C & exp. declining$^{a}$ & 0.01 & Calzetti\\
D & constant & 0.02 & SMC\\
E & constant & 0.004 & SMC\\
F & exp. declining$^{a}$ & 0.01 & SMC\\
\enddata
\tablenotetext{a}{Assuming $\tau=3\times 10^8\,{\rm yrs}$.}
\end{deluxetable}

\subsection{Measurement of \halpha~emission}\label{sec:halphaemission}

Rest-frame optical emission lines in $z>4$ galaxies are out of reach of current spectrographs. Specifically, the \halpha~emission provides a good tool to study the star-formation properties of galaxies in more detail. Fortunately, in the redshift range $4 < z < 5$, the \halpha~line falls in the Spitzer $3.6\,{\rm \mu m}$ filter, while the $4.5\,{\rm \mu m}$ filter lacks any strong emission lines. Therefore, the \iracA~color can be used to constrain the \halpha~line flux and its equivalent width (a proxy of recent stellar mass build up). This method leads to \halpha~emission properties that are statistically as accurate as derived from spectroscopic data \citep{FAISST16a}. Several such measurements have been carried out in the past with success \citep{SHIM11,STARK13,DEBARROS14,SMIT14,MARMOLQUERALTO16,FAISST16a,RASAPPU16,SMIT16,CAPUTI17,FAISST19b}. 

\begin{figure}
\includegraphics[width=1.00\columnwidth, angle=0]{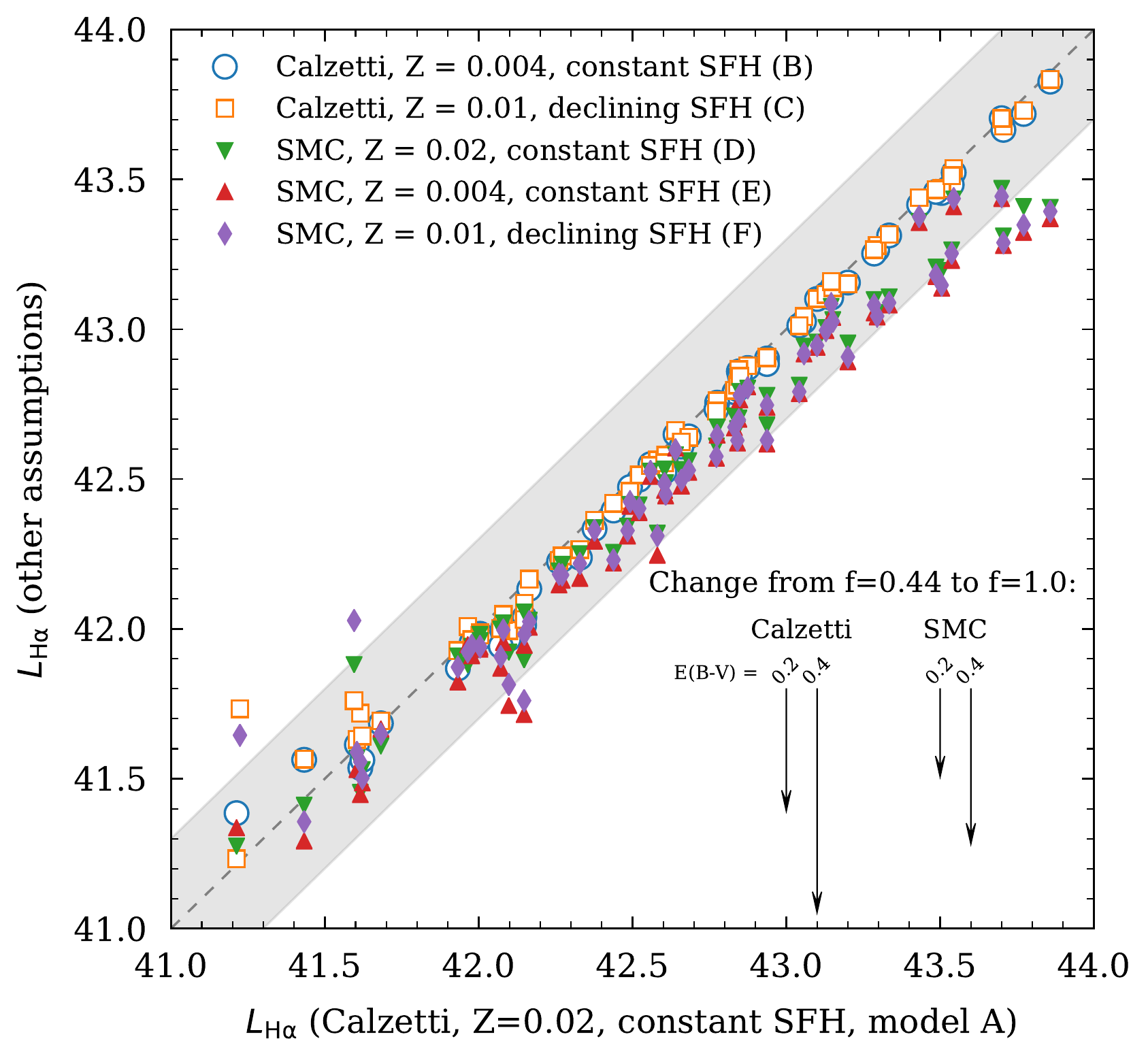}
\caption{Effect of different assumptions of the rest-frame optical continuum, reddening law, and $f$-factor (arrows show absolute decrease in luminosity from $f=0.44$ to $1.0$ for different stellar dust attenuations and reddening laws) on the measurement of the \halpha~luminosity. The gray band shows the 1-to-1 relation with $\pm0.3\,{\rm dex}$ margin. The different models for the continuum are labeled in the same way as in Table~\ref{tab:fitinput}). The $f$-factor (differential reddening between stellar continuum and nebular regions) has the strongest effect on the \halpha~luminosity measurements. \label{fig:halphasystematics}}
\end{figure}

About $55\%$ of the \textit{ALPINE} sample ($66$ galaxies) lie in this redshift range.
To measure the \halpha~luminosity and equivalent widths, we follow the same technique as outlined in \citet[][]{FAISST19b} (we refer to this paper for more technical details). In brief, this method makes an assumption on the rest-frame optical continuum to which emission lines are added in a consistent manner to reproduce the observed \iracA~colors of the galaxies. This approach is robust as it only depends on the slope of the rest-frame optical continuum, which is well defined and nearly independent of assumptions on age, metallicity, and star-formation history for galaxies younger than $\sim1\,{\rm Gyrs}$ (mostly the case at $z>4$).
To describe the rest-frame optical continuum, we use several basis stellar population models based on the \citet[][]{BRUZUALCHARLOT03} template library (see Table~\ref{tab:fitinput}). For the dust correction of the \halpha~emission, we assume the stellar $\ebmvs$~values derived by \texttt{LePhare}, which we convert to nebular extinction factors by assuming an $f$-factor\footnote{The $f$-factor, $f = \ebmvs/\ebmvn$, describes the differential dust reddening between the stellar continuum and nebular regions. Its value is largely unknown at $z>2$, but it is expected that $f$ approaches a value closer to unity at higher redshifts \citep{ERB06,REDDY10,KASHINO13,KOYAMA15,VALENTINO15,PUGLISI16,KASHINO17,FAISST19b}.} of $0.44$ as measured in local starburst galaxies \citep{CALZETTI00}. We also assume a Calzetti and SMC reddening law.
Furthermore, we assume an \nii~to \halpha~ratio of $0.15$, as expected for galaxies at $\logm = 10$ \citep{FAISST18}, to correct the blending of the \nii~and \halpha~lines.

\begin{figure}
\includegraphics[width=1\columnwidth, angle=0]{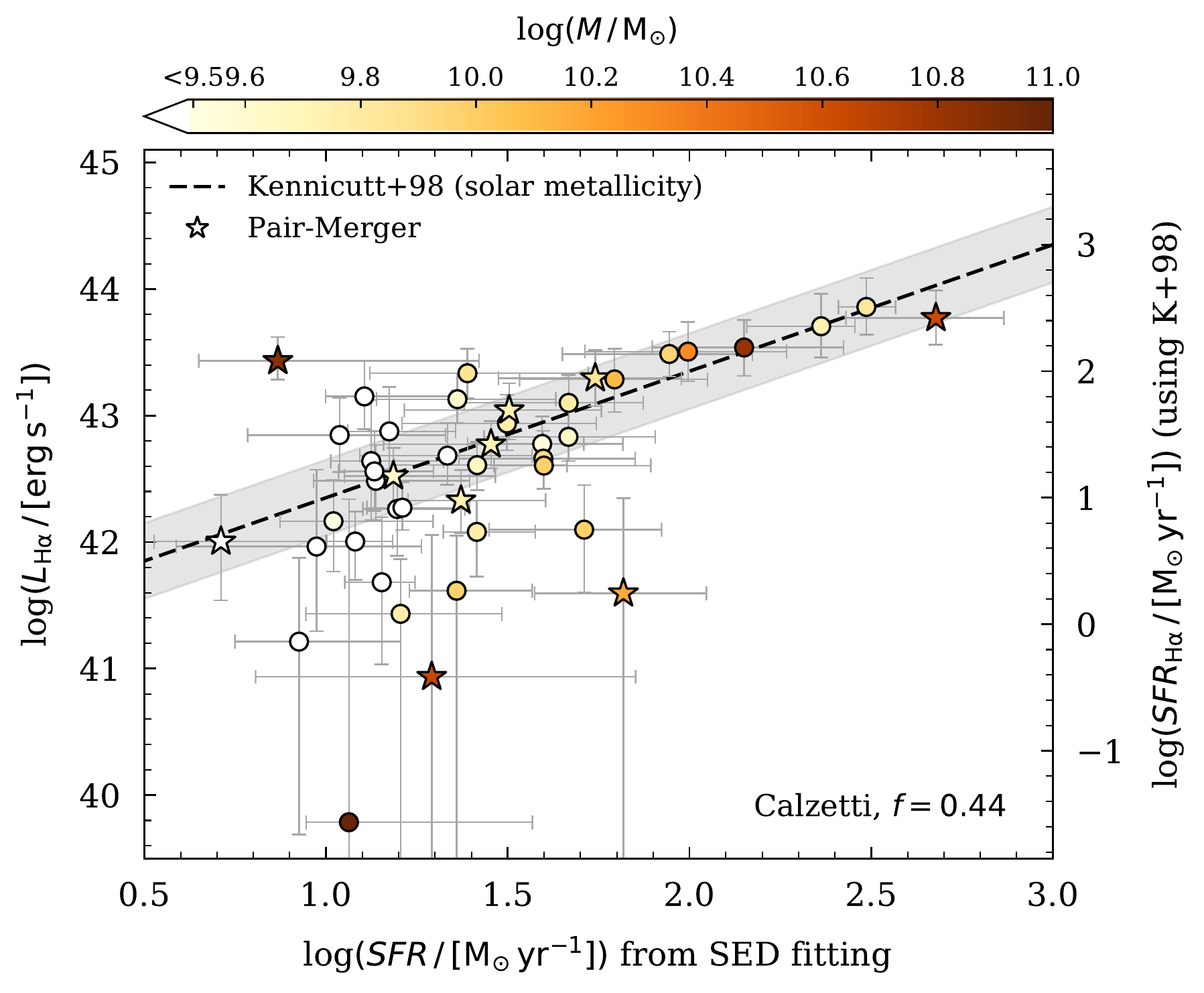}\\
\includegraphics[width=1\columnwidth, angle=0]{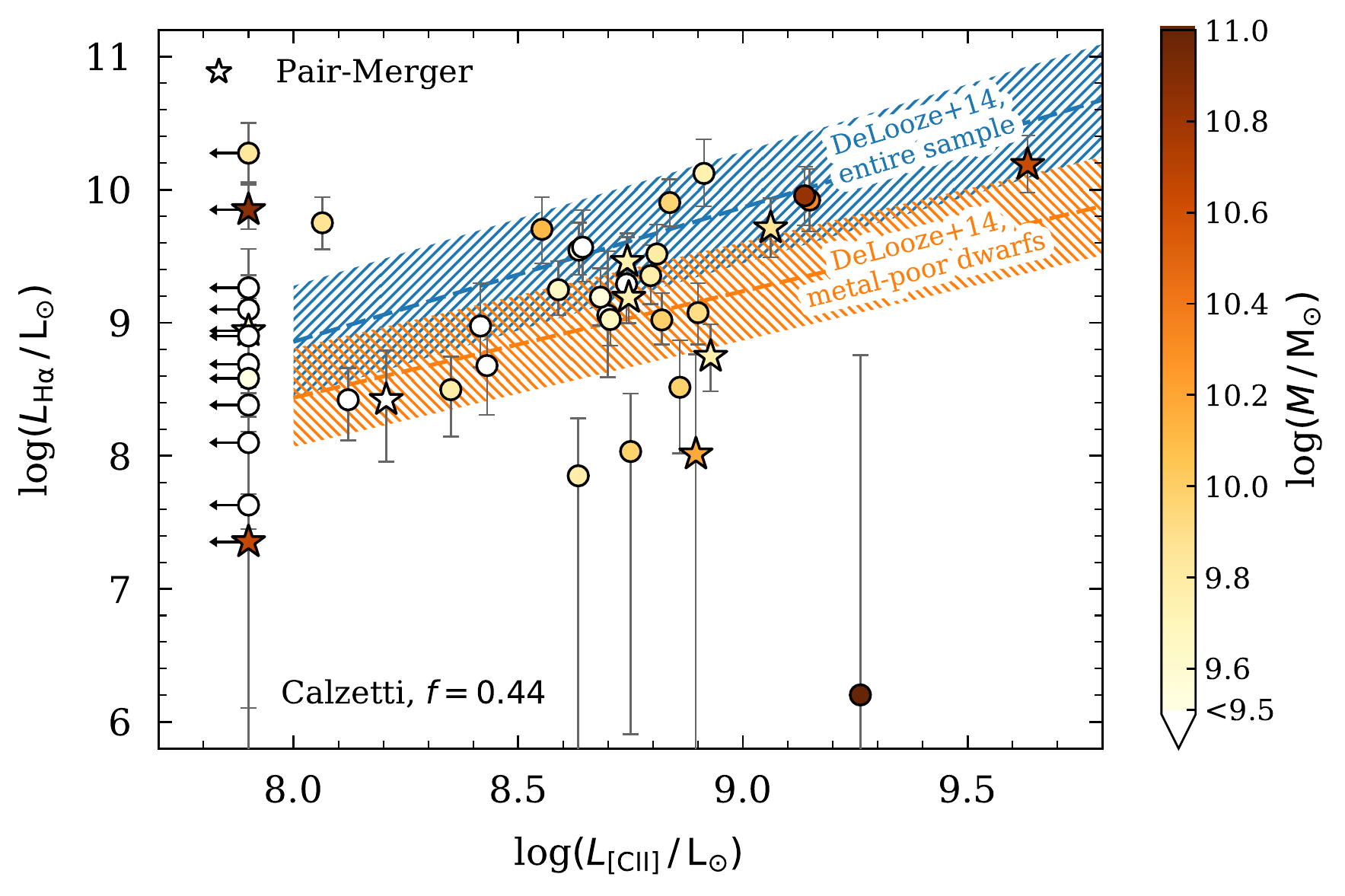}
\caption{\textit{Top:} Comparison of SED-derived SFRs to \halpha~luminosity (left $y$-axis) and \halpha-derived SFR (right $y$-axis). The latter are derived from \halpha~assuming the conversion factor by \citet[][]{KENNICUTT98} for solar metallicity (dashed line and $\pm0.3\,{\rm dex}$ margin, gray). Only galaxies with non-contaminated Spitzer photometry are shown. The symbols are color-coded by stellar mass. The star-symbols denote mergers based on the classification in \citet{ALPINE_LEFEVRE19}. The \halpha-dependent quantities are derived assuming a Calzetti reddening law, constant SFH, solar metallicity, and $f=0.44$ (see Figure~\ref{fig:halphasystematics} for effect of different assumptions). Above a SFR of $\sim13\, \Msol\,{\rm yr^{-1}}$, the two SFRs are comparable. Below that threshold, the scatter in the \halpha-derived SFRs increases due to low S/N of the $4.5\,{\rm \mu m}$ observations \citep[see][]{FAISST19b}.
\textit{Bottom:} Comparison of \halpha~and \Cii~luminosity. Shown are only galaxies with \halpha~measurements and non-contaminated Spitzer photometry. \Cii~undetected galaxies (at $<3.5\sigma$) are indicated with upper limits. The lines and dashed margins show the expected relation between \halpha~and \Cii~derived combining the relations from \citet[][]{KENNICUTT98} and \citet[][]{DELOOZE14} (for their entire sample and metal-poor dwarf galaxies, see text). Note that at lower \Cii~luminosities ($< 5\times 10^{8}\,{\rm L_{\odot}}$), the galaxies seem to be more consistent with the relation of local metal-poor dwarf galaxies.
\label{fig:halphaluminosity}}
\end{figure}

In Figure~\ref{fig:halphasystematics}, we show systematic uncertainties in the measurement of the \halpha~luminosity due to the assumptions in our model for the rest-frame optical continuum and the reddening law (models $A$ through $F$, see Table~\ref{tab:fitinput}), as well as the $f$-factor for $\ebmvs=0.2$ and $0.4$. It is evident that different assumptions in metallicity and SFH have a negligible impact on the measured \halpha~luminosity.
The choice of the reddening law matters as the \halpha~luminosity decreases by $\sim0.3\,{\rm dex}$ for galaxies at high \halpha~luminosities ($\logLha > 43.5$) assuming an SMC reddening law.
The $f$-factor is the largest uncertainty in this measurement method and will have to be pinned down by future observations with the JWST. For now, the assumed $f=0.44$ provides likely an \textit{upper} limit on the \halpha~luminosities. As shown by the arrows in Figure~\ref{fig:halphasystematics}, assuming an $f$-factor equal to unity (which is thought to be more likely based on observations at $z\sim2$) would decrease the \halpha~luminosities by up to $0.4\,{\rm dex}$ ($0.8\,{\rm dex}$) for a stellar dust reddening of $0.2$ ($0.4$) magnitudes. The correction in case of an SMC reddening law are $0.1-0.2\,{\rm dex}$ less.
We note that these factors also apply to \halpha-derived SFRs and any other quantity that depends linearly on the \halpha~luminosity.

The top panel of Figure~\ref{fig:halphaluminosity} compares the SFRs derived from SED fitting (Section~\ref{sec:sedfitting}) to the \halpha~luminosity and \halpha-derived SFRs for galaxies without contaminated Spitzer photometry. The latter is derived using the standard conversion factor given in \citet[][]{KENNICUTT98},
\begin{equation}\label{eq:kennicutt98}
    {\rm SFR\,(M_{\odot}\,yr^{-1})} = 4.5 \times 10^{-42}\, L_{\rm{H\alpha}}\,{\rm (erg\,s^{-1})},
\end{equation}
assuming solar metallicity and a Chabrier IMF. Assuming one-fifth solar metallicity, the inferred SFR is expected to be $\sim0.2\,{\rm dex}$ lower \citep{LY16}. As shown in \citet[][]{FAISST19b}, the uncertainty of this conversion factor is negligible compared to the impact of the uncertain $f$-factor.
The \halpha~luminosities in the \textit{ALPINE} sample range from $\sim 10^{41}\,\Lsol$ to $\sim 10^{44}\,\Lsol$ assuming $f=0.44$ and \citet{CALZETTI00} dust attenuation.
The \halpha-derived SFRs trace well the SED-derived SFRs above $\sim 13\,\Msol\,{\rm yr^{-1}}$. Below that value, we see a large scatter in \halpha~derived SFRs. which happens when the \halpha~emission becomes too faint to be measured reliably using the Spitzer broad bands. Specifically, this is the case roughly at $\logm = 9.5$, which corresponds to a $4.5\,{\rm \mu m}$ detection of less than $5\sigma$ \citep[see figure 3 and appendix in][]{FAISST19b}.

 The lower panel of Figure~\ref{fig:halphaluminosity} relates the \halpha~luminosity (here in units of solar luminosity) to the \Cii~luminosity measured by ALMA (also solar luminosity). In addition, we show the expected relation between \halpha~and \Cii~luminosity by combining Equation~\ref{eq:kennicutt98} with the linear relation between \Cii~and SFR derived by \citet[][]{DELOOZE14} from local and low-redshift galaxy samples,
\begin{equation}
    \log({\rm SFR}\,/\,[{\rm M_{\odot}\,yr^{-1}}]) = \alpha \times \log(L_{\rm [CII]}\,/\,{\rm L_{\odot}}) + \gamma,
\end{equation}{}
for different values of intercepts ($\gamma$) and slopes ($\alpha$). Specifically, we are showing the relation for their entire sample ($\alpha=1.01$, $\gamma=-6.99$, blue hatched) and the metal-poor dwarf galaxies ($\alpha=0.80$, $\gamma=-5.73$, red hatched).
For bright \Cii~galaxies ($L_{\rm [CII]}\gtrsim 5\times10^{8}\,{\rm L_{\odot}}$) we find a good agreement with the entire local sample. For lower \Cii~luminosities, the \halpha~measurements fall below this relation and are instead more consistent, although with a large scatter, with the relation of metal-poor local dwarf galaxies. The majority of \Cii~undetected galaxies align well with either relation, however, the uncertainty in the \halpha~measurements becomes substantial as galaxies fall below $\logm = 9.5$.

\begin{figure}[t!]
\includegraphics[width=1.0\columnwidth, angle=0]{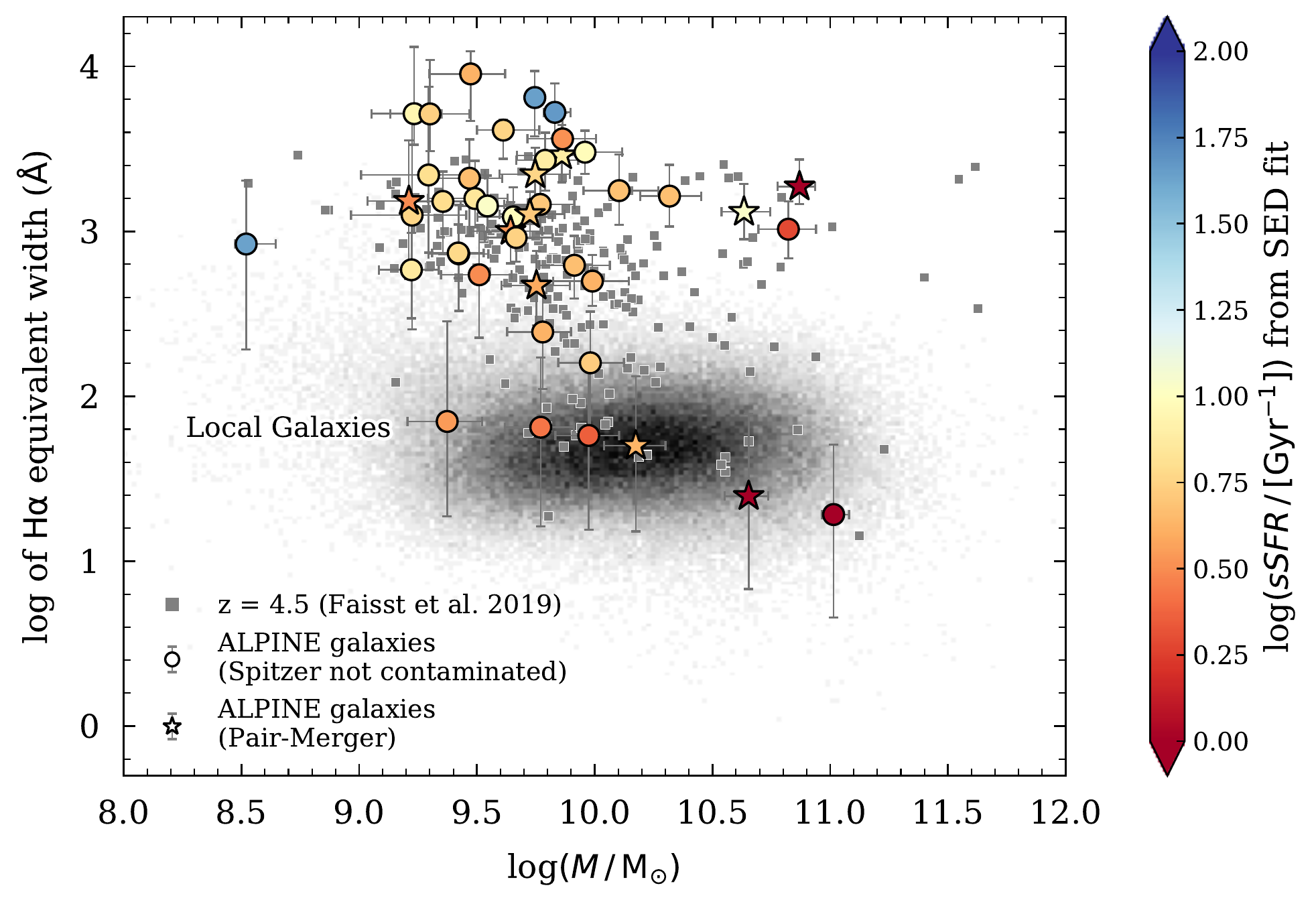}
\caption{The rest-frame \halpha~EW of our \textit{ALPINE} galaxies in context of local galaxies (gray cloud) and $z=4.5$ galaxies from \citet[][]{FAISST19b} (gray squares). The \halpha~EW is related to stellar mass and sSFR derived from SED fitting (color-coded).
The \textit{ALPINE} sample at $4 < z < 5$ builds a representative subsample of the general galaxy population at $z>4$, also in terms of \halpha~properties. Although with similar stellar masses as local galaxies, the high-redshift galaxies reside at significantly higher \halpha~EWs, which is naturally explained by their higher star formation.
Note the two galaxies with particularly low sSFR ($<1\,{\rm Gyr^{-1}}$, and consistently low \halpha~EWs of less than $30\,{\textrm \AA}$) that fall onto the massive end of the distribution of local galaxies. These galaxies are indicative for systems with evolved stellar populations at high redshifts with currently reduced star formation activity.  \label{fig:halphaew}}
\end{figure}

Figure~\ref{fig:halphaew} shows the rest-frame \halpha~EW distribution of our \textit{ALPINE} galaxies in the context of local galaxies (gray cloud) and other $z\sim4.5$ galaxies on the COSMOS field \citep[gray,][]{FAISST19b}. The \halpha~EW for the $z>4$ galaxies is derived consistently assuming a constant SFH, Calzetti reddening law, and $f=0.44$. The \textit{ALPINE} galaxies cover well the parameter space of the other $z\sim4.5$ galaxies, hence build a representative sample also in terms of \halpha~properties. For a fixed stellar mass, the high-redshift galaxies have higher \halpha~EWs compared to local galaxies, which is expected from a galaxy evolution point of view as galaxies at higher redshift are highly star forming.
Note that two galaxies in the \textit{ALPINE} sample have similar \halpha~EW values as massive ($\logm>10.5$) local galaxies. Consistently, also their sSFR are low ($<1\,{\rm Gyr^{-1}}$), which is indicative of them being systems with evolved stellar populations at high redshifts.

\section{Summary and Conclusions} \label{sec:end}

The early growth phase at redshifts $z=4-6$ marks an important time in which galaxies build up their stellar mass, enrich in metals and dust, and change their structure to transform into galaxies at the peak of SFR density or thereafter. For a better understanding of this interesting galaxy population, a multi-wavelength survey is crucial.
\textit{ALPINE} comprises a valuable set of $118$ galaxies at $4.4 < z < 5.9$ with unprecedented ALMA data at $\sim150\,{\rm \mu m}$ FIR wavelengths. Together with the ancillary data presented in this paper, it makes it the first large panchromatic survey to discover the formation and study the evolution of galaxies during the early growth phase.

Summarizing, the science enabling corner stone datasets of \textit{ALPINE} are:

\begin{itemize}
    \item[$-$] unprecedented ALMA observations to study the dust, gas, and outflow properties of the largest sample of galaxies to-date at $z>4$ \citep{ALPINE_BETHERMIN19},  \vspace{-0.3cm}
    
    \item[$-$] consistently calibrated deep spectroscopic observations at rest-frame UV wavelengths ($\S$\ref{sec:spectroscopy}) to study \lya~emission and absorption lines ($\S$\ref{sec:uvspectrameasurements}, \ref{sec:velocity}),\vspace{-0.3cm}
    
    \item[$-$] coherent ground-based (and space-based in ECDFS) imaging data from the optical to near-IR ($\S$\ref{sec:photometry}) for the measurement of various properties from SED fitting methods ($\S$\ref{sec:sedfitting}), including stellar masses and SFRs ($\S$\ref{sec:sedfittingmethod}), UV luminosities ($\S$\ref{sec:uvmags}), and UV continuum slopes to study stellar dust attenuation ($\S$\ref{sec:uvslopes}) ,\vspace{-0.3cm}
    
    \item[$-$] deep Spitzer imaging at $3.6\,{\rm \mu m}$ and $4.5\,{\rm \mu m}$ to measure \halpha~emission for $66$ galaxies between $4~<~z~<~5$ ($\S$\ref{sec:halphaemission}),
    \vspace{-0.3cm}
    
    \item[$-$] high-resolution HST/ACS imaging in $F814W$ for all galaxies and WFC3/IR imaging for a smaller fraction (less than $30\%$ with deep $F160W$ data) to study their resolved structure in connection with FIR \Cii~emission ($\S$\ref{sec:photometry}).

\end{itemize}{}

The \textit{ALPINE} sample is built upon several different selection methods (Section~\ref{sec:spectroscopyselection}, Figure~\ref{fig:redshiftdist}), hence contains a multitude of different spectroscopic properties.
Because of the requirement for spectroscopic confirmation, the sample is slightly biased towards brighter UV magnitudes (Section~\ref{sec:uvmags}, Figure~\ref{fig:muvdistributions}) and blue UV continuum slopes ($\Delta\beta\sim0.2$) compared to the average $4~<~z~<~6$ galaxy population (Section~\ref{sec:uvslopes}, Figure~\ref{fig:uvslopesmass}).
Nonetheless, stellar masses and SFRs, derived from the wealth of ancillary data, show that the \textit{ALPINE} sample is broadly representative of the $4 < z < 6$ galaxy population. 

The FIR \Cii~redshifts observed by ALMA allow us to set the systemic redshift of the galaxies in order to study velocity offsets of \lya~emission and several rest-frame UV absorption lines (Section~\ref{sec:uvspectrameasurements}).
From one galaxy at $z=4.57$ with optical \oii~measurements acquired from Keck/MOSFIRE, we show that the \oii~and FIR \Cii~redshifts are in excellent agreement, hence the latter likely is a good tracer of the systemic redshift derived by optical emission lines at lower redshifts (Section~\ref{sec:mosfireoii}, Figure~\ref{fig:mosfireoii}). In general concordance with studies at $z=2-3$ (using \halpha~to define the systemic redshift), we find that on average \lya~is red shifted ($\sim180\,{\rm km\,s^{-1}}$) with respect to the \Cii~line, while and the absorption lines are blue shifted ($\sim -230\,{\rm km\,s^{-1}}$). In \citet[][]{ALPINE_CASSATA19}, we perform a more detailed comparison to samples at lower redshifts and study the implication on the \lya~escape fraction in correlation with \lya~equivalent widths.
Stacking the spectra in bins of sSFR, we find larger velocity offsets of absorption lines with respect to systemic for galaxies with high sSFRs, which is indicative of stronger winds and outflows in these galaxies (Section~\ref{sec:velocity}, Figure~\ref{fig:spectra_ssfr_stack}). This finding is in agreement with the recent work by \citet[][]{ALPINE_GINOLFI19}, who show a broadening in the FIR \Cii~profiles in \textit{ALPINE} galaxies with high star formation.

Statistically, the SFRs derived from \halpha~emission via the \citet[][]{KENNICUTT98} relation for galaxies between $4 < z < 5$ agree well with the values derived from SED fitting, assuming a differential dust reddening factor of $f=0.44$ (Section~\ref{sec:halphaemission}, upper panel of Figure~\ref{fig:halphaluminosity}). However, we observe a considerable scatter for fainter galaxies ($\logm < 9.5$) due to the lower S/N of the Spitzer observations.
Thanks to the large sample size of \textit{ALPINE}, we are able, for the first time, to compare the \halpha~luminosity to the \Cii~luminosity (lower panel of Figure~\ref{fig:halphaluminosity}). Overall, we find \halpha~luminosities as expected from the local relation between $L_{\rm [CII]}$ and SFR from \citet[][]{DELOOZE14} (using their fit to the entire sample). However, we find that at low \Cii~luminosities ($< 5\times 10^{8}\,{\rm L_{\odot}}$), the \halpha~luminosities are generally lower than what is predicted by that relation. Instead, the \citeauthor{DELOOZE14} relation derived from a sample of metal-poor dwarf galaxies is a better fit for those galaxies. This might suggest a more complex relation between SFR and \Cii~luminosity driven by metallicity or other properties of the ISM.

\textit{ALPINE} is the beginning of a thorough exploration of galaxies at $z>4$. It builds the foundation onto which future follow-up observations can build on. In fact, several follow-up programs are being granted, some of which are already on going. These include
\textit{(i)} additional HST WFC3/IR observations of interacting \textit{ALPINE} galaxies (PI: Faisst),
\textit{(ii)} follow up observations of \nii~at $205\,{\rm \mu m}$ with ALMA for $9$ \textit{ALPINE} galaxies (PI: Faisst),
\textit{(iii)} high spatial resolution ($\sim0.15\arcsec$) observations of the brightest \textit{ALPINE} galaxies (PI: Ibar), and
\textit{(iv)} the follow up of four serendipitous objects at $z>4$ with NOEMA (PI: Loiacono \& B\'ethermin).
In addition, several JWST proposals are in preparation.

All ancillary data products (including catalogs, images, and spectra) will be made public accessible. In the Appendix~\ref{app:dataproducts}, we detail the layout of the catalogs including the measurements detailed in this paper. In Appendix~\ref{app:additionalfigures}, we show HST cutouts in ACS $F814W$ and WFC3/IR $F160W$ bands as well as the rest-frame UV spectra of all individual \textit{ALPINE} galaxies.

This paper completes a series of three papers presenting the \textit{ALPINE} survey \citep{ALPINE_LEFEVRE19} and the data processing \citep{ALPINE_BETHERMIN19}.

\vspace{1.5cm}
{\footnotesize
\textit{Acknowledgements:} We would like to thank numerous people for the exchange of data without which the \textit{ALPINE} ancillary data paper would not exist. Especially we would like to thank E. Vanzella for helping us gathering the spectra in the ECDFS field and O. Ilbert for useful discussions that improved the SED fitting results. We also thank the anonymous referee for the suggestions that improved this paper.
This paper is based on data obtained with the ALMA Observatory, under Large Program 2017.1.00428.L. ALMA is a partnership of ESO (representing its member states), NSF(USA) and NINS (Japan), together with NRC (Canada), MOST and ASIAA (Taiwan), and KASI (Republic of Korea), in cooperation with the Republic of Chile. The Joint ALMA Observatory is operated by ESO, AUI/NRAO and NAOJ.
This program receives funding from the CNRS national program Cosmology and Galaxies.
This work is based on observations and archival data  made with the Spitzer Space Telescope, which is operated by the Jet Propulsion Laboratory, California Institute of Technology, under a contract with NASA along with archival data from the NASA/ESA Hubble Space Telescope.
This  research  made  also  use of  the  NASA/IPAC  Infrared  Science  Archive  (IRSA), which  is  operated  by  the  Jet  Propulsion  Laboratory, California Institute of Technology, under contract with the National Aeronautics and Space Administration.
In parts  based  on  data  products  from  observations  made with  ESO  Telescopes  at  the  La  Silla  Paranal  Observatory  under  ESO  programme  ID  179.A-2005  and  on data  products  produced  by  TERAPIX  and  the  Cambridge  Astronomy  Survey  Unit  on  behalf  of  the  UltraVISTA  consortium.
Based  on  data  obtained  with the  European  Southern  Observatory  Very  Large  Telescope, Paranal, Chile, under Large Program 185.A-0791, and made available by the VUDS team at the CESAM data center, Laboratoire d'Astrophysique de Marseille, France.
This work is based on observations taken by the 3D-HST Treasury Program (GO 12177 and 12328) with the NASA/ESA HST, which is operated by the Association of Universities for Research in Astronomy, Inc., under NASA contract NAS5-26555.
Furthermore, this work is based on data from the  W.M.  Keck  Observatory  and  the  Canada-France-Hawaii  Telescope,  as  well  as  collected  at  the  Subaru Telescope and retrieved from the HSC data archive system, which  is  operated  by  the  Subaru  Telescope  and Astronomy  Data Center  at the  National Astronomical Observatory  of  Japan.   The  authors  wish  to  recognize and acknowledge the very significant cultural role and reverence  that  the  summit  of  Mauna  Kea  has  always had  within  the  indigenous  Hawaiian  community. We are most fortunate to have the opportunity to conduct observations  from  this  mountain.  
Finally, we would also like to recognize the contributions from all of the members of the COSMOS Team who helped in obtaining and reducing the large amount of multi-wavelength data  that  are  now  publicly  available  through  IRSA  at \url{http://irsa.ipac.caltech.edu/Missions/cosmos.html}.
A.C., F.P., M.T., C.G., and F.L. acknowledge the support from grant PRIN MIUR 2017. G.C.J. acknowledges ERC Advanced Grant 695671 ``QUENCH'' and support by the Science and Technology Facilities Council (STFC).
E.I. acknowledges partial support from FONDECYT through grant No. 1171710.
The Cosmic Dawn Center (DAWN) is funded by the Danish National Research Foundation under grant No. 140.
S.T. acknowledges support from the ERC Consolidator Grant funding scheme (project Context, grant No. 648179). LV acknowledges funding from the European Union's Horizon 2020 research and innovation program under the Marie Sklodowska-Curie Grant agreement No. 746119. D.R. acknowledges support from the National Science Foundation under grant numbers AST-1614213 and AST-1910107 and from the Alexander von Humboldt Foundation through a Humboldt Research Fellowship for Experienced Researchers.

}

\clearpage
\newpage

\appendix
\counterwithin{figure}{section}
\counterwithin{table}{section}

\section{Description of Published Data Products}\label{app:dataproducts}

The data presented in this paper are summarized in three different catalogs. 
\begin{itemize}
\item The \textit{main catalog}, which contains properties consistently measured for all the galaxies. These include general information (such as coordinates, redshifts, selection, morphological class), measurements performed on the spectra (such as \lya~redshift and properties as well as absorption line redshifts), measurements from SED fitting (including UV continuum slopes), and \halpha~line properties and SFRs.
\item The \textit{ECDFS photometry catalog}, which contains all the Galactic extinction corrected total photometry (magnitude, fluxes, and uncertainties) of the galaxies in the ECDFS field. This catalog is based on the \textit{3D-HST} catalog.
\item The \textit{COSMOS photometry catalog}, which contains all the Galactic extinction corrected total photometry (magnitude, fluxes, and uncertainties) of the galaxies in the COSMOS field. This catalog is based on the \textit{COSMOS2015} catalog.
\end{itemize}

The following Tables~\ref{tab:catmain}, \ref{tab:catphotcosmos}, and \ref{tab:catphotgs} summarize the columns of each of these three catalogs.
The catalogs can be downloaded in \textit{FITS} format at 
\begin{center}
\framebox{\alpinedataurl}
\end{center}
Note that the Tables~\ref{tab:catphotgs} and~\ref{tab:catphotcosmos} only show an excerpt of the description of the \textit{ECDFS} and \textit{COSMOS} photometry catalog. The full versions can be found at the link above.

\begin{center}
\begin{longtable}{ p{0.2\columnwidth} p{0.10\columnwidth} p{0.60\columnwidth}}
\caption{Column description of main catalog. Sections of this paper where the measurements are discussed are indicated.}\label{tab:catmain}\\[-0.2cm]
 \hline \hline
 Column & Unit & Description  \\
\hline
\endfirsthead
\multicolumn{3}{c}%
{\tablename\ \thetable\ -- \textit{Continued from previous page}} \\
 \hline \hline
 Column Name & Unit & Description  \\
\hline
\endhead
\hline \multicolumn{3}{r}{\textit{Continued on next page}} \\
\endfoot
\endlastfoot
\\[-0.2cm]
\multicolumn{3}{c}{\textbf{General information and selection}}\\
\texttt{ALPINE\_ID} & \textit{-} & Unique name for each galaxy in string format\\
\texttt{RA} & \textit{degrees} & Right-ascension in J2000 in degrees from either the COSMOS or 3D-HST catalog\\
\texttt{delta\_RA} & \textit{milli-arcsec} & Constant shift (to be added to \texttt{RA}) in right-ascension due to astrometric offset (see Section~\ref{sec:astrometry})\\
\texttt{DEC} & \textit{degrees} & Declination in J2000 in degrees from either the COSMOS or 3D-HST catalog\\
\texttt{delta\_DEC} & \textit{milli-arcsec} & Constant shift (to be added to \texttt{DEC}) in declination due to astrometric offset (see Section~\ref{sec:astrometry})\\
\texttt{field} & \textit{-} & Field name ($1={\rm ECDFS}$ or $2={\rm COSMOS}$)\\
\texttt{selection} & \textit{-} & Original selection. For galaxies in the ECDFS field, this can be \textit{vlt} or \textit{grapes}. For galaxies in COSMOS, possible selections are \textit{CHANDRA}, \textit{LBG}, \textit{NB1}$^a$, \textit{NB2}$^a$, \textit{excess}, \textit{photz}, or \textit{vuds}. See Section~\ref{sec:spectroscopyselection} and Table~\ref{tab:preselection} for details.\\
\texttt{z\_orig} & \textit{-} & Original redshift used for initial selection (this redshift is derived from \lya~or absorption lines).\\
\texttt{z\_cii} & \textit{-} & Redshift determined from FIR \Cii~emission lines \citep[see details in ][]{ALPINE_BETHERMIN19}. Is \textit{-99} if \Cii~is not detected at S/N$>3.5$.\\
\texttt{morph\_class} & \textit{-} & Morpho-kinematic classes from \citet{ALPINE_LEFEVRE19}. Only for galaxies with $>3.5\sigma$ \Cii~detection (else class set to $-99$). The classes are: (1) rotator; (2) pair-merger (major or minor); (3) extended dispersion dominated; (4) compact dispersion dominated; (5) too weak for assigning a class.  \\
%
%
%
%
\\[0.5cm]
\multicolumn{3}{c}{\textbf{Measurements on spectra (Section~\ref{sec:spectroscopy}) }}\\
\texttt{has\_twin} & \textit{-} & A flag set to $1$ of for a galaxy has been observed by Keck/DEIMOS and VUDS (two spectra available). If false, the flag is set to $0$.\\
\texttt{z\_lya} & \textit{-} & Redshift determined from peak of \lya~emission \citep[see details in][]{ALPINE_CASSATA19}. Is \textit{-99} if no redshift measured.\\
\texttt{lya\_ew} & $\AA$ & Observer-frame \lya~emission equivalent \citep[see details in][]{ALPINE_CASSATA19}. Is \textit{-99} if no equivalent width is measured.\\
\texttt{lya\_ew\_err} & $\AA$ & $1\sigma$ uncertainty on observer-frame \lya~emission equivalent \citep[see details in][]{ALPINE_CASSATA19}. Is \textit{-99} if no equivalent width is measured and $-1$ if no continuum measured (i.e., EW is upper limit).\\
\texttt{f\_lya} & $erg/s/cm^2$ & \lya~emission flux \citep[see details in][]{ALPINE_CASSATA19}. Is \textit{-99} if flux is measured.\\
\texttt{flag\_specpro} & \textit{-} & Visual flag for reliability of absorption redshift measurements. Set to $-99$ if not attempted, then $1$, $2$, and $3$ for least, medium, and most robust.\\
\texttt{z\_iswind} & \textit{-} & Redshift determined from IS$+$wind absorption lines (see Section~\ref{sec:uvspectrameasurements}). Set to \textit{-99} if no redshift measured.\\
\texttt{z\_iswind\_low} & \textit{-} & Lower $95\%$ percentile of redshift determined from IS$+$wind absorption lines. Set to \textit{-99} if no redshift measured.\\
\texttt{z\_iswind\_up} & \textit{-} & Upper $95\%$ percentile of redshift determined from IS$+$wind absorption lines. Set to \textit{-99} if no redshift measured.\\
\texttt{n\_lines\_iswind\_used} & \textit{-} & Number of lines used for IS$+$wind redshift measurement. We advice to generally only use galaxies with a value $>2$ together with $\texttt{flag\_specpro}>0$ for a conservative sample selection. \\
\texttt{z\_wind} & \textit{-} & Redshift determined from wind absorption lines (see Section~\ref{sec:uvspectrameasurements}). Set to \textit{-99} if no redshift measured.\\
\texttt{z\_wind\_low} & \textit{-} & Lower $95\%$ percentile of redshift determined from wind absorption lines. Set to \textit{-99} if no redshift measured.\\
\texttt{z\_wind\_up} & \textit{-} & Upper $95\%$ percentile of redshift determined from wind absorption lines. Set to \textit{-99} if no redshift measured.\\
\texttt{n\_lines\_wind\_used} & \textit{-} & Number of lines used for wind redshift measurement. We advice to generally only use galaxies with a value $>0$ together with $\texttt{flag\_specpro}>0$ for a conservative sample selection. \\
%
%
%
%
\\[-0.2cm]
\multicolumn{3}{c}{\textbf{Properties from SED fitting with \texttt{LePhare} (Sections~\ref{sec:sedfitting} and~\ref{sec:uvmags})}}\\
\texttt{ID\_photcat} & \textit{-} & ID in the photometric catalogs. This is the \textit{3D-HST} catalog for galaxies in ECDFS and the \textit{COSMOS2015} catalog for galaxies in COSMOS.\\
\texttt{chi2} & \textit{-} & $\chi^2$ value given by the \texttt{LePhare} fit.\\
\texttt{Nband} & \textit{-} & Number of bands used for SED fitting.\\
\texttt{ebmv} & \textit{mag} & $E(B-V)$ derived from SED fitting.\\
\texttt{logAge} & \textit{yr} & Logarithmic age\\
\texttt{logAge\_loweff1sig} & \textit{yr} & Lower $1\sigma$ limit on age in log\\
\texttt{logAge\_higheff1sig} & \textit{yr} & Upper $1\sigma$ limit on age in log\\
\texttt{logMstar} & $M_{\odot}$ & Logarithmic stellar mass\\
\texttt{logMstar\_loweff1sig} & $M_{\odot}$ & Lower $1\sigma$ limit on stellar mass in log\\
\texttt{logMstar\_higheff1sig} & $M_{\odot}$ & Upper $1\sigma$ limit on stellar mass in log\\
\texttt{logSFR} & $M_{\odot}\,yr^{-1}$ & Logarithmic SFR\\
\texttt{logSFR\_loweff1sig} & $M_{\odot}\,yr^{-1}$ & Lower $1\sigma$ limit on SFR in log\\
\texttt{logSFR\_higheff1sig} & $M_{\odot}\,yr^{-1}$ & Upper $1\sigma$ limit on SFR in log\\
\texttt{logsSFR} & $yr^{-1}$ & Logarithmic sSFR\\
\texttt{logsSFR\_loweff1sig} & $yr^{-1}$ & Lower $1\sigma$ limit on sSFR in log\\
\texttt{logsSFR\_higheff1sig} & $yr^{-1}$ & Upper $1\sigma$ limit on sSFR in log\\
\texttt{M\_FUV} & $mag$ & Absolute rest-frame UV magnitude measured in the GALEX FUV filter (corresponding approximately to rest-frame $1500\,{\textrm \AA}$) \\
\texttt{M\_FUV\_low1sig} & $mag$ & Lower $1\sigma$ limit on absolute rest-frame UV magnitude \\
\texttt{M\_FUV\_high1sig} & $mag$ & Upper $1\sigma$ limit on absolute rest-frame UV magnitude \\
%
%
%
%
%
\\[-0.2cm]
\multicolumn{3}{c}{\textbf{UV continuum slopes ($\beta$) with different dust reddening (Section~\ref{sec:uvslopes}) }}\\
\texttt{beta\_med\_calz} & $-$ & UV slope measured assuming Calzetti dust\\
\texttt{beta\_low1sig\_calz} & $-$ & Lower $1\sigma$ UV slope limit (Calzetti dust)\\
\texttt{beta\_high1sig\_calz} & $-$ & Upper $1\sigma$ UV slope limit (Calzetti dust)\\
\texttt{beta\_med\_smc} & $-$ & UV slope measured assuming SMC dust\\
\texttt{beta\_low1sig\_smc} & $-$ & Lower $1\sigma$ UV slope limit (SMC dust)\\
\texttt{beta\_high1sig\_smc} & $-$ & Upper $1\sigma$ UV slope limit (SMC dust)\\
\texttt{beta\_med\_comb} & $-$ & UV slope measured by marginalizing over Calzetti and SMC dust\\
\texttt{beta\_low1sig\_comb} & $-$ & Lower $1\sigma$ UV slope limit (Calzetti$+$SMC dust)\\
\texttt{beta\_high1sig\_comb} & $-$ & Upper $1\sigma$ UV slope limit (Calzetti$+$SMC dust)\\
%
%
%
%
%
\\[-0.2cm]
\multicolumn{3}{c}{\textbf{\halpha~measurements from Spitzer colors (using Model A, see Section~\ref{sec:halphaemission})}}\\
\texttt{spitzer\_cont} & $-$ & Spitzer photometry contamination flag. Set to $0$, $1$, and $2$ for no, slight, and heavy contamination, respectively.\\
\texttt{ewha\_med} & $\AA$ & Rest-frame \halpha~equivalent width assuming \citet{CALZETTI00} dust attenuation and $f=0.44$ \\
\texttt{ewha\_low} & $\AA$ & Lower $1\sigma$ limit of rest-frame \halpha~equivalent width \\
\texttt{ewha\_up} & $\AA$ & Upper $1\sigma$ limit of rest-frame \halpha~equivalent width \\
\texttt{log\_halum\_med} & $erg\,s^{-1}$ & Logarithmic \halpha~luminosity assuming \citet{CALZETTI00} dust attenuation and $f=0.44$ \\
\texttt{log\_halum\_low} & $erg\,s^{-1}$ & Lower $1\sigma$ limit of \halpha~luminosity in log \\
\texttt{log\_halum\_up} & $erg\,s^{-1}$ & Upper $1\sigma$ limit of \halpha~luminosity in log \\
\texttt{log\_sfrha\_med} & $M_{\odot}\,yr^{-1}$ & Logarithmic SFR based on \halpha~luminosity. Derived assuming \citep{KENNICUTT98} (solar metallicity), \citet{CALZETTI00} dust attenuation, and $f=0.44$ \\
\texttt{log\_sfrha\_low} & $M_{\odot}\,yr^{-1}$ & Lower $1\sigma$ limit of \halpha~based SFR in log \\
\texttt{log\_sfrha\_up} & $M_{\odot}\,yr^{-1}$ & Upper $1\sigma$ limit of \halpha~based SFR log \\[0.2cm]
%
%
%
\hline
\multicolumn{3}{l}{
$^a$Note that \textit{NB1} and \textit{NB2} stand for the narrow-band selection of galaxies at $z\sim4.5$ and $z\sim5.7$, respectively.
}\\
\end{longtable}
\end{center}

\begin{center}
\begin{longtable}{ p{0.2\columnwidth} p{0.10\columnwidth} p{0.60\columnwidth}}
\caption{Excerpt of the column description of the photometry catalog for galaxies in the ECDFS field (Section~\ref{sec:photometrygs}). Wavelengths, depths, and references are given in Table~\ref{tab:photogoods}.}\label{tab:catphotgs}\\[-0.2cm]
 \hline \hline
 Column & Unit & Description  \\
\hline
\endfirsthead
\multicolumn{3}{c}%
{\tablename\ \thetable\ -- \textit{Continued from previous page}} \\
 \hline \hline
 Column Name & Unit & Description  \\
\hline
\endhead
\hline \multicolumn{3}{r}{\textit{Continued on next page}} \\
\endfoot
\endlastfoot
%
\texttt{ALPINE\_ID} & \textit{-} & Unique name for each galaxy in string format\\
\texttt{id\_3dhst} & \textit{-} & Unique identification number in the \textit{3D-HST} catalog \\
\texttt{ra\_3dhst} & \textit{degrees} & Right-ascension as given in \textit{3D-HST}  catalog\\
\texttt{dec\_3dhst} & \textit{degrees} &  Declination as given in \textit{3D-HST} catalog\\
%
%
%
\\[-0.2cm]
\multicolumn{3}{c}{\textbf{Galactic extinction corrected total fluxes with $1\sigma$ uncertainty}}\\
\texttt{f\_f160w} & $\mu Jy$ & HST/WFC3 $F160$ flux and uncertainty\\
\texttt{e\_f160w} & $\mu Jy$ & .\\
... & ... & ...\\
%
%
\\[-0.2cm]
\multicolumn{3}{c}{\textbf{Galactic extinction corrected total magnitudes with $1\sigma$ uncertainty}}\\
\multicolumn{3}{c}{\textit{Note: given are $1\sigma$ limits (and magnitude uncertainties are set to $-1$) if fluxes are smaller than $1\sigma$ flux uncertainties.}}\\
\texttt{mag\_f160w} & \textit{mag} & HST/WFC3 $F160$ magnitude and error\\
\texttt{magerr\_f160w} & \textit{mag} & .\\
... & ... & ...\\
%
%
\hline
\end{longtable}
\end{center}

\begin{center}
\begin{longtable}{ p{0.2\columnwidth} p{0.10\columnwidth} p{0.60\columnwidth}}
\caption{Excerpt of the column description of the photometry catalog for galaxies in the COSMOS field  (Section~\ref{sec:photometrycosmos}). Wavelengths, depths, and references are given in Table~\ref{tab:photocosmos}.}\label{tab:catphotcosmos}\\[-0.2cm]
 \hline \hline
 Column & Unit & Description  \\
\hline
\endfirsthead
\multicolumn{3}{c}%
{\tablename\ \thetable\ -- \textit{Continued from previous page}} \\
 \hline \hline
 Column Name & Unit & Description  \\
\hline
\endhead
\hline \multicolumn{3}{r}{\textit{Continued on next page}} \\
\endfoot
\endlastfoot
%
%
\texttt{ALPINE\_ID} & \textit{-} & Unique name for each galaxy in string format\\
\texttt{id\_cosmos15} & \textit{-} & Unique identification number in the \textit{COSMOS2015} catalog\\
\texttt{ra\_cosmos15} & \textit{degrees} & Right-ascension as given in the \textit{COSMOS2015} catalog\\
\texttt{dec\_cosmos15} & \textit{degrees} & Declination as given in the \textit{COSMOS2015} catalog\\
%
%
%
%
\\[-0.2cm]
\multicolumn{3}{c}{\textbf{Galactic extinction corrected total fluxes with $1\sigma$ uncertainty}}\\
\texttt{Ks\_FLUX\_APER3} & $\mu Jy$ & CFHT/WIRCam $K_{\rm s}$-band flux and uncertainty \\
\texttt{Ks\_FLUXERR\_APER3} & $\mu Jy$ & .\\
... & ... & ...\\
%
%
\\[-0.2cm]
\multicolumn{3}{c}{\textbf{Galactic extinction corrected total magnitudes with $1\sigma$ uncertainty}}\\
\multicolumn{3}{c}{\textit{Note: given are $1\sigma$ limits (and magnitude uncertainties are set to $-1$) if fluxes are smaller than $1\sigma$ flux uncertainties.}}\\
\texttt{Ks\_MAG} & \textit{mag} & CFHT/WIRCam $K_{\rm s}$-band magnitude and uncertainty\\
\texttt{Ks\_MAGERR} & \textit{mag} & .\\
... & ... & ...\\
%
%
\hline
\end{longtable}
\end{center}

\newpage
\section{Additional Figures}\label{app:additionalfigures}

In the following, we show the imaging and spectroscopic data for the individual \textit{ALPINE galaxies}.

\begin{figure*}[t!]
\includegraphics[width=1.0\columnwidth, angle=0]{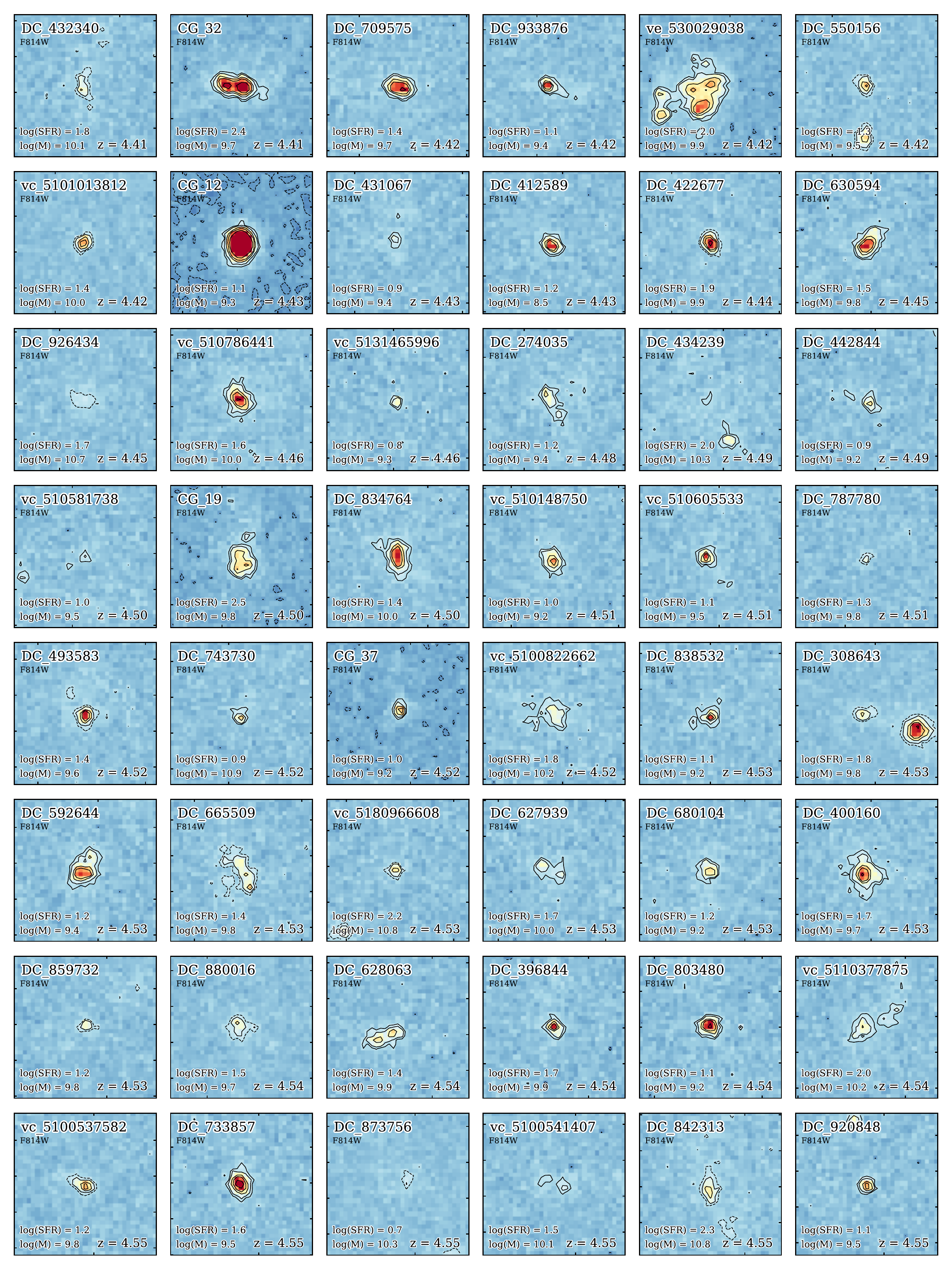}
\caption{$F814W$ cutouts sorted by redshift (part 1). The dashed contours show $-3\sigma$ levels and the solid contours show $3\sigma$, $5\sigma$, $10\sigma$, $15\sigma$, and $30\sigma$ levels. All cutouts are $2\arcsec$ on each side. \label{fig:f814w_cutouts1}}
\end{figure*}

\begin{figure*}[t!]
\includegraphics[width=1\columnwidth, angle=0]{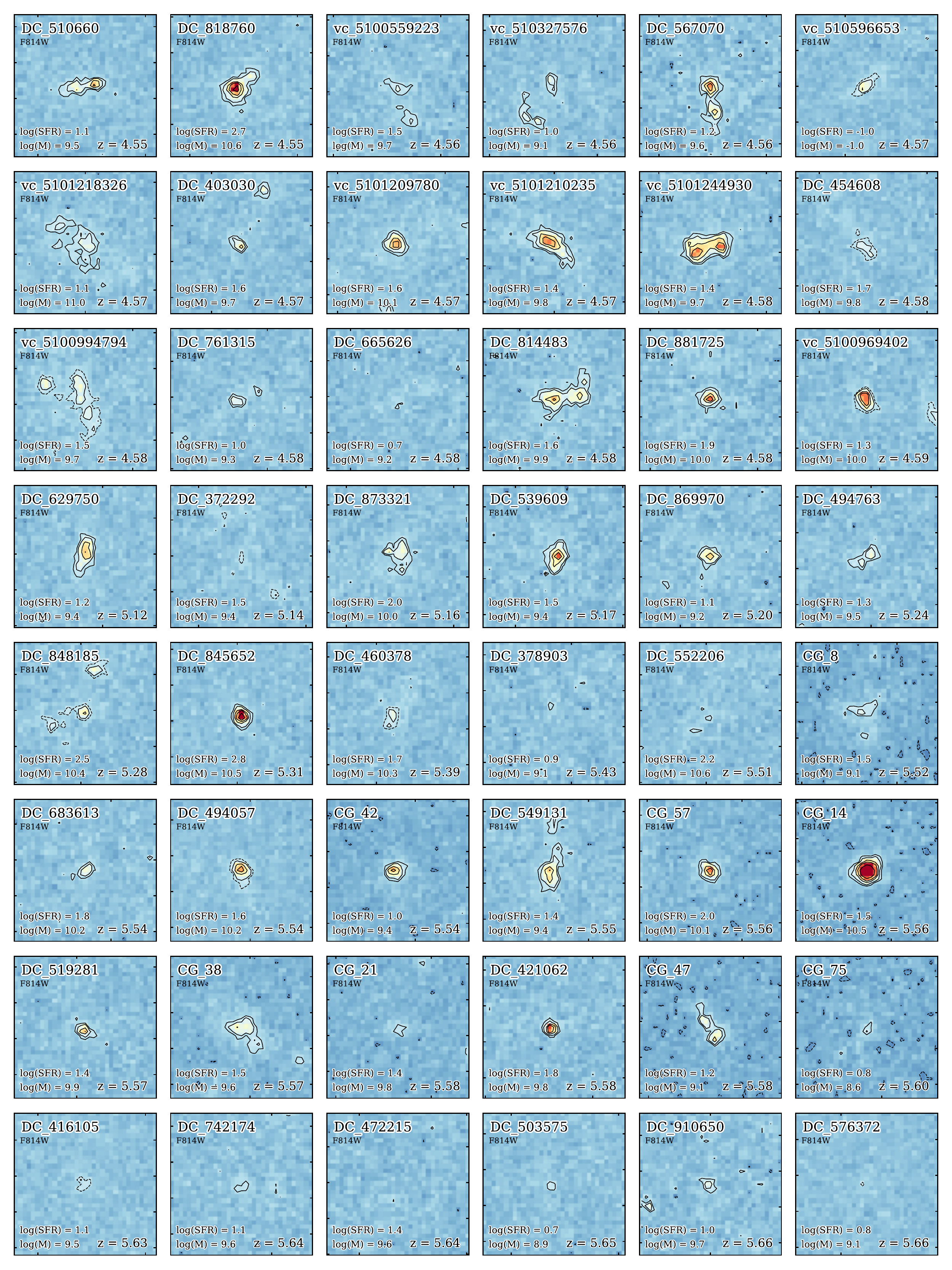}
\caption{$F814W$ cutouts sorted by redshift (part 2).  \label{fig:f814w_cutouts2}}
\end{figure*}

\begin{figure*}[t!]
\includegraphics[width=1\columnwidth, angle=0]{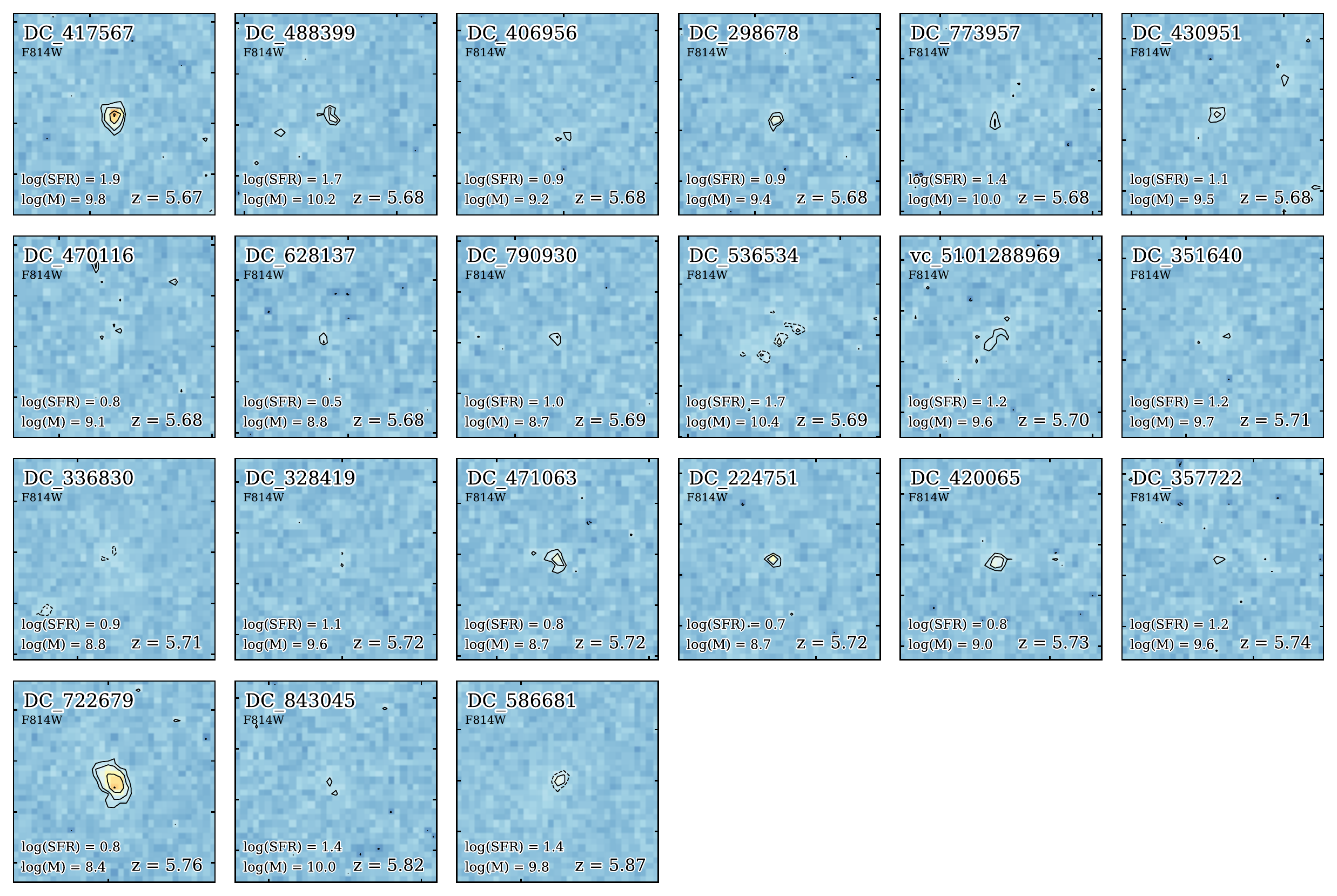}
\caption{$F814W$ cutouts sorted by redshift (part 3).  \label{fig:f814w_cutouts3}}
\end{figure*}

\begin{figure*}[t!]
\includegraphics[width=1\columnwidth, angle=0]{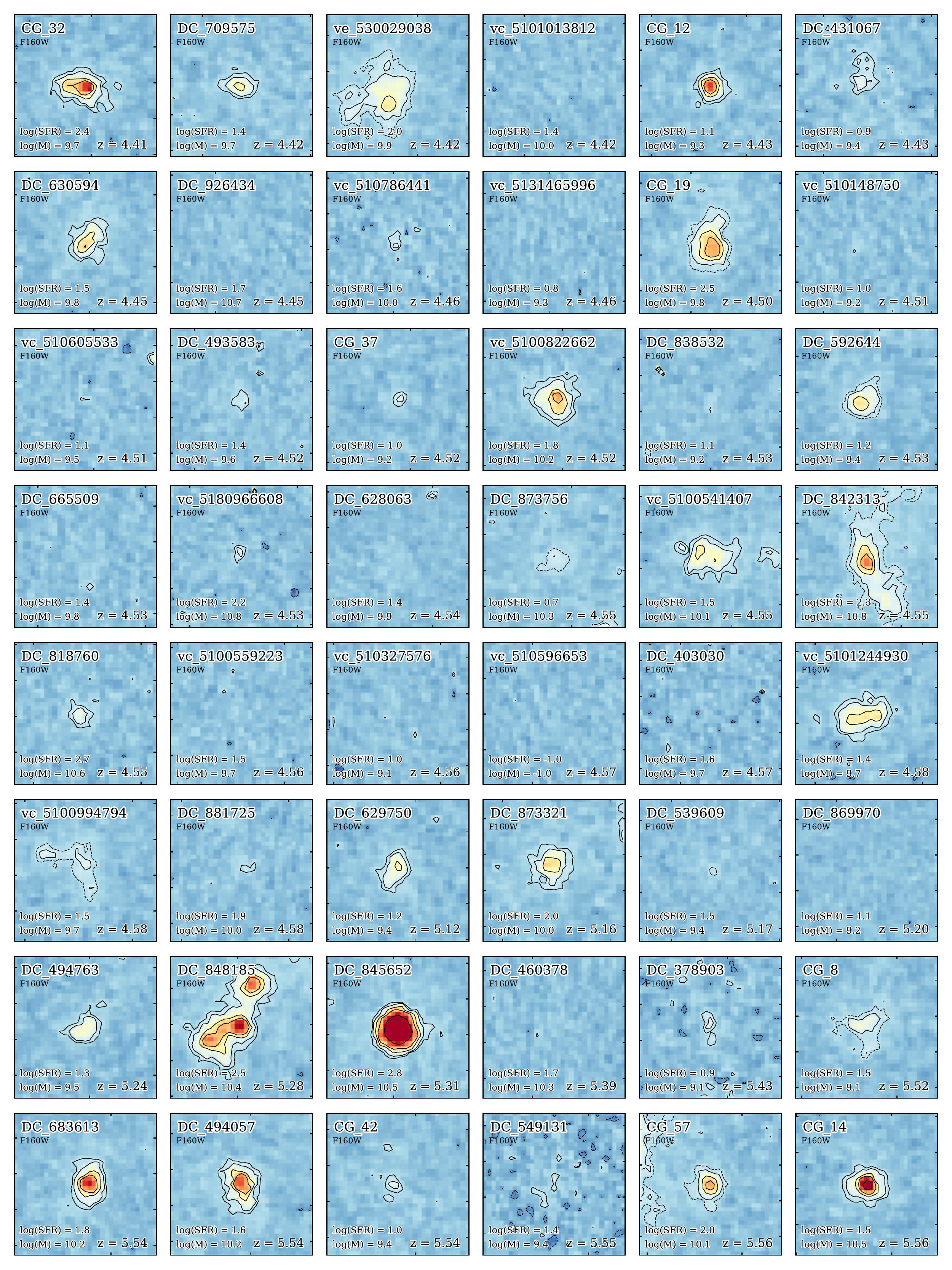}
\caption{$F160W$ cutouts sorted by redshift (part 1). The dashed contours show $-3\sigma$ levels and the solid contours show $3\sigma$, $5\sigma$, $10\sigma$, $15\sigma$, and $30\sigma$ levels.  All cutouts are $2\arcsec$ on each side. \label{fig:f160w_cutouts1}}
\end{figure*}

\begin{figure*}[t!]
\includegraphics[width=1\columnwidth, angle=0]{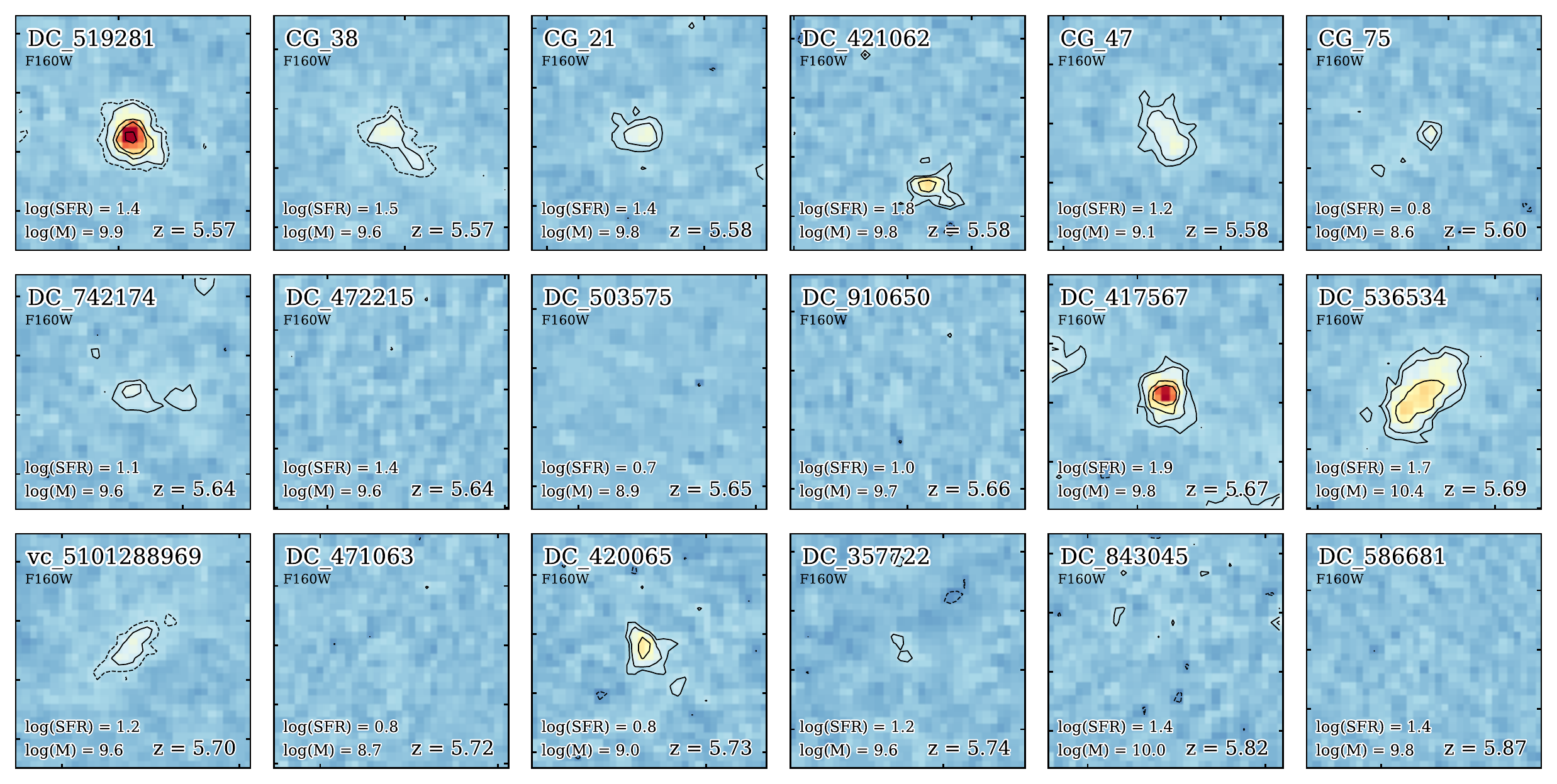}
\caption{$F160W$ cutouts sorted by redshift (part 2).  \label{fig:f160w_cutouts2}}
\end{figure*}

Figures~\ref{fig:f814w_cutouts1} through~\ref{fig:f814w_cutouts3} show $2\arcsec\times 2\arcsec$ cutouts of the HST \textit{F814W} band of each of the galaxies, sorted by increasing redshift. The redshift, stellar mass, and SFR is indicated. The dashed contours show $-3\sigma$ levels and the solid contours show $3\sigma$, $5\sigma$, $10\sigma$, $15\sigma$, and $30\sigma$ levels. The cutouts are oriented such that north is up and east is to the left.
Similarly, Figures~\ref{fig:f160w_cutouts1} and~\ref{fig:f160w_cutouts2} show the all the available HST $F160W$ data for the \textit{ALPINE} galaxies as of October 2019. The cutout size and the drawn $\sigma$-levels are the same as in the previous figures. Note that the HST program \textit{DASH} covers most of the galaxies, however, only a fraction is detected due to the low depth of these observations.

\begin{figure*}[t!]
\includegraphics[width=1.25\columnwidth, angle=90]{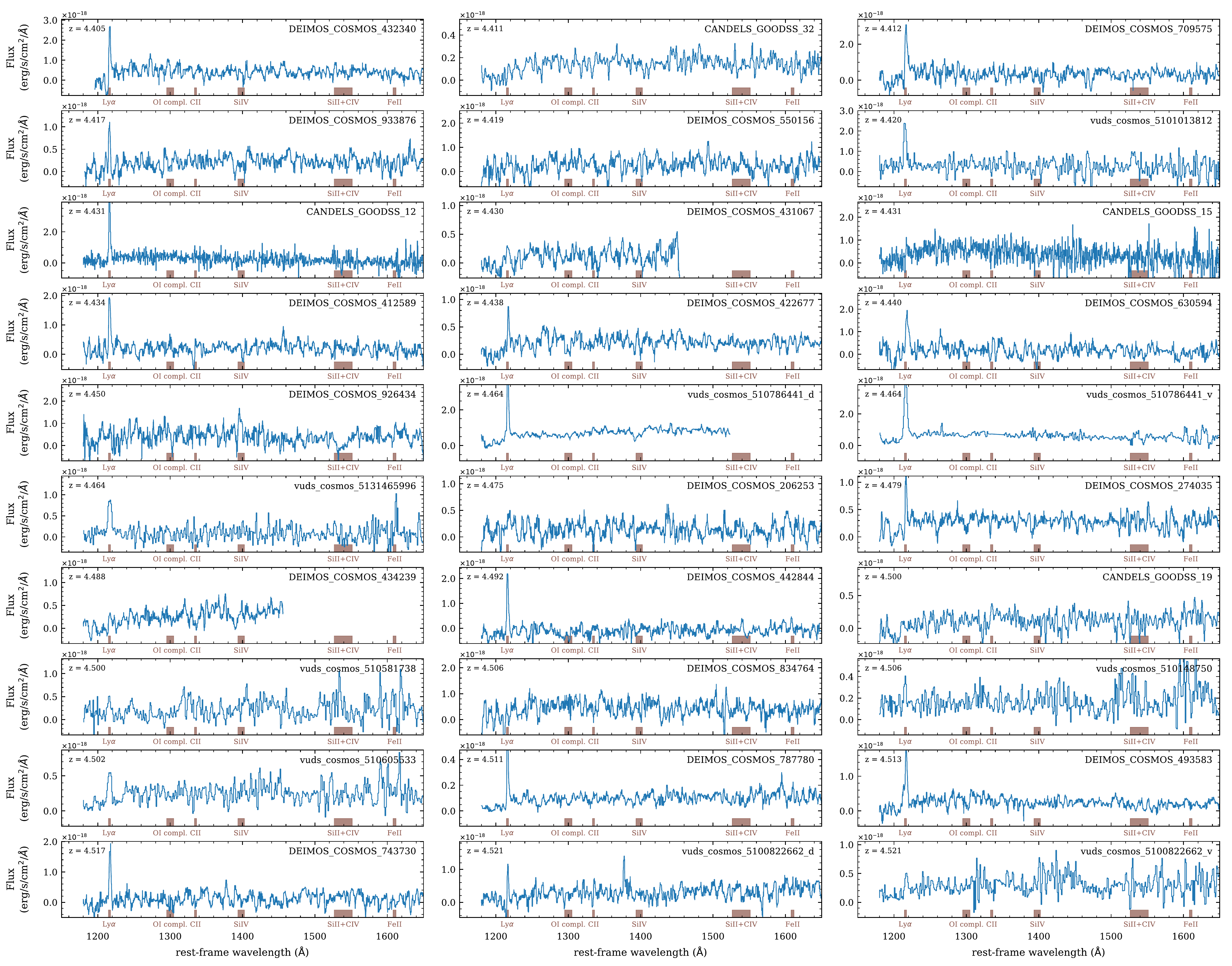}
\caption{Rest-frame UV spectra of all galaxies sorted by redshift (part 1).  \label{fig:allspectra1}}
\end{figure*}

\begin{figure*}[t!]
\includegraphics[width=1.25\columnwidth, angle=90]{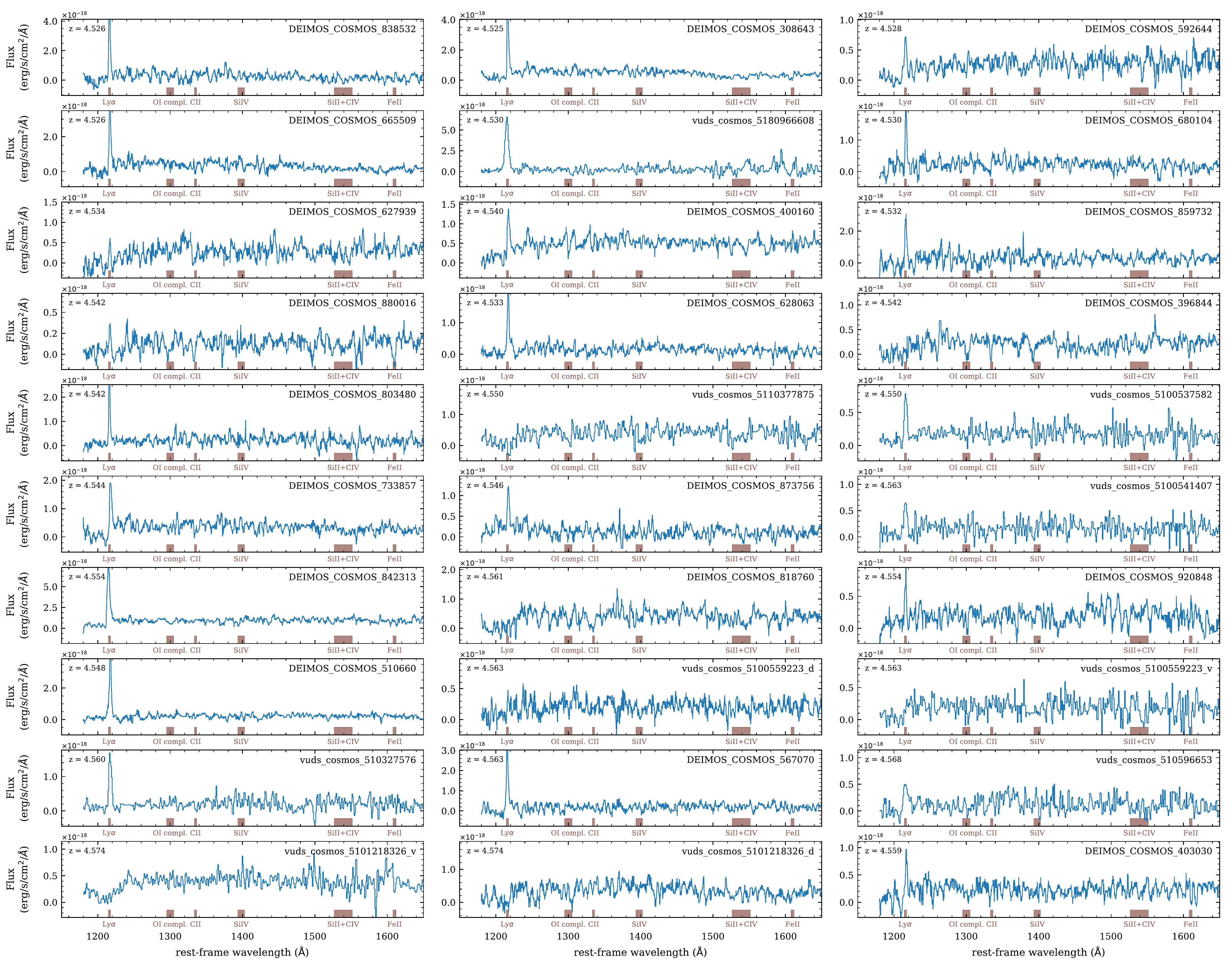}
\caption{Rest-frame UV spectra of all galaxies sorted by redshift (part 2).  \label{fig:allspectra2}}
\end{figure*}

\begin{figure*}[t!]
\includegraphics[width=1.25\columnwidth, angle=90]{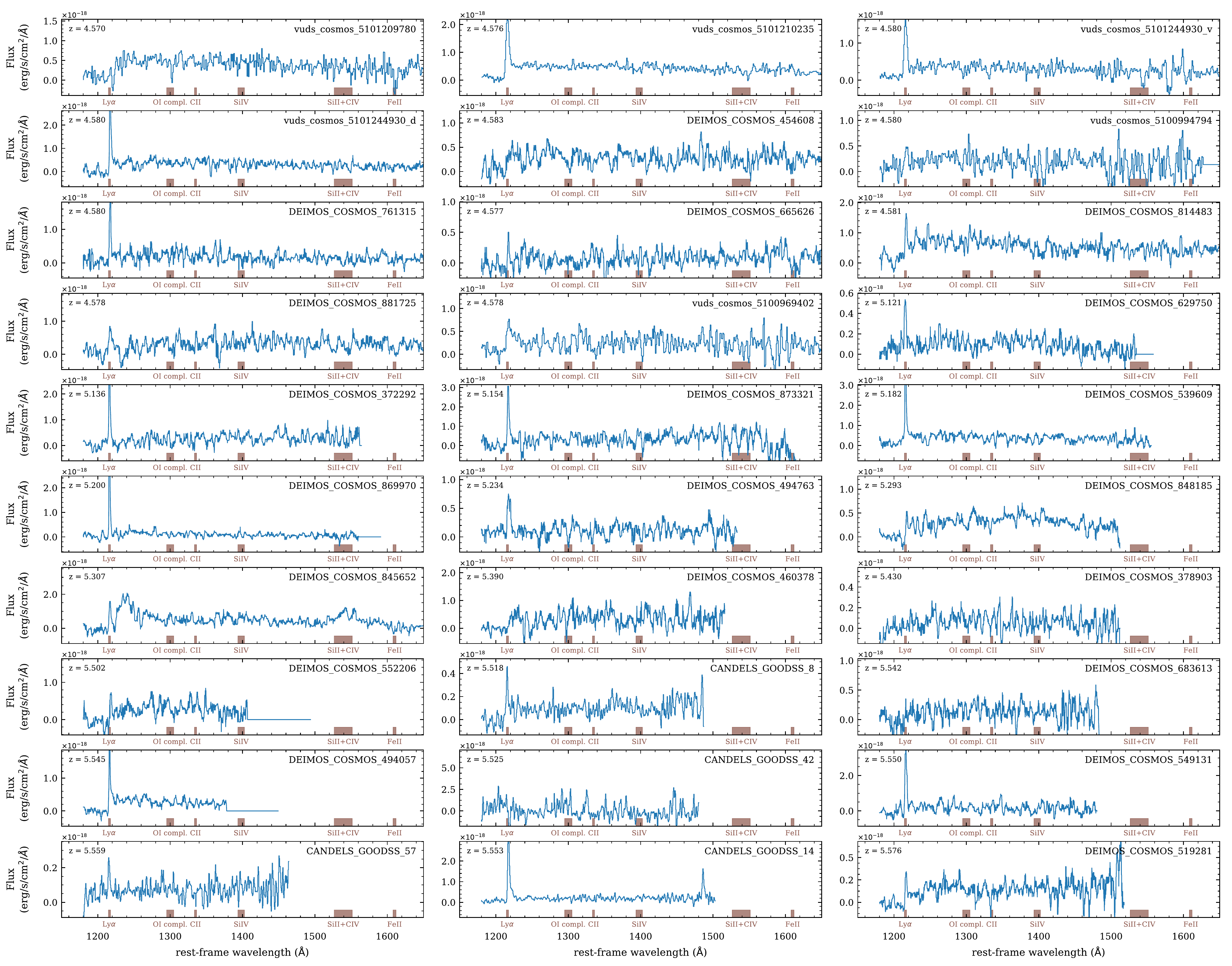}
\caption{Rest-frame UV spectra of all galaxies sorted by redshift (part 3).  \label{fig:allspectra3}}
\end{figure*}

\begin{figure*}[t!]
\includegraphics[width=1.25\columnwidth, angle=90]{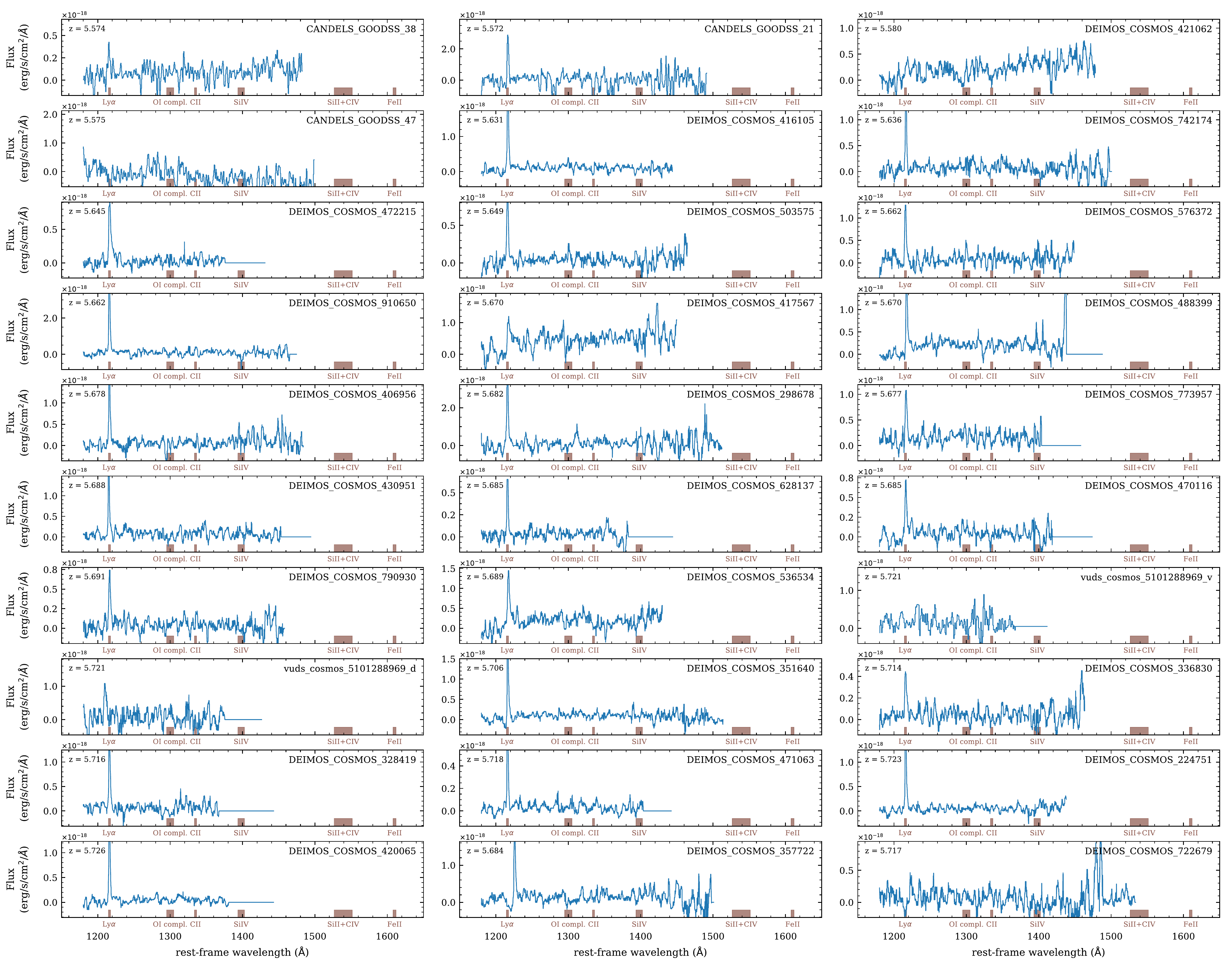}
\caption{Rest-frame UV spectra of all galaxies sorted by redshift (part 4).  \label{fig:allspectra4}}
\end{figure*}

\begin{figure*}[t!]
\includegraphics[width=1.25\columnwidth, angle=90]{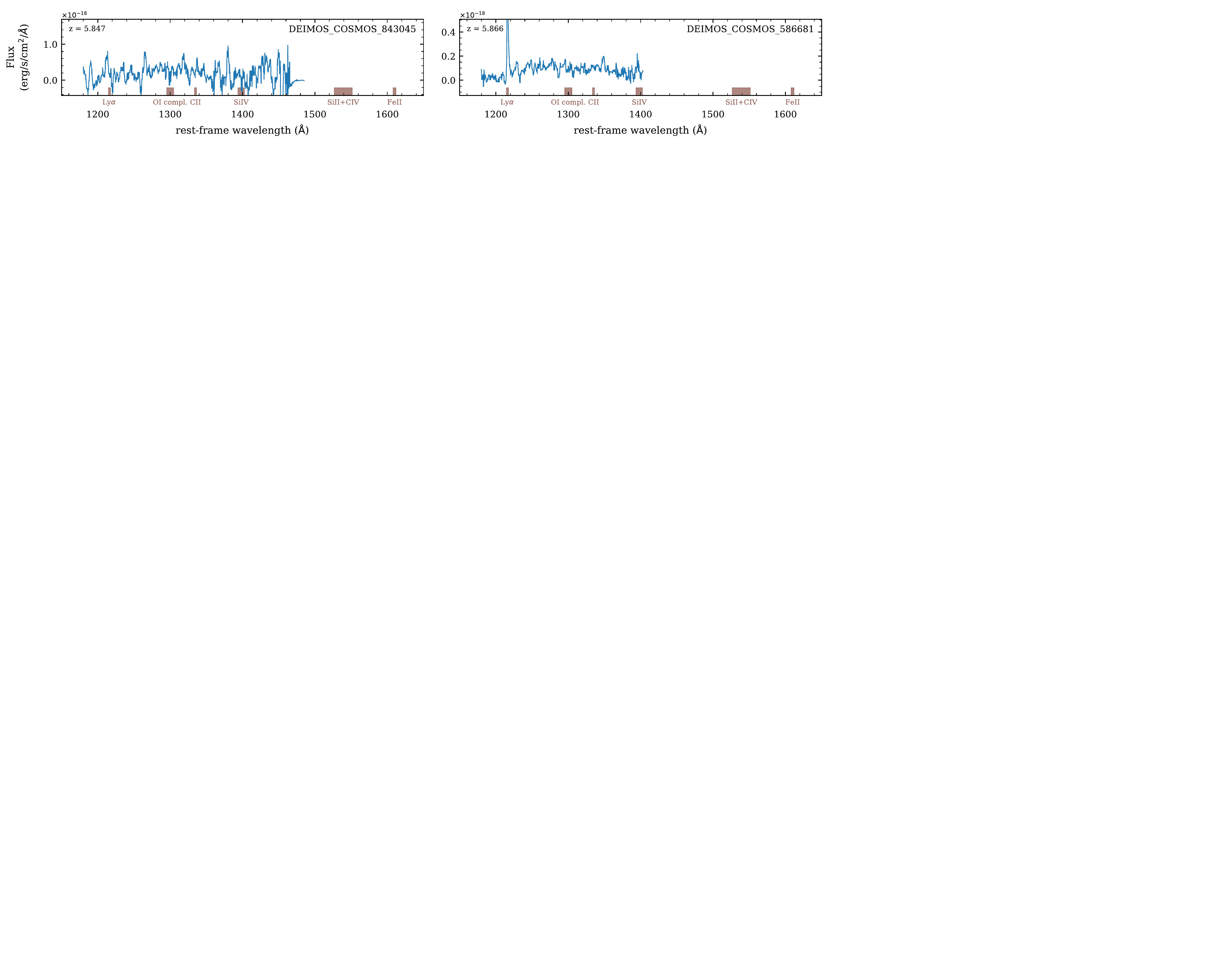}
\caption{Rest-frame UV spectra of all galaxies sorted by redshift (part 5).  \label{fig:allspectra5}}
\end{figure*}

Figures~\ref{fig:allspectra1} through~\ref{fig:allspectra5} show the rest-frame UV spectra for each \textit{ALPINE} galaxy sorted by increasing redshift. The spectra are smoothed with a Savitzky-Golay filter of size $2\,{\textrm \AA}$. Prominent emission lines as well as individual absorption lines and absorption line complexes are indicated by the dark-red bars (compare to Section~\ref{sec:velocity}). For some of the galaxies in COSMOS a spectrum obtained by VUDS and Keck/DEIMOS is available. The names of these galaxies have an appended ``\_v'' or ``\_d'', respectively. Note their different spectral resolution.







\bibliographystyle{aasjournal}
\bibliography{bibli.bib}




\end{document}